\documentclass[12pt]{article}
\usepackage{graphicx,subfigure,float,xcolor}
\usepackage{amsmath,amsthm,amssymb,bm}
\usepackage[round]{natbib}
\usepackage{enumitem}
\usepackage{multirow}
\usepackage{placeins}
\newcommand{\suit}[1]{\left( #1\right)}
\newcommand{\msuit}[1]{\left[ #1\right]}
\newcommand{\set}[1]{\left\{ #1\right\}}
\newcommand{\abs}[1]{\left| #1\right|}
\newcommand{\mI}{\mathcal{I}}

\newcommand{\bE}{\mathbb{E}}
\newcommand{\Var}{\textrm{Var}}
\newcommand{\Cov}{\textrm{Cov}}
\newcommand{\hatgamma}{\widehat{\gamma}}

\newcommand{\hatk}{\widehat{k}}

\newcommand{\mbX}{\boldsymbol{X}}
\newcommand{\mbY}{\boldsymbol{Y}}
\newcommand{\mbT}{\boldsymbol{T}}

\newcommand{\bolSigma}{\boldsymbol{\Sigma}}

\newcommand{\bolzeta}{\boldsymbol{\zeta}}

\newcommand{\tV}{\overline{V}}
 \newcommand{\tW}{\overline{W}}

\usepackage{url} 

\newcommand{\blind}{1}

\theoremstyle{plain}
\newtheorem{lemma}{Lemma}
\newtheorem{prop}{Proposition}
\newtheorem{theorem}{Theorem}

\theoremstyle{remark}
\newtheorem{remark}{Remark}

\begin{document}

\def\spacingset#1{\renewcommand{\baselinestretch}%
{#1}\small\normalsize} \spacingset{1}


\if1\blind
{
  \title{\bf High-dimensional inference for extreme value indices}
  \author{Liujun Chen \\
  Faculty of Business for Science and Technology, School of Management,\\ University of Science and Technology of China\\
    and \\
    Chen Zhou \\
    Econometric Institute, Erasmus University Rotterdam}
  \maketitle
} \fi

\if0\blind
{
  \bigskip
  \bigskip
  \bigskip
  \begin{center}
    {\LARGE\bf High-dimensional inference for extreme value indices}
\end{center}
  \medskip
} \fi

\bigskip
\begin{abstract}
  When applying multivariate extreme value statistics to analyze tail risk in compound events defined by a multivariate random vector, one often assumes that all dimensions share the same extreme value index. While such an assumption can be tested using a Wald-type test, the performance of such a test deteriorates as the dimensionality increases. 
    
 This paper introduces novel tests for comparing extreme value indices in high-dimensional settings, under both weak and general cross-sectional tail dependence.
We establish the asymptotic behavior of the proposed tests.
The proposed tests significantly outperform existing methods in high-dimensional scenarios in simulations.
We demonstrate real-life applications of the proposed tests for 
 two datasets previously assumed to have identical extreme value indices across all dimensions.
\end{abstract}

\noindent%
\vfill

\newpage
\spacingset{1.9} 

\sloppypar
\section{Introduction}

To analyze tail risks of compound events, i.e., an extreme event related to multiple dependent random variables, multivariate extreme value statistics provides a set of tools for modeling the tail region of a multivariate random vector. When the data are heavy tailed, the tail of each marginal distribution can be approximated by a Pareto distribution. The behavior of the tail is governed by the shape parameter of the Pareto distribution, which is commonly referred to as the extreme value index.

To simplify the multivariate model, in various domain science, it is commonly assumed that all marginal distributions share the same extreme value index. This assumption is foundational to several theoretical models, for example, the multivariate regular variation model proposed and applied in \cite{resnick2008extreme}, \cite{cai2011estimation} and  \cite{mainik2015portfolio}. In addition, this assumption is also adopted by spatial extremes models applied to meteorological extremes, see, e.g. \cite{buishand2008spatial}, \cite{fuentes2013nonparametric}, and \cite{Hector2023dis}. Such a maintained assumption, the equal extreme value indices hypothesis, needs to be tested before applying the aforementioned models.

The classical method for testing the equal extreme value indices hypothesis is via Wald-type tests, by combining the estimates of the extreme value indices for all dimensions. These tests enjoy favorable properties when the dimensionality of the data is low, see, e.g., \cite{kinsvater2016regional} and  \cite{daouia2023optimal}. However,  Wald-type tests exhibit unsatisfactory performance in high-dimensional scenarios, see, for example, our simulation study in Section \ref{sec:simu}.

Testing the equal extreme value indices hypothesis under a high-dimensional setting is therefore an important validation step before applying existing models with this maintained assumption to such data. For instance, \cite{kiriliouk2022estimating} estimated the probability of a multivariate ``failure set'' for the maximal wind speeds across all stations in the Netherlands; \cite{mainik2015portfolio} constructed an investment portfolio based on multivariate regular variation model using daily returns of the S\&P 500 stocks. All these applied studies assume equal extreme value indices across a large number of dimensions without a rigorous test.

In the field of high-dimensional statistics, it is known that traditional statistical methods, originally designed in a low dimensional context, often prove inadequate when applied to high-dimensional data. For instance, the  literature on the multivariate mean tests provide new testing methods in high-dimensional settings; see, e.g.,  \cite{tony2014two}, \cite{chang2017simulation} and \cite{giessing2023bootstrap}. We refer interested readers to \cite{Huang2022review} for a recent review of the mean tests problem in high-dimensional settings.

The ``dimensionality curse'' is more of a concern in extreme value statistics than in classical statistical problems such as the mean tests. Denote the dimensionality of the data as $p$ and the sample size of the data as $n$. High-dimensional statistics consider situations where $p=p(n)\to\infty$ as $n\to\infty$, sometimes allowing for $ \log(p)=O( n^{c})$ for some $0<c<1$, see e.g. \cite{fan2008sure} and \cite{wang2012quantile}. In extreme value statistics, the effective sample size, i.e. the number of observations used for estimation, is  much lower than $n$. For instance, in the peak-over-threshold approach, only the top $k$ observations are used; in the block-maxima approach, when considering disjoint blocks, $k$ block maximas are used. Theoretically, it is often required that $k:=k(n)$ satisfies $k\to\infty$ and $k/n\to 0$ as $n\to\infty$. When considering the ``dimensionality curse'', it is about comparing $p$ with the effective sample size $k$.   Such a situation urges to have suitable statistical inference methods in high-dimensional extremes.

In this paper,  we propose  novel testing procedures for  comparing  extreme value indices in a high-dimensional setting. 
 Consider independent and identically   distributed (i.i.d.) observations
$\mbX_1=(X_{1}^{(1)},\dots, X_{1}^{(p)})$, $\dots$, $\mbX_n=(X_{n}^{(1)},\dots, X_{n}^{(p)})$ drawn from a multivariate distribution function $F$ with  marginal distributions $F_1$, $\dots$, $F_p$. For all $j=1,\ldots, p$, assume that the distribution $F_j$ is heavy-tailed, i.e., there exists extreme value index $\gamma_j>0$ such that, for $x>0$, 
  $$
  \lim_{t\to\infty}\frac{1-F_j(tx)}{1-F_j(t)} = x^{-1/\gamma_j}, \quad j=1,\dots,p.
   $$
The first goal of this paper is to test the null hypothesis
$$
H_0:\gamma_j = \gamma_{j}^0, \quad \text{for all} \ j=1,\dots,p, 
$$  
where $\gamma_j^0$'s are prespecified positive values.  
    Additionally, we  extend our test procedure to 
 test whether the extreme value indices are identical across 
$p$ random variables,   that is,
$$
H_0^*:\gamma_1 = \cdots = \gamma_p,
$$
where the common extreme value index is not prespecified.

Our novel testing procedure is inspired  by    high-dimensional mean tests, with two major differences. Firstly, our analysis addresses a characteristic of the tail of marginal distributions, which differs largely from moderate level characteristics such as the mean. Secondly, our test procedure is based on estimating all marginal extreme value indices using the Hill estimator \citep{hill1975simple}. Unlike the sample mean, this estimator involves averaging the logarithms of order statistics, which are neither independent nor identically distributed. This complexity calls for novel proofs in  establishing the asymptotic theory of the test statistic in  high-dimensional settings.

There is a growing literature in studying extremes in high-dimensional settings. Recent advances in this area have primarily focused on modeling the sparse tail dependence structure of high-dimensional random vectors. For example, previous work has explored graphical modeling \citep{engelke2021learning, engelke2022structure, wan2023graphical, lederer2023extremes, engelke2025Extremal}, clustering analysis \citep{boulin2025high}, and principal component analysis \citep{butsch2025estimation}. These studies do not investigate  marginal behavior of the underlying high-dimensional random vector.  Other studies have addressed tail regression problems in high dimensions \citep{sasaki2024high, tang2024high}, where the focus is on univariate tail with high-dimensional covariates.

 In contrast, our focus lies in the marginal tail behavior of high-dimensional random vectors. 
 To the best of our knowledge,  there are  no existing methods tailored to address the testing problem associated with hypotheses $H_0$ or $H_0^*$ within  high-dimensional settings. The present paper reports  a first attempt on high-dimensional inference for extreme value indices,  with providing a few technical tools that can be used in future research. Different from  most studies in high-dimensional extremes, our tests allow for both sparse and  non-sparse cross-sectional tail dependence.

The rest of the paper is organized as follows. 
In Section \ref{sec:methodologies}, we introduce the test procedures for $H_0$. Section \ref{sec:equal:index} presents the test procedures for $H_0^*$.
A simulation study is carried out in Section \ref{sec:simu}. A real data
application is given in Section \ref{sec:application}. All the technical proofs are gathered in the Supplementary Material.

Throughout the paper, $a(t) \asymp b(t)$ means that both $\abs{a(t)/b(t)}$ and $\abs{b(t)/a(t)}$ are $O(1)$ as $t\to\infty$.  

\section{Test for $H_0$}\label{sec:methodologies}


To test the null hypothesis $H_0$, we  first construct a test statistic based on the well-known Hill estimator \citep{hill1975simple}, which is widely used for estimating positive extreme value indices. For each dimension $j \in \set{1,\dots,p}$, 
let $k_j$ be an intermediate sequence $k_j:= k_j(n)$ such that $k_j\to\infty$ and $k_j/n\to 0$ as $n\to\infty$.  The Hill estimator is defined as  
    \begin{equation}\label{eq:hill}
        \widehat{\gamma}_j(k_j):=\frac{1}{k_j}\sum_{i=1}^{k_j}\log \frac{X_{n-i+1,n}^{(j)}}{X_{n-k_j,n}^{(j)}},
    \end{equation}
       where $X_{1,n}^{(j)} \le \cdots \le  X_{n,n}^{(j)}$ are the order statistics of $\set{X_{1}^{(j)}, \dots, X_{n}^{(j)}}$. 
We introduce the test statistic 
$$
\mbT(k_1,\dots,k_p) = \max_{1\le j\le p} \sqrt{k_j}\abs{\frac{\hatgamma_j(k_j)}{\gamma_{j}^0}-1}.
$$
To establish the asymptotic behavior of the test statistic $\mbT(k_1,\dots,k_p)$ under $H_0$, we   assume some regularity conditions.  
\begin{enumerate}[label=(A)]
        \item\label{condition:SOC} There exist constants $\rho_j<0$  and  eventually positive or negative functions $A_j$, $j=1,\dots,p$,  such that as $t\to\infty$, $A_j(tx)/A_j(t)= x^{\rho_j}(1+o(1))$ uniformly for all $x>1$ and 
        $$
        \sup_{x>1}\abs{\frac{U_j(tx)}{U_j(t)} x^{-\gamma_j} - 1}=O\suit{1} A_j(t),
        $$
        where the $O(1)$ terms are uniform for $1\le j\le p$. 
    Here, $U_j(x) = F_j^{\leftarrow}(1-1/x)$ with $^\leftarrow$ denoting the left-continuous inverse function.
    \end{enumerate}
    Condition \ref{condition:SOC} is a typical second order condition in extreme value analysis to control the biases of  Hill estimators $\widehat{\gamma}_j$, see e.g. Chapter 2 of \cite{haan2006extreme}.
     In the high-dimensional setting, we require that the second-order conditions hold uniformly for $1\le j\le p$.
    The Condition \ref{condition:SOC} is equivalent to 
    \begin{equation}\label{eq:F:A}
      \sup_{x>1}\abs{\frac{1-F_j(tx)}{1-F_j(t)}x^{1/\gamma_j}-1} = O\suit{1}A_j\suit{\frac{1}{1-F_j(t)}},
    \end{equation}
    uniformly for $1\le j\le p$;
    see e.g. Theorem 2.3.9 of \cite{haan2006extreme}.

\subsection{Test under weak dependence}
In this subsection, we impose a sparsity condition on the tail dependence structure of $\mbX$, under which we show that the test statistic converges to a Gumbel distribution after proper normalization. 

 \begin{enumerate}[label=(B)]
            \setcounter{enumi}{0}
            \item\label{Eigens} 
        Define $\mbY = \suit{ Y^{(1)},\dots,Y^{(p)}}^\top$, where
$$
    Y^{(j)} = \sqrt{\frac{n}{k_j}} \suit{ \frac{1}{\gamma_j}\log \frac{X^{(j)}}{U_j(n/k_j)}-1}\mI\set{X^{(j)}>U_j(n/k_j)}, \quad j=1,\dots,p. 
$$
Let $\bolSigma = (\sigma_{ij})_{p\times p}$  denote the covariance matrix of $\mbY$.
  Assume that, for sufficiently large $n$,  $\max_{1\le i<j\le p} \abs{\sigma_{ij}} \le c<1$  and $\max_{1\le i\le p}\sum_{j=1}^p \sigma_{ij}^2\le C,$ for some $0<c<1, C>0$.
        \end{enumerate}

Condition \ref{Eigens} imposes sparsity restrictions on the covariance matrix of $\mbY$. This is comparable with restrictions on the covariance matrix in high-dimensional mean tests. For instance, for the high-dimensional mean tests, similar constraints are assumed on the covariance matrix of the original random vector $\mbX$, see e.g. \cite{tony2014two} and \cite{feng2022asymptotic}. 

The interpretation of Condition (B) is not straightforward: it involves marginal characteristics, (effective) sample size, and dependence structure. In fact, the limit of the covariance matrix $\bolSigma$ is only related to the tail dependence structure of $\mbX$. Specifically, when $k_i = k_j$, we have $\sigma_{ij} \to R_{ij}(1,1)$ as $n \to \infty$, where $R_{ij}(1,1)$ denotes the tail dependence coefficient \citep{sibuya1960bivariate} of the random pair $(X_i, X_j)$. See Section E of the Supplementary Material for a detailed discussion.  We also provide a set of sufficient conditions of Condition \ref{Eigens} in Section E
 of the Supplementary Material. The sufficient conditions segment conditions on the sparsity of tail dependence, the choice of $k_j$ and dimensionality.

Besides  Conditions \ref{condition:SOC} and \ref{Eigens}, we make assumptions on the choices of  
 the intermediate sequences $k_1,\dots,k_p$ and the dimension $p$.     Denote $k_{\min} = \min_{1\le j \le p} k_j$ and $k_{\max} = \max_{1\le j \le p}k_j$.

\begin{enumerate}[label=(C)]
    \item \label{condition:k:choice}  
    As $n\to\infty$, $p = p(n)\to\infty$ and 
\begin{align}
    \frac{k_{\min}}{\log^5 p} \to  \infty, \quad      \frac{\log k_{\max}}{\log p}  =O(1), \label{condition:upper:lower} \\
    \sqrt{\log p}\max_{1\le j \le p} \abs{\sqrt{k_j}A_j(n/k_j)} =  o(1). \label{condition:biases} 
\end{align}

\end{enumerate}

 Condition \eqref{condition:upper:lower} provides a lower bound for $k_{\min}$ and an upper bound for $k_{\max}$.  The requirement    $\log p = o(k_{\min}^{1/5})$ 
 is similar to the condition $\log p =o(n^{1/4})$ used in high-dimensional mean tests \citep{tony2014two}. 
The discrepancy arises from two aspects. First, the minimum effective sample size across the $p$ dimensions   is $k_{\min}$. 
 Second, the power is  $1/5$ rather than $1/4$, because we are dealing with the logarithms of heavy-tailed random variables instead of sub-Gaussian ones.  For instance, the logarithm of a Pareto distributed random variable follows an exponential distribution, which is not sub-Gaussian. As a consequence, the Hill estimator can be viewed as approximately the mean of exponentially distributed random variables rather than sub-Gaussian ones.

Condition  \eqref{condition:biases} 
  aims at controlling the biases in the Hill estimators  $\widehat{\gamma}_j(k_j), j=1,\dots,p$, uniformly. Note that in a univariate context, the assumption $\sqrt{ k_j}A_j(n/k_j) = o(1)$ is often invoked to assume away the asymptotic bias of estimators of the extreme value index.

\begin{remark} \label{remark:conditions}
    One example for $k_1,\dots,k_p$ satisfying Condition \ref{condition:k:choice}   can be given as follows. We assume that $\rho=: \max(\rho_1,\dots,\rho_p)<0$.
      Choose $k_1 =k_2=\cdots = k$ and  
    $k \asymp n^{\eta}$ as $n\to\infty$, with  $\eta< (-\rho)/(-\rho+1/2)$. Then Condition \ref{condition:k:choice} holds provided that  $ \log p = o(n^{\alpha})$, with $0<\alpha<\min\set{\eta/5,-2\rho(1-\eta)-\eta}$.  

\end{remark}

We  establish the asymptotic theory of the test statistic $\mbT(k_1,\dots,k_p)$ under $H_0$  in the following theorem.
\begin{theorem}\label{theorem:max:i}
    Assume that Conditions \ref{condition:SOC}, \ref{Eigens} and \ref{condition:k:choice} hold.
   Under the null hypothesis $H_0$, as $n\to\infty$,  for any $x\in \mathbb{R}$,
$$
\Pr(\mbT^2(k_1,\dots,k_p) - 2\log p + \log \suit{\log p}\le x)\to \exp\set{-\frac{1}{\sqrt{\pi}}\exp(-x/2)}.
$$
\end{theorem}

Theorem \ref{theorem:max:i} demonstrates that the test statistic $\mbT(k_1,\dots,k_p)$, upon appropriate transformation, converges to a Gumbel distribution, also recognized as the type I extreme value distribution. 
The limiting distribution in our theory is the same as that of the high-dimensional mean test statistic in \cite{tony2014two}. Intuitively, this follows from the fact that the test statistic $\mbT(k_1,\dots,k_p)$ is a maximum of $p$ estimation errors which are asymptotically normally distributed. Hence, obtaining the Gumbel distribution as a limit is in line with the classical extreme value theorem \citep{fisher1928limiting, gnedenko1943distribution}, despite that the $p$ estimation errors are neither exactly normally distributed, nor independent.

The proof of Theorem \ref{theorem:max:i}  faces technical challenges from extreme value statistics and high-dimensional statistics. 
Notably, the test statistic $\mbT(k_1,\dots,k_p)$ involves high order statistics at each dimension, which are neither independent nor identically distributed.  To handle that, we 
first  establish the asymptotic theory of the `pseudo'  test statistic 
$$
\widetilde{\mbT}(k_1,\dots,k_p):=\max_{1\le j \le p}\abs{ \sqrt{k_j} \suit{ \frac{\widetilde{\gamma}_j(k_j)}{\gamma_j^0}-1 } },
$$
where $\widetilde{\gamma}_j(k_j)$ is the  `pseudo' Hill estimator,  defined as 
$$
\widetilde{\gamma}_j(k_j) =  \frac{ \sum_{i=1}^n \set{ \log  X_i^{( j)} -\log U_j(n/k_j)}\mI\set{X_i^{(j)}\ge U_j(n/k_j)}  }{ \sum_{i=1}^n \mI\set{X_i^{(j)}\ge U_j(n/k_j)}}.
$$
Since both the numerator and denominator of $\widetilde{\gamma}_j(k_j)$ are sums of i.i.d. random variables, 
we can use \cite{zaitsev1987gaussian}
to establish the asymptotic theory of  $\widetilde{\mbT}(k_1,\dots,k_p)$. This technique is widely used in the study of the maximum of high-dimensional random vectors; see, for example, \cite{cai2013two}, \cite{tony2014two}, \cite{ma2021global}, \cite{feng2022high}, and \cite{tang2022conditional}. 
Secondly, we demonstrate that the difference between 
 $\widetilde{\mbT}(k_1,\dots,k_p)$ and $\mbT(k_1,\dots,k_p)$ is negligible. The Bahadur-Kiefer process \citep{kiefer1967bahadur} is utilized as a pivotal tool for  this claim.

On the basis of Theorem \ref{theorem:max:i}, we construct the following  test procedure,   referred to as the {\it Gumbel test}.   Define  
$
q_{\alpha} = -\log(\pi) - 2\log \suit{\log  \set{1/(1-\alpha)}},
$
which is the $(1-\alpha)$ quantile of the limit Gumbel distribution.
We reject $H_0$ if and only if $\mbT(k_1,\dots,k_p)\ge   c_{\alpha} $, where $c_{\alpha} = \suit{2\log p -\log(\log p)+q_{\alpha}}^{1/2}$.

\subsection{Test under general dependence}

One main drawback of the proposed Gumbel test is that it requires the sparsity of the covariance matrix of $\mbY$, as specified in Condition \ref{Eigens}. This condition is roughly equivalent to the sparsity of the tail dependence structure of $\mbX$. In practical applications such as climate extremes or financial risk analysis,  strong cross-sectional dependence in the tail is common.  This sparsity assumption may fail.   To  broaden the applicability of our test, we propose a multiplier bootstrap procedure that directly approximates the  distribution of the test statistic under the null, without relying on the sparsity assumption. 

The detailed procedure of our proposed multiplier {\it bootstrap test} is given as follows:
\begin{enumerate}
    \item[(I).] Independent from the data $\mbX_1,\dots, \mbX_n$,   we generate a sequence of independent $N(0,1)$ random variables $\xi_1,\dots,\xi_n$. 
    \item[(II).] Using   $\xi_1,\dots,\xi_n$  as multipliers, we construct the multiplier bootstrap statistic 
       $$
        \mbT^B(k_1,\dots,k_p) = \max_{1\le j\le p} \frac{1}{\sqrt{k_j}\gamma_j^0} \abs{\sum_{i=1}^{n} \xi_i\suit{\log X_{i}^{(j)} - \log X_{n-k_j,n}^{(j)}-\gamma_j^0} \mI\suit{X_i^{(j)}>X_{n-k_j,n}^{(j)} } }.
       $$
    \item[(III).] For a given significance level $\alpha\in \suit{0,1}$, compute the critical value $c_{\alpha}^B$ as the conditional $1-\alpha$ quantile of $\mbT^B(k_1,\dots,k_p)$ given the data $\mbX_1,\dots, \mbX_n$. 
    \item[(IV).] We reject $H_0$ if  $ \mbT(k_1,\dots,k_p)\ge c_{\alpha}^B$. 
\end{enumerate}

To establish the asymptotic behavior of the bootstrap test,    we modify the assumptions on the choices of  
 the intermediate sequences $k_1,\dots,k_p$ and the dimension $p$. 
\begin{enumerate}[label=(C$ ^\prime$)]
    \item \label{condition:k:choice:boot}  
    As $n\to\infty$, $p = p(n)\to\infty$ and 
\begin{align*}
    \frac{k_{\min}}{\log^7 p} \to  \infty, \quad \frac{k_{\min}}{\log^5 n}\to\infty,  \quad  \frac{\log k_{\max}}{\log p}  =O(1),   \\
    \sqrt{\log p}\max_{1\le j \le p} \abs{\sqrt{k_j}A_j(n/k_j)} =  o(1). 
\end{align*}

\end{enumerate}

\begin{theorem}\label{Theorem:size:boot}
    Assume that Conditions \ref{condition:SOC}  and \ref{condition:k:choice:boot}  hold. Then, under $H_0$, as $n\to\infty$,
    $$
        \Pr( \mbT(k_1,\dots,k_p)\ge c_{\alpha}^B)\to \alpha. 
    $$
\end{theorem}

In the construction of the multiplier bootstrap  statistic,  the multipliers are multiplied with non-i.i.d. observations in each dimension. This differs from classical multiplier bootstrap in the high-dimensional mean tests. Nevertheless, the 
 proposed bootstrap procedure still achieves asymptotic validity without requiring the  sparsity Condition \ref{Eigens}.
 This flexibility, however, necessitates  a slightly stronger requirements on the minimum effective sample size $k_{\min}$, as in Condition 
\ref{condition:k:choice:boot}.

The proof of Theorem \ref{Theorem:size:boot} follows a similar framework as that of Theorem 1. Instead of using \cite{zaitsev1987gaussian}, the proof relies 
  on the central limit theorems for  high-dimensional random vectors \citep{chernozhuokov2022improved} to handle the joint distribution of $p$ estimation errors. These results have also been  widely used in the high-dimensional inference problems, see for example, the mean tests \citep{chang2017simulation,xue2020distribution}, the covariance matrix test \citep{chang2017comparing}, and the martingale difference test \citep{chang2023testing}.

\subsection{Power analysis}
We first analyze the power of the Gumbel test.  Denote  
$$
 \delta_j =  \frac{\gamma_j}{\gamma_{j}^0}-1, \quad j=1,\dots,p.  
 $$  
We consider the following local alternative hypothesis $H_1$, 
$$
H_1: \quad \max_{1\le j\le p}  \abs{\sqrt{k_j}\delta_j} \ge \sqrt{\lambda \log p},
$$
for some constant $\lambda>2$.

 \begin{theorem}\label{theorem:power:i}
     Assume the same conditions as in Theorem \ref{theorem:max:i}.
     Under the local alternative hypothesis $H_1$,  we have that,  
     $$
 \lim_{n\to\infty}\Pr\suit{\mbT(k_1,\dots,k_p)\ge c_{\alpha} } = 1.
     $$
 \end{theorem}

 We then analyze the power of the  bootstrap test.  
 We consider the alternative hypothesis 
$$
\widetilde{H}_1: \quad   \sqrt{k_{\min}}\max_{1\le j\le p} \abs{\delta_j} \ge  \sqrt{\lambda \log p},
$$
for some constant $\lambda>2$. The alternative hypothesis  $\widetilde{H}_1$ is slightly stronger than $H_1$.

 \begin{theorem}\label{theorem:power}
     Assume the same conditions as in Theorem \ref{Theorem:size:boot}.
     Under the local alternative hypothesis $\widetilde{H}_1$,  we have that,  
     $$
 \lim_{n\to\infty}\Pr\suit{\mbT(k_1,\dots,k_p)\ge  c_{\alpha}^B} = 1.
     $$
 \end{theorem}

 \section{Test for equal extreme value indices hypothesis $H_0^*$}\label{sec:equal:index}

 In this subsection, we adapt the test procedure in Section \ref{sec:methodologies} to test the equal  extreme value indices hypothesis
 $$
 H_0^*:\gamma_1 = \cdots = \gamma_p.
 $$ 
 Under the null hypothesis, we estimate the common extreme value index by 
 $$
 \overline{\gamma} = \frac{1}{p}\sum_{j=1}^p \widehat{\gamma}_j(k_j),
 $$
 where $\widehat{\gamma}_j(k_j)$ is the Hill estimator in \eqref{eq:hill}. We then adaptively consider the test statistic
 $$
 \mbT_*(k_1,\dots,k_p)= \max_{1\le j\le p} \sqrt{k_j}\abs{\frac{\hatgamma_j(k_j)}{\overline{\gamma}}-1}.
 $$ 

To establish the asymptotic theory of the test statistic $\mbT_*(k_1,\dots,k_p)$, we require choosing $k_j$, $j=1,\dots, 
 p$ at a similar level.
\begin{enumerate}[label=(D)]
             \item\label{eq:k:choice:other} Choose $k_j, j=1,\dots,p$, such that, as $n\to\infty$,
             $$
             \begin{aligned}
                 c_{L} \le\frac{\min_{1\le j\le p} k_j}{k} \le \frac{\max_{1\le j\le p} k_j}{k}\le c_{U},  \\
             \end{aligned}
             $$
     where $0<c_{L}\le c_{U}$ are positive constants, and $k$ is an intermediate sequence such that as $n\to\infty$, $k\to\infty$ and $k/n\to 0$.           \end{enumerate}
 Note that, $\overline{\gamma}$ is the average of all  $p$ Hill estimators with effective sample sizes $k_j$, $j=1,\dots, p$.  Ensuring that $k_j, j=1,\dots, p$
   are comparable across dimensions prevents that the asymptotic behavior of 
   $\overline{\gamma}$ is  dominated by those Hill estimators with the lowest level of  $k_j$.

\subsection{Test under weak dependence}

We establish  the asymptotic theory of the test statistic $\mbT_*(k_1,\dots,k_p)$ under 
the sparsity Condition \ref{Eigens}.
 \begin{theorem}\label{theorem:identical:size}
     Assume that Conditions \ref{condition:SOC}, \ref{Eigens}, \ref{condition:k:choice} and \ref{eq:k:choice:other} hold. 
     Then  under $H_0^*$, as $n\to\infty$,
$$
\Pr(\mbT^2_*(k_1,\dots,k_p)-2\log p + \log \suit{\log p}\le x)\to \exp\set{-\frac{1}{\sqrt{\pi}}\exp(-x/2)}.
$$
 \end{theorem}
  The  asymptotic behavior of  $\mbT_*(k_1,\dots,k_p)$ is identical to that of 
  $\mbT(k_1,\dots,k_p)$, 
   since $\overline{\gamma}(k_1,\dots,k_p)$ converges to $\gamma_0$ at a rate faster than $(k \log p)^{1/2}$ as shown in the following proposition.
 
 \begin{prop}\label{theorem:average}
     Assume the  same conditions  as in Theorem \ref{theorem:identical:size}. Then under $H_0^*$,  as $n\to\infty$, 
     $$
     \overline{\gamma} -\gamma_0 = o_P\suit{\frac{1}{ \sqrt{ k\log p}}}.
     $$
     \end{prop}

\subsection{Test under general dependence}
 
 To avoid the  sparsity Condition \ref{Eigens}, 
 we propose the following multiplier bootstrap procedure  for $H_0^*$:
\begin{enumerate}
    \item[(I).] Independent from the data $\mbX_1,\dots, \mbX_n$,   we generate a sequence of independent $N(0,1)$ random variables $\xi_1,\dots,\xi_n$. 
    \item[(II).] Using   $\xi_1,\dots,\xi_n$  as multipliers, we construct the multiplier bootstrap statistic 
       $$
        \mbT_*^B(k_1,\dots,k_p) = \max_{1\le j\le p} \frac{1}{\sqrt{k_j} \overline{\gamma} } \abs{\sum_{i=1}^{n} \xi_i\suit{\log X_{i}^{(j)} - \log X_{n-k_j,n}^{(j)}- \overline{\gamma} } \mI\suit{X_i^{(j)}>X_{n-k_j,n}^{(j)} } }.
       $$
    \item[(III).] For a given significance level $\alpha\in \suit{0,1}$, compute the critical value $c_{\alpha,*}^B$ as the conditional $1-\alpha$ quantile of $\mbT^B_*(k_1,\dots,k_p)$ given the data $\mbX_1,\dots, \mbX_n$. 
    \item[(IV).] We reject $H_0$ if $ \mbT_*(k_1,\dots,k_p)\ge c_{\alpha,*}^B$. 
\end{enumerate}
 
 The consistency of the multiplier bootstrap procedure  is established in the following theorem.
  \begin{theorem}\label{theorem:identical:size:boot}
     Assume that Conditions \ref{condition:SOC},   \ref{condition:k:choice:boot} and \ref{eq:k:choice:other} hold. 
     Then  under $H_0^*$, as $n\to\infty$,
     $$
     \Pr\suit{\mbT_*(k_1,\dots,k_p)\ge c_{\alpha,*}^B}\to \alpha.
     $$
 \end{theorem}

 \section{Simulation}\label{sec:simu}

\subsection{Simulation setting}\label{sec:simu:set}

 In this section, we present a simulation study to illustrate the finite sample performance of our testing procedure for the null hypothesis $H_0: \gamma_j = \gamma_j^0$, for all $1\le j\le p$.  
 Without loss of generality, we shall always take $\gamma_j^0 = 1$, $j=1,\dots,p$. 
 We choose the true extreme value indices of the data generating processes under the null hypothesis and alternative hypothesis as follows.
 Under the null hypothesis, $\gamma_j = \gamma_j^0$, $j=1,\dots,p$. Under the alternative hypothesis,  $ (\gamma_1,\dots,\gamma_p)^\top$  has $m$ entries that differ from $1$. The indices of such entries $\mathcal{S}$ are uniformly drawn from the set $\set{1,\dots,p}$.  
 In this study, we take $m = \lfloor p^{1/3} \rfloor$ where $\lfloor x\rfloor$ denotes the largest integer that is  smaller than or equal to $x$. For each $j\in \mathcal{S}$,  the deviation $\delta_j$ is set to be either  $ 2\sqrt{\log p/k_j}$ or $- 2\sqrt{\log p/k_j}$ with equal probability. For $j \in \mathcal{S}^c$, we set $\gamma_j = 1$. 

The samples are generated from the following models.

 \begin{enumerate}
     \item[(M1)] Let $M^{(i)} = (M^{(i,1)}, M^{(i,2)})^\top$, $i=1,\dots,\lceil p/2 \rceil$ be  i.i.d.  random vectors following a 
      bivariate  Cauchy distribution with scale matrix $ \left[ \begin{array}{cc}
         1 & 0.7 \\
         0.7 & 1
      \end{array} \right]$. Here, $\lceil x \rceil$ denotes the smallest integer that is larger than or equal to $x$.
       For $j=1,\dots,p$, define 
      $$
      \begin{aligned}
         \widetilde{X}^{(j)} = \left\{ 
             \begin{array}{ll}
                 M^{( \lceil j/2 \rceil,1)},  & \ \text{if}\ j \  \text{is odd}, \\
                 M^{( \lceil j/2 \rceil,2)},  & \ \text{if}\ j \  \text{is even}.
             \end{array} 
         \right.
      \end{aligned}
      $$ 
      Then  we transform the marginal distribution of $ \widetilde{X}^{(j)}$ to a  Student-t distribution with degree of freedom $1/\gamma_j$,   by  
      $$
 X^{(j)} =  \text{St}_{1/\gamma_j}^{-1}\set{\text{St}_{1}\suit{\widetilde{X}^{(j)}}}, \quad j=1,\dots,p,
      $$
     where $\text{St}  _{v}(\cdot)$ denotes the cumulative distribution function of a Student-t distribution with degree of freedom $v$.
 \item [(M2)] We generate $(\widetilde{X}^{(1)}, \dots,  \widetilde{X}^{(p)}) $ from a Gaussian copula with correlation matrix $\boldsymbol{R}$, where $(\boldsymbol{R})_{ij} =0.5^{|i-j|}$, $i,j=1,\dots,p$.  Then we transform the marginal distribution of $ \widetilde{X}^{(j)}$ to  a Pareto   distribution with shape parameter $1/\gamma_j$, by  
 $$
 X^{(j)} = (1- \widetilde{X}^{(j)} )^{-\gamma_j}, \quad j=1,\dots, p. 
 $$
 \item[(M3)] We generate $(\widetilde{X}^{(1)}, \dots,  \widetilde{X}^{(p)})$ from a Gumbel copula with parameter $\theta=2$.        Then  we transform the marginal distribution of $ \widetilde{X}^{(j)}$ to a  Student-t distribution with degree of freedom $1/\gamma_j$,   by  
      $$
 X^{(j)} =  \text{St}_{1/\gamma_j}^{-1}\suit{\widetilde{X}^{(j)}}, \quad j=1,\dots,p.
      $$
 \item[(M4)] Let $Z_1, Z_2$ be independent random variables, each  following a Fr\'echet distribution  with shape parameter $1$, where the distribution function is given by $\text{Fr}_1(x) = \exp(-x^{-1})$.  We generate $U_1, \dots, U_p$  independently from the uniform distribution on the interval $[0,1]$. Define 
 			$$
 				X^{(j)} =  \max(U_jZ_1, (1-U_j)Z_2), \quad j=1,\dots, p.  
 			$$ 
 \end{enumerate}
Note that, models (M1) and (M2) correspond to weak dependence, while models (M3) and (M4) correspond to strong dependence.

 We compare the performance of the Gumbel test,  the bootstrap test and  the Wald-type test.  
  The Wald-type test statistic $\mbT_W$  for $H_0$  is defined as 
 $$
 \mbT_{W} =:\mbT_{W}(k_1,\dots,k_p) = \bolzeta^\top \widetilde{\bolSigma}^{-1}\bolzeta,
 $$
 where $\bolzeta$ is a $p$-dimensional vector with component 
 $\bolzeta_j = \sqrt{k_j} \suit{\hatgamma_j(k_j)/\gamma_{j}^0 -1}$,
 and 
 $\widetilde{\bolSigma}$ is  a $p\times p$ matrix with $\widetilde{\bolSigma}_{ij}$ being the sample estimation of the tail dependence coefficient $R_{ij}(1,1)$, i.e., 
 $$
 \widetilde{\bolSigma}_{ij} = \frac{1}{k}\sum_{s=1}^n \mI\suit{X_s^{(i)}> X_{n-k,n}^{(i)}, X_s^{(j)}> X_{n-k,n}^{(j)}},
 $$
 see e.g. \cite{drees1998best}.
 We then reject the null hypothesis when $\mbT_{W} >\chi^2_{p,1-\alpha}$.  The asymptotic theory of $\mbT_W$ is only established for  fixed $p$. Although there are no theoretical guarantees for $\mbT_W$ in high-dimensional settings, we implement the Wald-type test as a benchmark for comparison. The Wald-type test is not applicable to model (M4) because the matrix $\widetilde{\bolSigma}$ is often non-invertible due to strong dependence among the random variables.

 Under each model, the random vector $\mbX$ is generated with sample size $n=2000$ and dimension $p=50$, $100$ and $150$.  
 For $j=1,\dots,p$, the number of tail observations $k_j$ is set uniformly to $k_j = k$. Two different values of $k$ are considered: 
 $k=90$ and $k=120$.
 The type I error and the power of the tests are calculated from 2000 replications.

\subsection{Simulation results}

 \begin{table}[htbp]    
     \centering
     \setlength{\tabcolsep}{2pt}
     \begin{tabular}{cc|ccc|ccc}
     \hline
       & & \multicolumn{3}{c}{$k=90$} & \multicolumn{3}{c}{$k=120$}  \\
       & & $p=50$     &  $p=100$        & $p=150$      &   $p=50$    & $p=100$      &   $p=150$  \\
             \hline\multirow{3}{*}{(M1)}       
 & Gumbel & 0.0545 & 0.0675 & 0.079 & 0.061 & 0.069 & 0.08 \\ 
   & Bootstrap & 0.032 & 0.0305 & 0.0305 & 0.039 & 0.0435 & 0.0405 \\ 
   & Wald & 0.1095 & 0.1765 & 0.2885 & 0.119 & 0.2025 & 0.3125 \\ 
      \hline\multirow{3}{*}{(M2)} 
    & Gumbel & 0.061 & 0.0585 & 0.0845 & 0.056 & 0.0635 & 0.07 \\ 
   & Bootstrap &0.0295 & 0.025 & 0.031 & 0.033 & 0.0315 & 0.038\\ 
   & Wald & 0.1985 & 0.311 & 0.4995 & 0.214 & 0.3575 & 0.565 \\ 
  \hline\multirow{3}{*}{(M3)}
   & Gumbel & 0.029 & 0.0275 & 0.0255 & 0.0245 & 0.019 & 0.0305 \\ 
   & Bootstrap & 0.049 & 0.0465 & 0.0535 & 0.042 & 0.0435 & 0.045 \\ 
   & Wald & 0.3555 & 0.7795 & 0.971 & 0.317 & 0.6865 & 0.942 \\ 
   \hline\multirow{2}{*}{(M4)}
    & Gumbel & 0.0135 & 0.005 & 0.003 & 0.0095 & 0.006 & 0.005 \\ 
   & Bootstrap & 0.058 & 0.051 & 0.049 & 0.048 & 0.049 & 0.0555 \\ 
           \hline
     \end{tabular}
     \caption{Type I errors of the tests  with $\alpha=0.05$.}
     \label{table:EmpiricalSize}
 \end{table}

 The type I error of the tests, i.e., the rejection rates under the null hypothesis,  are  displayed in Tables \ref{table:EmpiricalSize}.  We observe that, the type I error of the Wald-type test $\mbT_{W}$  exceeds the significance level $\alpha = 0.05$ substantially, indicating that the Wald-type test performs unsatisfactorily in high-dimensional settings.

For models (M1) and (M2), both tests are theoretically valid, and the difference in rejection rates is mainly a finite-sample effect. Note that, the Gumbel test relies on three layers of asymptotic approximations: (i) the  Hill estimator is approximated by the  mean of exponentially distributed random
variables; 
 (ii) a high-dimensional Gaussian approximation  is applied to normalized i.i.d. sums, and (iii) the maximum of (nearly) independent Gaussian components is approximated by  a Gumbel limit. The bootstrap test is based on a parallel sequence of approximations, replacing the final Gumbel approximation with a bootstrap approximation to the distribution of the max-type statistic.
 
By contrast, models (M3) and (M4) involve substantial dependence among the limiting Gaussian components. Under strong dependence, the realized test statistics tend to be smaller. As a result, the approximation (iii) fails and  the Gumbel test becomes conservative. By contrast,
 the bootstrap procedure better adapts to the underlying dependence structure, leading to the correct size control.

The power of the tests, i.e., the rejection rates under the alternative hypothesis, is presented in Table \ref{table:Power}. Results are excluded if the corresponding type I error deviates substantially from the significance level ($\alpha = 0.05$). Overall, the multiplier bootstrap test performs well across all scenarios. Additionally, when the data exhibits weak dependence (models (M1) and (M2)), the Gumbel test demonstrates relatively higher power compared to the multiplier bootstrap test.

 \begin{table}[htbp]  
     \centering
     \setlength{\tabcolsep}{2pt}
     \begin{tabular}{ll|ccc|ccc}
     \hline
       & & \multicolumn{3}{c}{ $k=90$} & \multicolumn{3}{c}{$k=120$} \\
       & & $p=50$     &  $p=100$        & $p=150$      &   $p=50$    & $p=100$      &   $p=150$   \\
       \hline\multirow{2}{*}{(M1)}   
 & Gumbel & 0.976 & 0.9995 & 1.000 & 0.963 & 0.999 & 0.9995 \\ 
   & Bootstrap & 0.931 & 0.993 & 0.992 & 0.941 & 0.9925 & 0.9955 \\ 
   \hline\multirow{2}{*}{(M2)}
 & Gumbel & 0.988 & 0.999 & 0.9995 & 0.9915 & 1.000 & 0.9995 \\ 
   & Bootstrap & 0.951 & 0.9835 & 0.9895 & 0.984 & 0.9995 & 0.9975 \\  
    \hline\multirow{1}{*}{(M3)} 
   & Bootstrap & 0.996 & 0.9985 & 1.000 & 0.9935 & 0.999 & 1.000 \\
    \hline\multirow{1}{*}{(M4)} 
   & Bootstrap & 0.9355 & 1.000 & 1.000 & 0.9985 & 0.9995 & 1.000 \\
    \hline
     \end{tabular}
     \caption{Power of the tests   with  $\alpha=0.05$.}
     \label{table:Power}
 \end{table}

We also compare the computational time of the Gumbel test and the multiplier bootstrap test.  The results are reported in Table \ref{table:time}. The Gumbel test is considerably faster than the multiplier bootstrap test. This computational advantage is consistent across the considered settings.

\begin{table}[htbp]
	     \centering
     \setlength{\tabcolsep}{2pt}
     \begin{tabular}{ll|ccc|ccc}
     \hline
       & & \multicolumn{3}{c}{ $k=90$} & \multicolumn{3}{c}{$k=120$} \\
       & & $p=50$     &  $p=100$        & $p=150$      &   $p=50$    & $p=100$      &   $p=150$   \\
       \hline\multirow{2}{*}{(M1)}   
 & Gumbel & 0.01 & 0.02 & 0.03 & 0.01 & 0.02 & 0.03 \\ 
   & Bootstrap & 5.79 & 11.3 & 16.79 & 6.34 & 12.36 & 18.37 \\ 
   \hline\multirow{2}{*}{(M2)}
 & Gumbel & 0.01 & 0.02 & 0.03 & 0.01 & 0.02 & 0.03 \\ 
   & Bootstrap & 5.79 & 11.28 & 16.8 & 6.34 & 12.36 & 18.37\\  
    \hline\multirow{2}{*}{(M3)} 
   & Gumbel & 0.01 & 0.02 & 0.03 & 0.01 & 0.02 & 0.03 \\
  & Bootstrap& 6.54 & 12.8 & 19.25 & 7.09 & 13.88 & 20.83\\
    \hline\multirow{2}{*}{(M4)} 
   & Gumbel & 0.01 & 0.02 & 0.03 & 0.01 & 0.02 & 0.03\\
   & Bootstrap & 6.63 & 12.97 & 19.76 & 7.38 & 14.36 & 21.57\\
    \hline
     \end{tabular}
     \caption{Average computation time of the tests under the null hypothesis (in seconds), with the experiment implemented on an Intel Xeon Gold 6252 CPU.}
     \label{table:time}
\end{table}

\subsection{Sensitivity Analysis}

First, we evaluate the sensitivity of the power with respect to the number of deviations $m$. We repeat the analysis by setting $m=1$ instead of $m=\lfloor p^{1/3} \rfloor$, so that only one component of $(\gamma_1,\ldots,\gamma_p)$ differs from 1. All other settings remain the same as in Section~\ref{sec:simu:set}. The results are reported in Table~\ref{table:Power:m1}. We find that the proposed test still achieves considerable power under very sparse alternatives, although the power is naturally lower than that in less sparse settings.

 \begin{table}[htbp]  
     \centering
     \setlength{\tabcolsep}{2pt}
     \begin{tabular}{ll|ccc|ccc}
     \hline
       & & \multicolumn{3}{c}{ $k=90$} & \multicolumn{3}{c}{$k=120$} \\
       & & $p=50$     &  $p=100$        & $p=150$      &   $p=50$    & $p=100$      &   $p=150$   \\
       \hline\multirow{2}{*}{(M1)}   
 & Gumbel & 0.67 & 0.82 & 0.85 & 0.67 & 0.76 & 0.8 \\ 
   & Bootstrap & 0.6 & 0.69 & 0.7 & 0.62 & 0.67 & 0.69\\ 
   \hline\multirow{2}{*}{(M2)}
 & Gumbel & 0.87 & 0.93 & 0.72 & 0.67 & 0.9 & 0.93 \\ 
   & Bootstrap & 0.78 & 0.83 & 0.62 & 0.61 & 0.84 & 0.86\\  
    \hline\multirow{1}{*}{(M3)} 
   & Bootstrap & 0.73 & 0.84 & 0.87 & 0.76 & 0.8 & 0.85 \\
    \hline\multirow{1}{*}{(M4)} 
   & Bootstrap & 0.94 & 0.89 & 0.99 & 0.88 & 0.92 & 0.93 \\
    \hline
     \end{tabular}
     \caption{Power of the tests   with $m=1$ and   $\alpha=0.05$.}
     \label{table:Power:m1}
 \end{table}

\begin{figure}[htbp]
  \includegraphics[width=1\textwidth]{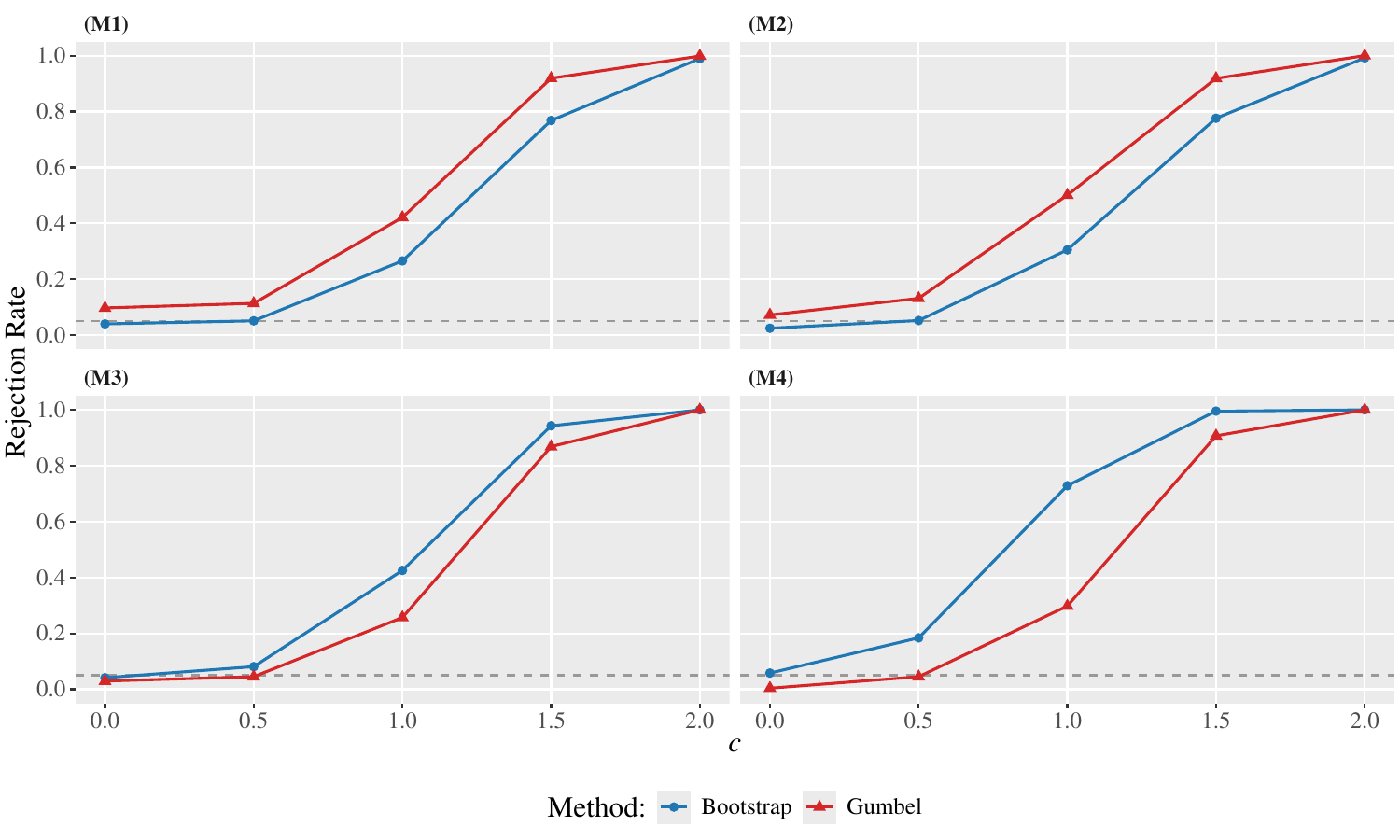}	
       \caption{The rejection rates of the multiplier bootstrap test and the Gumbel test for different values of $c$. }
       \label{figure:varydelta}
  \end{figure}

Next, we evaluate the sensitivity of the power with respect to the magnitude of the deviation signals $\delta_j$. Specifically, under the alternative hypothesis, we set
$$
\delta_j = \pm c\sqrt{\log p / k}, \quad j \in \mathcal{S},
$$
where the sign is chosen with equal probability and $c$ varies from 0 to 2. We consider $k=90$ and $p=150$, while all other settings remain the same as in Section~\ref{sec:simu:set}. The results are shown in Figure  \ref{figure:varydelta}. We observe that the power of the proposed test increases with $c$ and approaches one as the signal strength becomes stronger.

 \section{Application}\label{sec:application}
 We apply our developed methods to two datasets to test whether the extreme value indices are constant over the $p$ dimensions. These datasets have been analyzed by \cite{kiriliouk2022estimating} under the assumption of identical extreme value indices.

We  compare our methods  with the Wald-type test     \citep{kinsvater2016regional}, 
$$
 \mbT_{W}^* =:\mbT_{W}^*(k_1,\dots,k_p) = \suit{\bolzeta^*}^\top \widetilde{\bolSigma}^{-1}\bolzeta^*,
 $$
 where $\bolzeta^*$ is a $p$-dimensional vector with component 
 $\bolzeta_j^* = \sqrt{k_j} \suit{\hatgamma_j(k_j)/\overline{\gamma}(k_1,\dots,k_p) -1}$,
 and 
 $\widetilde{\bolSigma}$ is  defined as in the simulation study. We then reject the null hypothesis when $\widetilde{\mbT}_{W} >\chi^2_{p-1,1-\alpha}$.
 We  choose a  constant  $k_j$ over the $p$-dimensions throughout the application, that is $k_1=\cdots =k_p = k$.

\begin{figure}[htbp]
     \centering
     \subfigure{
     \includegraphics[width =0.48\textwidth]{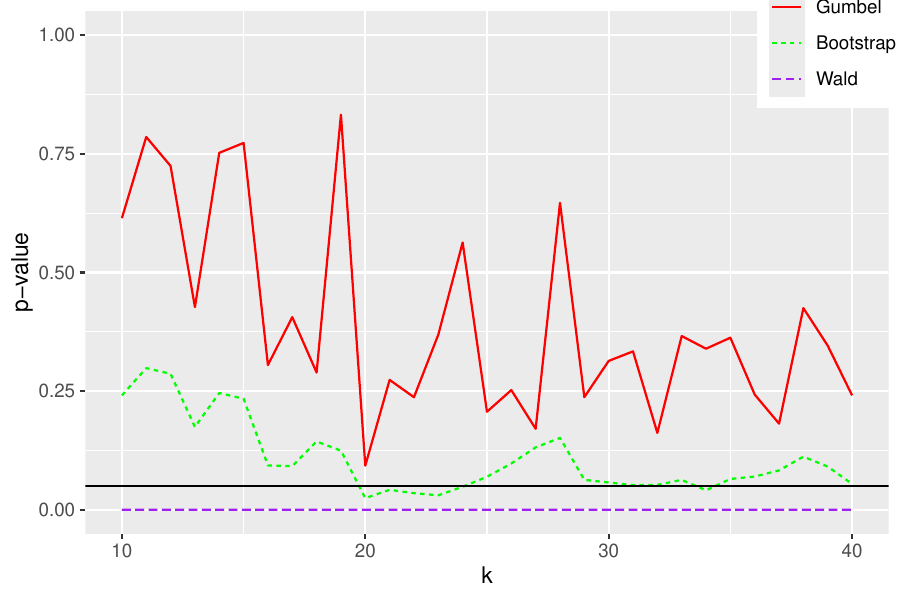}}
     \subfigure{
         \includegraphics[width =0.48\textwidth]{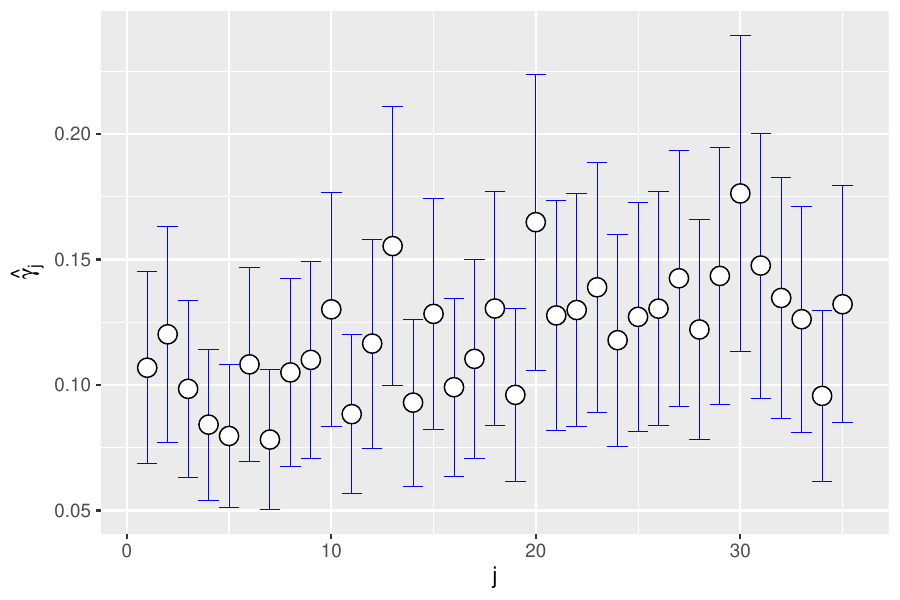}
     }
     \caption{Left: the $p$-values of the tests against different levels of $k$  for the wind gust data. The black horizontal line indicates the significance level, $\alpha = 0.05$. Right: the Hill estimates $\widehat{\gamma}_j$ with 95\% confidence intervals for the wind gust data. }
     \label{figure:application}
 \end{figure}

 The first dataset consists of the daily maximal speeds of wind gust  in the Netherlands    for $p=35$ different stations during the winter months (October through March) from 2015 to 2019, with $n  = 911$ observations\footnote{This dataset is  available from the Royal Netherlands Meteorological Institute (KNMI), https://climexp.knmi.nl/. }. 
 We test the constancy of the extreme value indices over the $p$ stations. 
 The obtained $p$-values against various levels of $k$ are shown in the left panel of  Figure \ref{figure:application}. 
 Employing the Wald-type test, we would reject the null hypothesis at the 5\% significance level. Conversely, 
the Gumbel test  suggests not rejecting the null hypothesis for all levels of $k$. The multiplier bootstrap test produces $p$-values that lie between those of the Wald-type and Gumbel tests, and  suggests retaining the null hypothesis for most levels of $k$.  We plot the Hill estimates, along with a 95\% confidence interval for $k = 30$, in the right panel  of  Figure \ref{figure:application}. We observe no apparent differences across these Hill estimates.

\begin{figure}[htbp]
     \centering
     \subfigure{
     \includegraphics[width =0.48\textwidth]{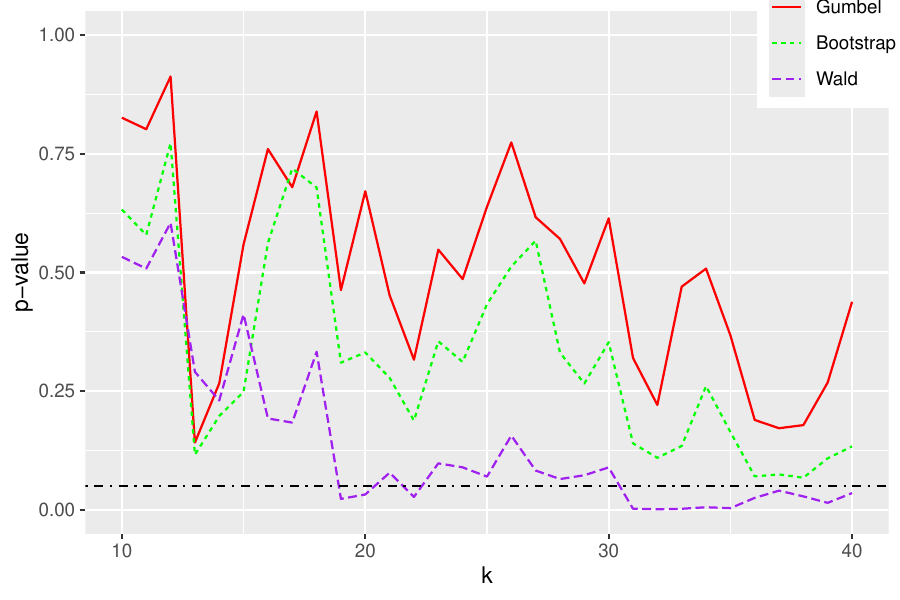}}
     \subfigure{
         \includegraphics[width =0.48\textwidth]{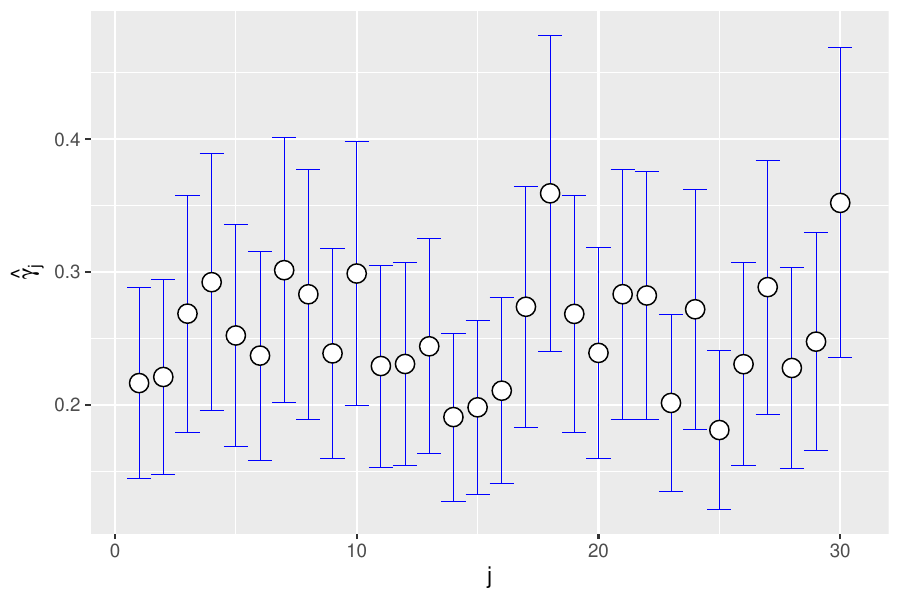}
     }
     \caption{Left: the $p$-values of the tests against different levels of $k$  for the portfolio data. The black horizontal line indicates the significance level, $\alpha = 0.05$.  Right: the Hill estimates $\widehat{\gamma}_j$ with  95\% confidence intervals for the portfolio data.}
     \label{figure:application:portfolio}
 \end{figure}

 We also analyze a dataset containing daily loss returns from $30$ different portfolios spanning 2010 to 2019, resulting in 1258 observations\footnote{The dataset is  downloaded from the Kenneth French Data Library.}.  The obtained $p$-values against various levels of $k$ are shown in the  left panel of  Figure \ref{figure:application:portfolio}. 
 The  Gumbel test and the multiplier bootstrap test produce $p$-values  higher than the $0.05$  for all levels of $k$.   Consequently, the result suggests that the null hypothesis should not be rejected at the 5\% significance level. 
 The Wald-type test will reject the null hypothesis for some $k$ values and not reject it for other values of $k$.  We also display the Hill estimates with a 95\% confidence interval for $k=35$
  in the right panel of Figure \ref{figure:application:portfolio}, and observe no substantial differences across these Hill estimates.
 
\section*{Acknowledgments}
The authors are grateful to the Editor, Associate Editor, and two anonymous referees for their valuable suggestions. 
In particular, their encouragement for exploring the non-sparse case has  greatly improved this work. Liujun Chen’s research was partially supported by the National Key R\&D Program of China(2024YFA1012200), and the National Natural Science Foundation of China grants 12301387 and 12471279. The authors report there are
no competing interests to declare.

 \newpage 
\appendix
\setcounter{equation}{0}   
\renewcommand{\theequation}{S\arabic{equation}}
\setcounter{prop}{0}   
\renewcommand{\theprop}{S\arabic{prop}}
\setcounter{lemma}{0}   
\renewcommand{\thelemma}{S\arabic{lemma}}

\setcounter{page}{1}   
\renewcommand{\thepage}{S\arabic{page}}

\begin{center}
\LARGE
\textbf{Supplementary Material}	
\end{center}

\section{Proof of Theorem \ref{theorem:max:i}}

Recall that, 
$$
\widetilde{\gamma}_j(k_j) =  \frac{ \sum_{i=1}^n \set{ \log  X_i^{( j)} -\log U_j(n/k_j)}\mI\set{X_i^{(j)}\ge U_j(n/k_j)}  }{ \sum_{i=1}^n \mI\set{X_i^{(j)}\ge U_j(n/k_j)}}.
$$
Write
$$
\begin{aligned}
    \mbT^2(k_1,\dots,k_p) = &\max_{1\le j \le p} \set{ \sqrt{k_j} \suit{ \frac{\widehat{\gamma}_j(k_j)}{\gamma_j} -1} }^2\\
    =& \max_{1\le j \le p}\set{ \sqrt{k_j} \suit{ \frac{\widetilde{\gamma}_j(k_j)}{\gamma_j}-1 }  + \sqrt{k_j}\suit{\frac{\widehat{\gamma}_j(k_j) - \widetilde{\gamma}_j(k_j)}{\gamma_j}} }^2,
\end{aligned}
$$
By Lemma \ref{lemma:max:inequality} below, we have that, 
$$
\begin{aligned}
  &  \abs{ \mbT^2(k_1,\dots,k_p)  -  \max_{1\le j \le p}\set{ \sqrt{k_j} \suit{ \frac{\widetilde{\gamma}_j(k_j)}{\gamma_j}-1 } }^2 }  \\
    \le & \max_{1\le j \le p}\set{ \sqrt{k_j}\suit{\frac{\widehat{\gamma}_j(k_j) - \widetilde{\gamma}_j(k_j)}{\gamma_j}} }^2 \\
    &+ 2\max_{1\le j \le p} \set{ \sqrt{k_j}\abs{\frac{\widehat{\gamma}_j(k_j) - \widetilde{\gamma}_j(k_j)}{\gamma_j}} }\max_{1\le j\le p} \set{ \sqrt{k_j} \abs{ \frac{\widetilde{\gamma}_j(k_j)}{\gamma_j}-1 } }.
\end{aligned}
$$
Theorem \ref{theorem:max:i}   follows from the following two lemmas.

\begin{lemma}\label{lemma:tu:tilde}
    Assume that Conditions \ref{condition:SOC}, \ref{Eigens} and \ref{condition:k:choice} hold.  Then, as $n\to\infty$, for every $x>0$,
   $$
   \begin{aligned}
   \Pr \msuit{\max_{1\le j \le p}\set{ \sqrt{k_j} \suit{ \frac{\widetilde{\gamma}_j(k_j)}{\gamma_j}-1 } }^2 - 2\log p + \log \suit{\log p}\le x }\to & \exp\set{-\frac{1}{\sqrt{\pi}}\exp(-x/2)}, \\
   \Pr \msuit{\max_{1\le j \le p}\set{ \sqrt{k_j} \suit{ \frac{\widetilde{\gamma}_j(k_j)}{\gamma_j}-1 } } \le  \suit{2\log p - \log \suit{\log p}+ x}^{1/2} }\to & \exp\set{-\frac{1}{2\sqrt{\pi}}\exp(-x/2)}.
   \end{aligned}
   $$ 
\end{lemma}

\begin{lemma}\label{lemma:gamma:diff}
Assume that Conditions \ref{condition:SOC} and \ref{condition:k:choice} hold.  Then, as $n\to\infty$,  
    $$
    \max_{1\le j \le p} \abs{ \sqrt{k_j}\suit{\frac{\widehat{\gamma}_j(k_j) - \widetilde{\gamma}_j(k_j)}{\gamma_j}}} =o_P(1/\sqrt{\log p}).
    $$
\end{lemma}

\subsection{Preliminary Lemmas}
In this subsection, we gather some lemmas that are useful for  proving Lemma \ref{lemma:tu:tilde} and Lemma \ref{lemma:gamma:diff}.

\begin{lemma}\label{lemma:max:inequality}
  Let $a_i$ and $b_i$, $i=1,\dots,n $, be two real sequences. Then, 
  $$
  \abs{\max_{1\le i\le n} a_i^2 - \max_{1\le i\le n} b_i^2} \le 2\max_{1\le i\le n} \abs{a_i} \max_{1\le i\le n }\abs{a_i-b_i} +\max_{1\le i\le n} \suit{a_i-b_i}^2.
  $$
\end{lemma}
\begin{proof}[Proof of Lemma \ref{lemma:max:inequality}]
We first consider the case $\max_{1\le i\le n}\abs{a_i} \ge \max_{1\le i\le n}\abs{b_i}$. In this case, we have that, 
$$
\begin{aligned}
    \abs{\max_{1\le i\le n} a_i^2 - \max_{1\le i\le n} b_i^2} =  & \max_{1\le i\le n} a_i^2 - \max_{1\le i\le n} b_i^2 \\
    =& \max_{1\le i\le n}(b_i+a_i-b_i)^2 - \max_{1\le i\le n} b_i^2\\
    =& \max_{1\le i\le n} \set{b_i^2+(a_i-b_i)^2+2 b_i(a_i-b_i) } - \max_{1\le i\le n} b_i^2\\
    \le & \max_{1\le i \le n} (a_i-b_i)^2  +2 \max_{1\le i\le n} |b_i| \max_{1\le i\le n} \abs{a_i-b_i}\\
    \le &\max_{1\le i \le n} (a_i-b_i)^2  +2 \max_{1\le i\le n} |a_i| \max_{1\le i\le n} \abs{a_i-b_i}.
\end{aligned}
$$

Next, we consider the case   $\max_{1\le i\le n}\abs{a_i} < \max_{1\le i\le n}\abs{b_i}$. In this case, we have that, 
$$
\begin{aligned}
    \abs{\max_{1\le i\le n} a_i^2 - \max_{1\le i\le n} b_i^2} =& \max_{1\le i\le n} b_i^2 - \max_{1\le i\le n} a_i^2 \\
    =& \max_{1\le i\le n}(a_i+b_i-a_i)^2  - \max_{1\le i\le n} a_i^2 \\
    =& \max_{1\le i\le n}\set{a_i^2 +(b_i-a_i)^2 +2a_i(b_i-a_i)}  - \max_{1\le i\le n} a_i^2 \\
    \le &  \max_{1\le i \le n} (a_i-b_i)^2  +2 \max_{1\le i\le n} |a_i| \max_{1\le i\le n} \abs{a_i-b_i} . 
\end{aligned}
$$
The proof is then complete.
\end{proof}

\begin{lemma}[Bernstein's inequality, see \cite{shorack1986empirical}, page 855]\label{Bernstein}
    Let $Z_1,\dots,Z_n$ be independent random variables with $|Z_i|\le M$ almost surely and $\bE(Z_i) = \mu$ for all $1\le i\le n$, where $M>0, \mu\in\mathbb{R}$.  Then, for any $\varepsilon>0$,  
    $$
\Pr\suit{\abs{\frac{1}{n}\sum_{i=1}^n Z_i - \mu}>\varepsilon} \le 2 \exp\set{-\frac{n\varepsilon^2}{2\frac{1}{n}\sum_{i=1}^n \mathrm{Var}(Z_i) +2 M\varepsilon/3}}.
    $$
\end{lemma}

\begin{lemma}[Bonferroni inequality]
    Let $B = \cup_{t=1}^p B_t$. For any $s<[p/2]$,
    $$
    \sum_{t=1}^{2s}(-1)^{t-1} E_t \le \Pr(B)\le \sum_{t=1}^{2s-1}(-1)^{t-1} E_t,
    $$
    where 
    $$
    E_t = \sum_{1\le i_1\le \cdots \le i_t\le p}\Pr\suit{B_{i_1}\cap \cdots \cap B_{i_t}}.
    $$
\end{lemma}

\begin{lemma}\label{lemma:bound:k}
    Assume that Conditions \ref{condition:SOC} and \ref{condition:k:choice} hold. Then as $n\to\infty$,
   $$
  \max_{1\le j \le p} \sqrt{k_j}\abs{\frac{\widetilde{k}_j}{k_j}-1}=O_P(\sqrt{\log p}),
   $$
where
$$
\widetilde{k}_j = \sum_{i=1}^n \mI\set{X_i^{(j)}\ge U_j(n/k_j)}.
$$
\end{lemma}
\begin{proof}[Proof of Lemma \ref{lemma:bound:k}]
Denote 
$$
\begin{aligned}
    Z_i^{(j)} = & \frac{n\sqrt{ k_j/\log p }}{k_j}\mI\set{X_i^{(j)}\ge U_j(n/k_j)}.
\end{aligned}
$$
Then 
$$
\frac{1}{n}\sum_{i=1}^n  Z_i^{(j)}  =  \frac{\widetilde{k}_j}{k_j} \sqrt{k_j/\log p}.
$$
Obviously, we have that,  $Z_i^{(j)} \le  n /\sqrt{k_j\log p} $ almost surely, and
$$
\begin{aligned}
    \bE Z_i^{(j)} = &\sqrt{k_j/\log p},  \\
    \textrm{Var} (Z_i^{(j)})  =&  \frac{n^2 }{k_j \log p} \frac{k_j}{n}\suit{1-\frac{k_j}{n}} = \frac{n}{\log p}\suit{1-\frac{k_j}{n}} .
\end{aligned}
$$

By applying Lemma \ref{Bernstein} with $\mu = \sqrt{k_j/\log p}$ and $M = n/\sqrt{k_j\log p }$, we have that, for any $M_0>0$, 
$$
\begin{aligned}
    \Pr\suit{ \sqrt{k_j}\abs{ \frac{\widetilde{k_j}}{k_j}-1}> M_0 \sqrt{\log p}} & =   \Pr\suit{\abs{\frac{1}{n}\sum_{i=1}^n Z_i^{(j)}-\sqrt{k_j/\log p}}>M_0}\\
    &\le 2 \exp\suit{-\frac{nM_0^2}{2\frac{n}{\log p}+\frac{2 n}{3\sqrt{k_j\log p}} M_0  }}\\
    &=2 \exp\suit{ -   \log p\frac{M_0^2}{ 2  + \frac{2}{3}M_0 \sqrt{\frac{\log p}{k_j}} } }.
\end{aligned}
$$
  It follows that, for any $M_0>0$,
  $$
  \begin{aligned}
    \Pr\suit{ \max_{1\le j \le p} \sqrt{k_j}\abs{ \frac{\widetilde{k_j}}{k_j}-1}>M_0 \log p} &\le 2p \max_{1\le j\le p} \exp\suit{ -   \log p\frac{M_0^2}{ 2  + \frac{2}{3}M_0 \sqrt{\frac{\log p}{k_j}} } }.
  \end{aligned}
  $$
  By    Condition    \ref{condition:k:choice}, we have that, as $n\to\infty$,
$$
\begin{aligned}
   \max_{1\le j \le p} \frac{\log p}{k_j} =& o(1), \\
\end{aligned}
$$
Thus,  by taking $M_0>2$, we have that,  as $n\to\infty$, 
$$
2p \max_{1\le j\le p} \exp\suit{ -   \log p\frac{M_0^2}{ 2  + \frac{2}{3}M_0 \sqrt{\frac{\log p}{k_j}} } } \to 0, 
$$
and hence 
$$
\Pr\suit{ \max_{1\le j \le p} \sqrt{k_j}\abs{ \frac{\widetilde{k}_j}{k_j}-1}>M_0\log p} \to 0. 
$$
Then Lemma \ref{lemma:bound:k} holds.
\end{proof}

\begin{lemma}\label{Lemma:uniform:distribution}
    Assume that Conditions \ref{condition:SOC} and \ref{condition:k:choice} hold. Then as $n\to\infty$,
    $$
\sup_{1\le j \le p}\sqrt{k_j}\abs{\frac{U_j(n/k_j)}{ X_{n-k_j,n}^{(j)}}-1} = O_{p}(\sqrt{\log  p}).
    $$
\end{lemma}
\begin{proof}[Proof of Lemma \ref{Lemma:uniform:distribution}]

   Define $\Gamma_i^{(j)} = 1-F_j(X_i^{(j)}), i=1,\dots,n, j=1,\dots,p$. 
   Let $\Gamma_{1,n}^{(j)}\le \dots \le \Gamma_{n,n}^{(j)}$
   denote the order statistics of $\Gamma_1^{(j)},\dots, \Gamma_{n}^{(j)}$, $j=1,\dots,p$. 
   Then 
   $$
X_{n-k_j,n}^{(j)} = U_j\suit{\frac{1}{\Gamma_{k_j+1,n}^{(j)}}}, \quad  j=1,\dots,p.
   $$

    First, we show that,
\begin{equation}\label{eq:uniform:dis:joint}
    \max_{1\le j \le p} \sqrt{k_j}\abs{\frac{n\Gamma_{k_j+1,n}^{(j)}}{k_j}-1}=O_P(\sqrt{\log p}).
\end{equation}
For any $M_0>0$, we have that, 
$$
\begin{aligned}
    \Pr\suit{  \max_{1\le j\le p} \sqrt{k_j}\abs{\frac{n \Gamma_{k_j+1,n}^{(j)} }{k_j}-1} \ge \sqrt{\log p } M_0}\le& p\max_{1\le j\le p}\Pr\suit{ \sqrt{k_j}\abs{\frac{n \Gamma_{k_j+1,n}^{(j)} }{k_j}-1} \ge \sqrt{\log p } M_0}\\
    = & p\max_{1\le j \le p}  \Pr\suit{\abs{\Gamma_{k_j+1,n}^{(j)} - \frac{k_j}{n} } \ge M_0 \sqrt{\frac{\log p}{k_j}}\frac{k_j}{n}}.
\end{aligned}
$$
Note that, $\Gamma_{k_j+1,n}^{(j)}$ follows a Beta distribution $B(k_j+1,n-k_j)$, see e.g. Chapter 3.1 of \cite{shorack1986empirical}.  
Thus, by \cite{skorski2023bernstein}, we have that,  for any $x>0$, 
\begin{equation}\label{eq:Bernstein:beta}
    \Pr\suit{ \abs{\Gamma_{k_j+1,n}^{(j)}-\frac{k_j+1}{n} }> x}\le \exp\suit{-\frac{x^2}{2(a_n+b_nx/3)} }+\exp\suit{-\frac{x^2}{2a_n}}, 
\end{equation}
where 
$$
\begin{aligned}
    a_n =& \frac{(k_j+1)(n-k_j)}{(n+1)^2(n+2)} \sim \frac{k_j}{n^2},\\ 
    b_n =& \frac{2(n-k_j-1)}{(n+1)(n+3)} \sim \frac{2}{n}. \\
\end{aligned}
$$
By applying \eqref{eq:Bernstein:beta} with $x = M_0 \sqrt{\frac{\log p}{k_j}}\frac{k_j}{n}$, we have that,  as $n\to\infty$, 
$$
\begin{aligned}
    \frac{x^2}{a_n} \sim & M_0^2\log p, \\\
    \frac{x^2}{xb_n}\sim & \frac{M_0}{2}\sqrt{k_j\log p}.
\end{aligned} 
$$
Thus, by taking $M_0$ sufficiently large, we have that, as $n\to\infty$, 
$$
\begin{aligned}
p\max_{1\le j \le p}\Pr\suit{ \abs{\Gamma_{k_j+1,n}^{(j)}-\frac{k_j+1}{n} }>  M_0 \sqrt{\frac{\log p}{k_j}}\frac{k_j}{n}} = o(1).
\end{aligned}
$$
Hence, \eqref{eq:uniform:dis:joint} holds.

By  \eqref{eq:uniform:dis:joint}, we have that, as $n\to\infty$, with probability tending to 1, 
$$
\min_{1\le j \le p} \frac{1}{\Gamma_{k_j+1,n}^{(j)}} \to\infty.
$$ 
Thus, by \eqref{eq:uniform:dis:joint} and  Condition \ref{condition:SOC}, we have that,  as $n\to\infty$,
$$
\begin{aligned}
 \sqrt{k_j}\abs{\frac{U_j(n/k_j)}{ X_{n-k_j,n}^{(j)}}-1} =&  \sqrt{k_j}\abs{\frac{U_j(n/k_j)}{ U_j(1/\Gamma_{k_j+1,n}^{(j)})}-1} \\
 =& \sqrt{k_j}\abs{  \suit{\frac{n\Gamma_{k_j+1,n}^{(j)}}{k_j}}^{\gamma_j}-1   +O_P\suit{1}A_j(n/k_j) },
\end{aligned}
$$
where the $O_P(1)$ term is uniform for all $1\le j\le p$. 
By  applying the mean-value theorem to the function $f_j(x) = (1+x)^{\gamma_j}$, we have that, 
$$
\begin{aligned}
    \suit{\frac{n\Gamma_{k_j+1,n}^{(j)}}{k_j}}^{\gamma_j}-1  = & \gamma_j \suit{\frac{n\Gamma_{k_j+1,n}^{(j)}}{k_j}-1  }+ \frac{\gamma_j(\gamma_j-1)}{2}(1+\zeta_j)^{\gamma_j-2} \suit{\frac{n\Gamma_{k_j+1,n}^{(j)}}{k_j}-1  }^2,
\end{aligned}
$$
where $\zeta_j$ is between $0$ and $\frac{n\Gamma_{k_j+1,n}^{(j)}}{k_j}-1$. Thus, by \eqref{eq:uniform:dis:joint}, we have that, 
$$
\suit{\frac{n\Gamma_{k_j+1,n}^{(j)}}{k_j}}^{\gamma_j}-1 =\gamma_j \suit{\frac{n\Gamma_{k_j+1,n}^{(j)}}{k_j}-1  } +o_P\suit{1}\frac{\log p}{k_j},
$$
where the $o_p(1)$ term is uniform for all $1\le j \le p$.  It follows that, as $n\to\infty$,
\begin{equation}\label{eq:expansion:UX}
    \begin{aligned}
        &\max_{1\le j \le p}\sqrt{k_j}\abs{\frac{U_j(n/k_j)}{ X_{n-k_j,n}^{(j)}}-1} \\
        \le &\max_{1\le j \le p}\sqrt{k_j}\gamma_j \abs{\frac{n\Gamma_{k_j+1,n}^{(j)}}{k_j}-1 } +\max_{1\le j \le p} \abs{\sqrt{k_j}A_j(n/k_j)} +o(1)\max_{1\le j \le p} \sqrt{k_j} \frac{\log p}{k_j}  \\
        =&O_P(1)\sqrt{\log p},
    \end{aligned}
\end{equation}
by  \eqref{eq:uniform:dis:joint} and   Condition  \ref{condition:k:choice}. The proof is then complete.
\end{proof}

\begin{lemma}\label{Lemma:kiefer}
    Let $\Gamma_1,\dots, \Gamma_n$ be i.i.d. random variables following uniform distributions on the interval $[0,1]$,
    and  $\Gamma_{1,n}\le \dots \le \Gamma_{n,n}$
    denote the order statistics of $\Gamma_1,\dots, \Gamma_{n}$. Let $k = k(n)$ be an intermediate sequence such that,  $k\to\infty, k/n\to 0$, as $n\to\infty$. 
    Define
$$
R_{n}  = \Gamma_{k+1,n} - \frac{k}{n} + \frac{1}{n}\sum_{i=1}^n \mI\suit{ \Gamma_i\le \frac{k}{n}} -\frac{k}{n},
$$ 
Then, for  sufficiently large $n$, 
$$
\Pr\suit{ \abs{R_n}>C_1 \frac{k^{2/5}}{n}  }  \le  C_2\exp(-C_3k^{1/5}),
$$
where $C_1>0, C_2>0$, and $C_3>0$ are positive constants   not depending on $n$.

\end{lemma}
\begin{remark}
    A similar, but somewhat different result has been shown in 
  \cite{gribkova2012bahadur}, i.e., as $n\to\infty$,
  $$
  \Pr\suit{ \abs{R_n}> C_1 \frac{k^{1/4} \log^{3/4} (n)}{n} } = O(n^{-c}), 
$$
for any constant $c>0$.
\end{remark}

\begin{proof}[Proof of Lemma \ref{Lemma:kiefer}]
    Denote
    $\tau_{n} = k/n$ and $\widetilde{k} = \sum_{i=1}^n \mI\suit{ \Gamma_i\le \tau_{n} }$.
We first define a set 
$$
\mathcal{F} = \set{ \abs{\widetilde{k}- k} < C_4 k^{1/2+\varepsilon}},
$$
where $\varepsilon = 1/10$ and $C_4>0$ is a positive constant. 
By using the Bernstein's inequality (Lemma \ref{Bernstein}), we have that, for sufficiently large $n$, 
\begin{equation}\label{eq:kiefer:set}
    \begin{aligned}
        \Pr\suit{ \abs{\widetilde{k}- k} \ge C_4 k^{1/2+\varepsilon} }\le & \exp\suit{ -\frac{1}{2}\frac{C_4^2 k^{1+2\varepsilon}}{k+\frac{1}{3} C_4k^{1/2+\varepsilon} } }  \le C_5\exp(-C_6k^{1/5}),
    \end{aligned}
\end{equation}
for some $C_5>0, C_6>0$ not depending on $n$. 
It follows that,  
$$
\Pr\suit{\mathcal{F}}\ge 1-C_5\exp(-C_6k^{1/5}).
$$

Without loss of generality, we assume that $k+1\le \widetilde{k}$ (the case $k+1>\widetilde{k}$ can be handled in a similar way). 
 Write 
$$
\begin{aligned}
    R_n =  \Gamma_{k+1,n} - \tau_{n}\frac{k+1}{\widetilde{k}+1}+R_n^\prime,
\end{aligned}
$$
where 
$$
\begin{aligned}
    R_n^\prime  = \tau_{n} \frac{k+1}{\widetilde{k}+1} - \tau_{n} +\frac{\widetilde{k}-k}{n}  =\frac{(\widetilde{k}-k)^2}{n(\widetilde{k}+1)}+\frac{\widetilde{k}-k}{n(\widetilde{k}+1)}.
\end{aligned}
$$
On the set $\mathcal{F}$, we have that, as $n\to\infty$,
$$
R_n^\prime =O\suit{1}\frac{k^{2\varepsilon}}{n}  =o(1)\frac{k^{2/5}}{n}.
$$
Thus, the term $R_n^\prime $ is of a negligible order for our
purposes and it suffices to show that, as $n\to\infty$,
\begin{equation}\label{eq:kiefer:gamma}
    \Pr\suit{ \abs{\Gamma_{k+1,n} - \tau_{n}\frac{k+1}{\widetilde{k}+1}}>C_1 \frac{k^{2/5}}{n}  }  \le C_7\exp(-C_8 k^{1/5}),
\end{equation}
for some $C_7>0, C_8>0$ not depending on $n$.

Conditional on $\widetilde{k}$, by Lemma 6.1 of \cite{gribkova2012bahadur},  we have that, the order statistic $\Gamma_{k+1,n}$ is distributed as the $(k+1)$-th order statistic $\Gamma_{k+1, \widetilde{k}}^\prime$ of an independent sample $U_1^\prime, \dots, U_{\widetilde{k}}^\prime$ of random variables uniformly distributed on the interval $(0, \tau_n)$. Its expectation is 
$$
\bE\suit{\Gamma_{k+1,\widetilde{k}}^\prime} = \tau_{n} \frac{k+1}{\widetilde{k}+1}.
$$
Then, we can write
$$
\begin{aligned}
    & \Pr\suit{ \abs{\Gamma_{k+1,n} - \tau_{n}\frac{k+1}{\widetilde{k}+1}}>C_1 \frac{k^{2/5}}{n}}\\ 
    =& \Pr\suit{ \abs{\Gamma_{k+1,\widetilde{k}}^\prime - \tau_{n}\frac{k+1}{\widetilde{k}+1}}>C_1 \frac{k^{2/5}}{n}  }\\
    =&  \Pr\suit{ \Gamma_{k+1,\widetilde{k}}^\prime > \tau_{n}\frac{k+1}{\widetilde{k}+1}+C_1 \frac{k^{2/5}}{n}  } + \Pr\suit{ \Gamma_{k+1,\widetilde{k}}^\prime < \tau_{n}\frac{k+1}{\widetilde{k}+1}-C_1 \frac{k^{2/5}}{n}  } \\
    =&: P_{1} +P_{2}.
\end{aligned}
$$

We start with dealing with $P_1$. 
Denote 
$$
S_{n}  =\sum_{i=1}^{\widetilde{k}} \mI\set{U_{i}^\prime \le  \tau_{n} q_{n}}
$$
where
$$
\begin{aligned}
    q_{n} =& \min\suit{\frac{k+1}{\widetilde{k}+1}+C_1  k^{-3/5},1}  .
\end{aligned}
$$

If $q_{n} = 1$, then $P_{1} = 0$, and  the inequality we need is valid trivially. Then, we focus on the case $q_{n}<1$. The probability $P_{1}$ equals to 
$$
\begin{aligned}
    \Pr\suit{S_{n}<k+1} = & \Pr\suit{\frac{S_{n}}{\widetilde{k}} - q_{n}<\frac{k+1}{\widetilde{k}}-\frac{k+1}{\widetilde{k}+1}-  C_1 k^{-3/5}  }.
\end{aligned}
$$
On the set $\mathcal{F}$, we have that, as $n\to\infty$,
$$
\frac{k+1}{\widetilde{k}}-\frac{k+1}{\widetilde{k}+1} = \frac{k+1}{\widetilde{k}(\widetilde{k}+1)} = o(k^{-3/5}).
$$
 Then, this term can be omitted for our analysis purposes. 

By applying inequality (2.2) of \cite{hoeffding1963probability}, we obtain that,  
$$
\begin{aligned}
    \Pr\suit{\frac{S_{n}}{\widetilde{k}} - q_{n}<-C_1k^{-3/5} } \le   \exp\suit{ - \frac{C_1^2}{2}\frac{\widetilde{k} k^{-6/5}}{q_n(1-q_n)} }.
\end{aligned}
$$
On the set $\mathcal{F}$, we have that, for sufficiently large $n$,
$$
1-q_{n}  = 1-\frac{k+1}{\widetilde{k}+1} - C_1k^{-3/5}= \frac{\widetilde{k}-k}{\widetilde{k}+1} -C_1k^{-3/5}          \le C_4 k^{-1/2+\varepsilon},
$$
and hence
$$
\frac{C_1^2}{2}\frac{\widetilde{k} k^{-6/5}}{q_{n}(1-q_{n})} \ge \frac{C_1^2}{2C_4}\frac{\widetilde{k} k^{-6/5}}{ k^{-1/2+\varepsilon}} = \frac{C_1^2}{2C_4} \widetilde{k} k^{-4/5}  =\frac{C_1^2}{2C_4} \frac{\widetilde{k}}{k} k^{1/5}\ge C_9 k^{1/5},
$$
for some $C_9>0$ not depending on $n$.
Then, we have that, for sufficiently large $n$, 
$$
P_{1} \le C_{10}\exp(-C_9 k^{1/5}),
$$
for some $C_{10}>0$ not depending on $n$.
Similarly, we can show that, $P_{2} \le C_{11}\exp(-C_{12}k^{1/5})$. Thus,  \eqref{eq:kiefer:gamma} holds and the proof is then complete. 
\end{proof}

\subsection{Proof of Lemma \ref{lemma:tu:tilde}}

We only prove the first statement. The second statement can be proved in a similar way.
Recall
$$
\begin{aligned}
    \widetilde{\mbT}^2(k_1,\dots,k_p)=& \max_{1\le j\le p} k_j \suit{\frac{\widetilde{\gamma}_j(k_j)}{\gamma_{j}}-1}^2\\
     = &\max_{1\le j\le p} k_j  \set{\frac{\sum_{i=1}^n \frac{1}{\gamma_j}\log \suit{\frac{X_i^{(j)}}{U_j(n/k_j)}} \mI\suit{X_i^{(j)}>U_j(n/k_j)} -  \widetilde{k}_j  }{ \widetilde{k}_j  }}^2\\
     =& \max_{1\le j \le p} \frac{\msuit{\sqrt{k_j}\set{\frac{1}{k_j}\sum_{i=1}^n \frac{1}{\gamma_j}\log \suit{\frac{X_i^{(j)}}{U_j(n/k_j)}} \mI\suit{X_i^{(j)}>U_j(n/k_j)}-  \widetilde{k}_j/k_j} }^2}{ \suit{\widetilde{k}_j/k_j}^2 } \\
     =& \max_{1\le j \le p} \frac{\msuit{\sum_{i=1}^n \frac{1}{\sqrt{k_j}} \set{ \frac{1}{\gamma_j}\log \suit{\frac{X_i^{(j)}}{U_j(n/k_j)}} - 1}\mI\suit{X_i^{(j)}>U_j(n/k_j)} }^2}{ \suit{\widetilde{k}_j/k_j}^2  } \\
     =& \max_{1\le j \le p} \frac{\set{\sum_{i=1}^n \frac{1}{\sqrt{n}} Y_i^{(j)}  }^2}{ \suit{\widetilde{k}_j/k_j}^2},
\end{aligned}
$$
where 
$$
Y_i^{(j)} =\sqrt{\frac{n}{k_j}} \suit{ \frac{1}{\gamma_j}\log \frac{X_i^{(j)}}{U_j(n/k_j)}-1}\mI\suit{X_i^{(j)}>U_j(n/k_j)}, \quad j=1,\dots,p, \quad  i=1,\dots,n.
$$
Obviously, 
$$
\frac{\max_{1\le j \le p} \set{\sum_{i=1}^n  \frac{Y_i^{(j)}}{\sqrt{n}}}^2 }{\max_{1\le j \le  p}  (\widetilde{k}_j/k_j)^2} \le \widetilde{\mbT}^2(k_1,\dots,k_p) \le \frac{\max_{1\le j \le p} \set{\sum_{i=1}^n  \frac{Y_i^{(j)}}{\sqrt{n}}}^2 }{\min_{1\le j \le  p}  (\widetilde{k}_j/k_j)^2}.
$$
By Lemma \ref{lemma:bound:k} and    Condition  \ref{condition:k:choice}, we have that, 
$$
 (\widetilde{k}_j/k_j)^2 = \suit{1+O_P(1)k_j^{-1/2}\sqrt{\log p}}^2 = 1+o_P(1)(1/\log p),
$$
uniformly for all $1\le j\le p$. Thus,
it suffices to show that, as $n\to\infty$,
\begin{equation}\label{eq:proof:Y:max:norm}
    \Pr\set{\max_{1\le j \le p} \suit{\sum_{i=1}^n  \frac{Y_i^{(j)}}{\sqrt{n}} }^2 -2\log p +\log \log (p)\le x  } \to \exp\suit{-\frac{1}{\sqrt{\pi}}\exp\suit{-\frac{x}{2}}}.
\end{equation}

Consider a truncted version of $Y_i^{(j)}$,
    $$
  \begin{aligned}
  	    \widetilde{Y}_i^{(j)}   	    :=&  Y_i^{(j)}   I\suit{  Y_i^{(j)}\le \tau_j^\prime} \\= & \sqrt{\frac{n}{k_j}} \suit{ \frac{1}{\gamma_j}\log \frac{X_i^{(j)}}{U_j(n/k_j)}-1}\mI\suit{U_j(n/k_j)<X_i^{(j)}\le \tau_j }  \\
  \end{aligned}
    $$
    where 
    \begin{equation}\label{s:def:tau}
    	    \begin{aligned}
                	   \tau_j^\prime =& 2 \sqrt{\frac{n}{k_j}} \log(k_j+p), \\
    	   \tau_{j} =& U_j(n/k_j)\exp(\gamma_j +3\gamma_j\log(k_j+p)), \\
    \end{aligned}
    \end{equation} 
  We intend to show that,  as $n\to\infty$,
    \begin{equation}\label{eq:set}
       \begin{aligned}
 \Pr\suit{ \max_{1\le i\le n}  X_i^{(j)}>\tau_j, \quad \text{for some} \  j\in  \set{1,\dots,p} } 
    \to  0.
       \end{aligned}
    \end{equation}
Note that, 
    $$
    \begin{aligned}
         &  \Pr\suit{ \max_{1\le i\le n}  X_i^{(j)}>\tau_j, \quad \text{for some} \  j\in  \set{1,\dots,p} }  \\
        \le & np \max_{1\le j \le p} \Pr\suit{X^{(j)} > \tau_j } \\
        =&  \max_{1\le j \le p} k_jp  \frac{\overline{F}_j\set{ U_j(n/k_j)\exp\suit{\gamma_j+3\gamma_j\log(k_j+\rho)}  } }{\overline{F}_j\set{U_j(n/k_j)}}, \\
    \end{aligned}
    $$
where $\overline{F}^{(j)}  =1-F^{(j)}$. 
By the Potter's inequality \citep{potter1942mean} and Condition \ref{condition:SOC}, 
we have that, with   $0<\varepsilon< \frac{1}{3}\min_{1\le j\le p} \gamma_j^{-1}$,   as $n\to\infty$,
$$
\begin{aligned}
    &\max_{1\le j \le p} k_jp  \frac{\overline{F}^{(j)}\set{ U_j(n/k_j)\exp\suit{\gamma_j+3\gamma_j\log(k_j+\rho)}}   }{\overline{F}^{(j)}\set{U_j(n/k_j)}} \\
    \le & (1+\varepsilon)   \max_{1\le j \le p} k_jp   \exp\set{\suit{-1+\varepsilon\gamma_j}\suit{1+3\log(p+k_j )   }}  \\ 
    \le & (1+\varepsilon)  \max_{1\le j \le p} k_jp \exp\set{-\frac{2}{3}\suit{1+3\log(p+k_{j}) } } \\
    \to & 0.
\end{aligned}
$$
Thus, \eqref{eq:set} holds.  By  \eqref{eq:set}, we have that, 
\begin{equation}\label{s:trunc:Y}
\begin{aligned}
	&\Pr\suit{\abs{\max_{1\le j\le p}\suit{\sum_{i=1}^n  \frac{\widetilde{Y}_i^{(j)}}{\sqrt{n}}}^2  - \max_{1\le j\le p}\suit{ \sum_{i=1}^n \frac{Y_i^{(j)}}{\sqrt{n}}}^2}> 0} \\
    \le & \Pr\suit{ \max_{1\le i\le n}  X_i^{(j)}>\tau_j, \quad \text{for some} \  j\in  \set{1,\dots,p} }  \\
    \to & 0.
\end{aligned}	
\end{equation}
Thus,   we can prove \eqref{eq:proof:Y:max:norm} provided that,
\begin{equation}\label{proof:eq:normalization}
    \Pr \suit{\max_{1\le j \le p} \suit{\sum_{i=1}^n \frac{\widetilde{Y}_i^{(j)}}{\sqrt{n}}}^2 - 2\log p +\log \log p\le x}\to \exp\suit{-\frac{1}{\sqrt{\pi}}\exp\suit{-\frac{x}{2}}}.  
\end{equation}

Denote 
$$
V_j = \sum_{i=1}^n \frac{\widetilde{Y}_i^{(j)}}{\sqrt{n}}.
$$

Under Condition \ref{condition:k:choice}, by 
using calculations similar to those in the proof of Lemma \ref{lemma:bound:W} (see below),  we obtain that,   $
\bE V_j = o(1/\sqrt{\log p})$ and $\text{Var} (V_j)=1+o(1/\log p)$, as $n\to\infty$, uniformly for all $1\le j\le p$. 
Denote $x_p = 2\log p - \log \log (p) +x$. By the Bonferroni inequality,  we have for any $s\le [p/2]$,
$$
\sum_{t=1}^{2s} (-1)^{t-1} E_t\le \Pr\suit{\max_{1\le j\le p}|V_j|>\sqrt{x_p}} \le  \sum_{t=1}^{2s-1} (-1)^{t-1} E_t,
$$
where 
$$
E_t =\sum_{1\le j_1 <\dots<j_t\le p} \Pr\suit{|V_{j_1}|>\sqrt{x_p},\dots,|V_{j_t}|>\sqrt{x_p}}.
$$
Let $(W_1,\dots,W_p)^\top$ be  a Gaussian random vector with the same  covariance structure as the random vector $\suit{\widetilde{Y}_1^{(1)},\dots, \widetilde{Y}_1^{(p)}}^\top$. 
By Theorem 1.1 in \cite{zaitsev1987gaussian}, we have that, for any $\lambda\ge 0$ and  $t>0$,
$$
\begin{aligned}
   &\abs{ \Pr\suit{|V_{j_1}|>\sqrt{x_p},\dots,|V_{j_t}|>\sqrt{x_p}} -\Pr \suit{\min_{j \in \set{j_1,\dots,j_t}}\abs{W_j}>x_p- \lambda} }   \\
    \le & c_1 t^{5/2}\exp\suit{-\frac{\lambda}{c_2 t^{5/2} \max_{1\le j\le p}\tau_j^\prime/\sqrt{n}}},   
\end{aligned}
$$
where $c_1, c_2>0$ are positive constants.
We take $\lambda = \varepsilon (\log p)^{-1/2}$ where $\varepsilon  \to 0$ sufficiently slow.  Then, by  Condition \ref{condition:k:choice}, we have that, as $n\to\infty$, for any fixed $t>0$, 
$$
\begin{aligned}
    t^{5/2} \exp\suit{-\frac{\lambda }{c_2 t^{5/2} \tau_n/\sqrt{n}}} =   t^{5/2} \exp\suit{-\frac{\varepsilon \sqrt{ k_{\min} } }{c_2 t^{5/2}(\log p)^{1/2} \log(p+k_{\max} )}}= p^{-c},
\end{aligned}
$$
for any large $c>0$.
It follows that, for any fixed $s>0$, 
$$
\begin{aligned}
 \Pr\suit{\max_{1\le j\le p}|V_j|>x_p}  
    \le  \sum_{t=1}^{2s-1}(-1)^{t-1} \sum_{1\le j_1\le \cdots\le j_t\le p} \Pr \set{\min_{j \in \set{j_1,\dots,j_t}}\abs{W_j}>x_p- \varepsilon (\log p)^{-1/2}}+o(1),
\end{aligned}
$$
and
$$
\begin{aligned}
    \Pr\suit{\max_{1\le j\le p}|V_j|>x_p}  
    \ge  \sum_{t=1}^{2s}(-1)^{t-1} \sum_{1\le j_1\le \cdots\le j_t\le p} \Pr \set{\min_{j \in \set{j_1,\dots,j_t}}\abs{W_j}>x_p- \varepsilon (\log p)^{-1/2}}-o(1).
\end{aligned}
$$

The rest of the proofs are similar to that of Lemma 6 in \cite{tony2014two}, and thus omitted.  We have then proved \eqref{proof:eq:normalization} and then \eqref{eq:proof:Y:max:norm} and  Lemma \ref{lemma:tu:tilde}.

\subsection{Proof of Lemma \ref{lemma:gamma:diff}}

Define 
$$\widehat{\overline{F}}_{j,n}(x) =\frac{1}{n}\sum_{i=1}^n \mI\suit{X_i^{(j)} \ge x}.
$$ 
Using integration by parts, we have that, 
    $$
    \begin{aligned}
        \widehat{\gamma}_j(k_j) = &\frac{1}{k_j}\sum_{i=1}^n \log \frac{X_{i}^{(j)}}{X_{n-k_j,n}^{(j)}}\mI\suit{X_i^{(j)}\ge X_{n-k_j,n}^{(j)}} \\
        =& -\frac{n}{k_j} \int_{X_{n-k_j,n}^{(j)}}^\infty \log \frac{v}{X_{n-k_j,n}^{(j)}}d \widehat{\overline{F}}_{j,n}(v) \\
        =& \frac{n}{k_j}\int_{X_{n-k_j,n}^{(j)}}^\infty \widehat{\overline{F}}_{j,n}(v)\frac{1}{v}dv.
    \end{aligned}
    $$
Similarly, we have that, 
    $$
    \widetilde{\gamma}_j(k_j) = \frac{n}{\widetilde{k}_j}\int_{U_j(n/k_j)}^\infty \widehat{\overline{F}}_{j,n}(v)\frac{1}{v}dv.
    $$
It follows that, 
$$
\begin{aligned}
    &\sqrt{k}_j\suit{\frac{\widehat{\gamma}_j(k_j)-\widetilde{\gamma}_j(k_j)}{\gamma_j}} \\
    =&\frac{1}{\gamma_j} \sqrt{k_j}\int_{X_{n-k_j,n}^{(j)}}^{U_j(n/k_j)}\frac{n}{k_j} \widehat{\overline{F}}_{j,n}(v)\frac{1}{v}dv+ \sqrt{k_j}\suit{\frac{\widetilde{k}_j}{k_j} -1}  \frac{ \widetilde{\gamma_j}(k_j)}{\gamma_j} \\
    =& I_{1,n,j} +I_{2,n,j}.
\end{aligned}
$$
We are going to show that $I_{1,n,j}$ and $I_{2,n,j}$ are related to the tail quantile process and tail empirical process, respectively. Then, $I_{1,n,j} +I_{2,n,j}$ is related to the Bahadur-Kiefer process \citep{kiefer1967bahadur}.

We start with   $I_{1,n,j}$. Note that,
$$
\begin{aligned}
    \frac{1}{\gamma_j}\int_{X_{n-k_j,n}^{(j)}}^{U_j(n/k_j)}\frac{n}{k_j} \widehat{\overline{F}}_{j,n}(v)\frac{1}{v}dv = \frac{1}{\gamma_j} \int_{1}^{U_j(n/k_j)/X_{n-k_j,n}^{(j)} } \frac{n}{k_j} \widehat{\overline{F}}_{j,n}(xX_{n-k_j,n}^{(j)})\frac{dx}{x}.
\end{aligned}
$$ Define 
$$
g_j(x) = \widehat{\overline{F}}_n\suit{xX_{n-k_j,n}^{(j)}}\frac{1}{x}.
$$ By the mean-value theorem to the function $y:\mapsto\int_{1}^{y} g_j(x)dx$, we have that, 
$$
I_{1,n,j} = g_j(\zeta_j) \frac{1}{\gamma_j}\suit{\frac{ U_j(n/k_j)}{X_{n-k_j,n}^{(j)}} -1 },
$$
where $\zeta_j$ is  between $1$ and $U_j(n/k_j)/X_{n-k_j,n}^{(j)} $. Since $g_j(x)$
is a monotone function of $x$ and $g_j(1)=1$, we have that,
$$
\abs{g_j(\zeta_j)-1}\le \abs{\frac{n}{k_j}  \widehat{\overline{F}}_{j,n}(U_j(n/k_j))-1} = \abs{\frac{\widetilde{k}_j}{k_j}-1}=O_P(1)\sqrt{\frac{\log p}{k_j}},
$$
by Lemma \ref{lemma:bound:k}.
Combining with Lemma \ref{Lemma:uniform:distribution}, we have that, 
$$
\begin{aligned}
    I_{1,n,j} =& \frac{1}{\gamma_j}\set{\frac{ U_j(n/k_j)}{X_{n-k_j,n}^{(j)}} -1} +O_{p}\suit{1}\frac{\log p }{k_j}\\
             = & \frac{1}{\gamma_j}\set{\frac{ U_j(n/k_j)}{X_{n-k_j,n}^{(j)}} -1} + o_P\suit{\frac{1}{\sqrt{\log p}}},
\end{aligned}
$$
by  Condition   \ref{condition:k:choice}.

Recall that, 
$$
X_{n-k_j,n}^{(j)} = U_j\suit{\frac{1}{\Gamma_{k_j+1,n}^{(j)}}}, \quad  j=1,\dots,p,
$$
where $\Gamma_{k_j+1,n}^{(j)}$ is defined in the proof of Lemma \ref{Lemma:uniform:distribution}.
By \eqref{eq:expansion:UX}, we have that,  as $n\to\infty$, 
$$
\begin{aligned}
    I_{1,n,j} = &  \sqrt{k_j} \suit{\frac{n\Gamma_{k_j+1,n}^{(j)}}{k_j}-1} +o_P\suit{\frac{1}{\sqrt{\log p}}}, 
\end{aligned}
$$
where the $o_P(\cdot)$ term  is uniform for $1\le j\le p$.

For $I_{2,n,j}$,  we have that, as $n\to\infty$,  
$$
\begin{aligned}
    I_{2,n,j} =& \sqrt{k_j}\suit{\frac{\widetilde{k}_j}{k_j} -1}  \frac{ \widetilde{\gamma_j}(k_j)}{\gamma_j} \\
    =&\sqrt{k_j}\suit{\frac{\widetilde{k}_j}{k_j} -1} \suit{1+o_p(1)\sqrt{\frac{\log p}{k_j}} } \\
    =& \sqrt{k_j}\suit{\frac{\widetilde{k}_j}{k_j} -1} + o_P(1)\suit{\frac{\log p}{\sqrt{k_j}} } \\
    = & \sqrt{k_j}\suit{\frac{\widetilde{k}_j}{k_j} -1} + o_P(1)\suit{\frac{1}{\sqrt{\log p}}}, \\
    =& \sqrt{k_j}\set{\frac{1}{k_j}\sum_{i=1}^n \mI\suit{ \Gamma_i^{(j)}\le \frac{k_j}{n}} -1} + o_P(1)\suit{\frac{1}{\sqrt{\log p}}},
\end{aligned}
$$
where $ \Gamma_i^{(j)}$
is defined in the proof of Lemma \ref{Lemma:uniform:distribution}.
Here, the second equality follows from Lemma  \ref{lemma:tu:tilde}; the second equailty follows from Lemma \ref{lemma:bound:k}; the fourth equality follows from Condition 
\ref{condition:k:choice}.

Combining the results of $I_{1,n,j}$ and $I_{2,n,j}$, we have that,  as $n\to\infty$, 
$$
\begin{aligned}
   &\max_{1\le j \le p} \sqrt{k}_j\abs{\frac{\widehat{\gamma}_j(k_j)-\widetilde{\gamma}_j(k_j)}{\gamma_j}} \\
=& \max_{1\le j \le p} \abs{ \sqrt{k_j} \suit{\frac{n\Gamma_{k_j+1,n}^{(j)}}{k_j}-1}+ \sqrt{k_j}\set{\frac{1}{k_j}\sum_{i=1}^n \mI\suit{ \Gamma_i^{(j)}\le \frac{k_j}{n}} -1}} +o_P\suit{\frac{1}{\sqrt{\log p}}}.
\end{aligned}
$$

Denote 
$$
R_j  = \sqrt{k_j} \suit{\frac{n\Gamma_{k_j+1,n}^{(j)}}{k_j}-1}+ \sqrt{k_j}\set{\frac{1}{k_j}\sum_{i=1}^n \mI\suit{ \Gamma_i^{(j)}\le \frac{k_j}{n}} -1}.
$$
By Lemma \ref{Lemma:kiefer},  we have that, for sufficiently large $n$, 
$$
\Pr\suit{ \abs{R_j}> C_1 \frac{n}{\sqrt{k_j}}  \frac{k_j^{2/5}}{n}}  \le C_2\exp(-C_3 k_j^{1/5}), \quad j=1,\dots,p.
$$
By Condition \ref{condition:k:choice}, we have that, as $n\to\infty$, 
$$
\begin{aligned}
    \frac{n}{\sqrt{k_j}}  \frac{k_j^{2/5}}{n} = k_j^{-1/10}  =o\suit{\frac{1}{\sqrt{\log p}}}.
\end{aligned}
$$
It follows that, for any $\varepsilon>0$, as $n\to\infty$, 
$$
\begin{aligned}
    \Pr \suit{ \max_{1\le j \le p} \abs{R_j}>\frac{\varepsilon}{\sqrt{\log p}} } \le & p\max_{1\le j \le p}\Pr\suit{\abs{R_j}>\frac{\varepsilon}{\sqrt{\log p}}} \\
    \le & p \max_{1\le j \le p}\Pr\suit{ \abs{R_j}> C_1 \frac{n}{\sqrt{k_j}}  \frac{k_j^{2/5}}{n}} \\
    \le& C_2 p\max_{1\le j \le p}\exp(-C_3 k_j^{1/5}) \\
    =&o(1),
\end{aligned}
$$
by Condition    \ref{condition:k:choice}.
Hence, we have that, as $n\to\infty$,
$$
\max_{1\le j \le p} \abs{ \sqrt{k_j} \suit{\frac{n\Gamma_{k_j+1,n}^{(j)}}{k_j}-1}+ \sqrt{k_j}\set{\frac{1}{k_j}\sum_{i=1}^n \mI\suit{ \Gamma_i^{(j)}\le \frac{k_j}{n}} -1}} =o_P\suit{\frac{1}{\sqrt{\log p}}}.
$$
The proof is then complete.

\section{Proof of Theorem \ref{Theorem:size:boot}}

Define 
$$
\begin{aligned}
        V_i^{(j)} = &\sqrt{\frac{n}{k_j}} \suit{ \log X_{i}^{(j)} - \log X_{n-k_j,n}^{(j)}-\gamma_j}\mI\suit{X_i^{(j)}>X_{n-k_j,n}^{(j)} }, \\
         W_i^{(j)} =& \sqrt{\frac{n}{k_j}} \suit{ \log X_{i}^{(j)} - \log  U_j(n/k_j)-\gamma_j}\mI\suit{X_i^{(j)}>U_j(n/k_j) }.\\
\end{aligned}
$$

\begin{lemma}\label{lemma:bound:W}
      Assume that Conditions \ref{condition:SOC} and \ref{condition:k:choice:boot} hold.  Then, as $n\to\infty$,  
      $$
\max_{1\le j\le p}  \abs{\frac{1}{\sqrt{n}}\sum_{i=1}^n W_i^{(j)}} =O_P(\sqrt{\log p}).
      $$
\end{lemma}
\begin{proof}[Proof of Lemma \ref{lemma:bound:W}] 
Similar to the proof of Lemma \ref{lemma:tu:tilde}, define 
$$
\tW_i^{(j)} =  \sqrt{\frac{n}{k_j}} \suit{ \log X_{i}^{(j)} - \log  U_j(n/k_j)-\gamma_j}\mI\suit{ U_j(n/k_j)<X_i^{(j)}\le \tau_j },
$$
where $\tau_j$ is defined in \eqref{s:def:tau}. By \eqref{eq:set}, we have that, 
  as $n\to\infty$, 
$$
\begin{aligned}
        &\Pr\suit{\max_{1\le j\le p}\abs{\frac{1}{\sqrt{n}}\sum_{i=1}^n  W_i^{(j)} - \frac{1}{\sqrt{n}}\sum_{i=1}^n  \tW_i^{(j)} }>0} \\
        \le & \Pr\suit{ \max_{1\le i\le n}  X_i^{(j)}>\tau_j, \quad \text{for some} \  j\in  \set{1,\dots,p} } 
        \to &  0.
\end{aligned}
$$
Thus, it suffices to show that,  
\begin{equation}\label{s:eq:bound:tw}
    \max_{1\le j\le p}  \abs{\frac{1}{\sqrt{n}}\sum_{i=1}^n \tW_i^{(j)}}   =\max_{1\le j\le p}  \abs{\frac{1}{n}\sum_{i=1}^n  \sqrt{n}\tW_i^{(j)}} =O_P(\sqrt{\log p}).
\end{equation}
We first calculate the mean and variance of $\sqrt{n}\tW_i^{(j)}$.  By  Condition \ref{condition:SOC}, we have that,
$$
\begin{aligned}
    & \bE \suit{ \log X_{i}^{(j)} - \log  U_j(n/k_j)-\gamma_j}\mI\suit{ U_j(n/k_j)<X_i^{(j)}\le \tau_j }\\
    =&  - \int_{U_j(n/k_j)}^{\tau_j} \log y \ d \overline{F}_j(y)-\set{\gamma_j + \log  U_j(n/k_j)}   \set{k_j/n-\overline{F}_j(\tau_j)} \\
    =& \int_{U_j(n/k_j)}^{\tau_j} \frac{\overline{F}_j(y)}{y}dy - \set{\overline{F}_j(\tau_j) \log \tau_j -\overline{F}_j(U_j(n/k_j)) \log U_j(n/k_j)) } \\
    &\quad-\set{\gamma_j + \log  U_j(n/k_j)} \suit{k_j/n-\overline{F}_j(\tau_j)} \\
    =&\int_{U_j(n/k_j)}^{\tau_j} \frac{\overline{F}_j(y)}{y}dy -  \overline{F}_j(\tau_j) \log \frac{\tau_j}{U_j(n/k_j)}  -\gamma_j \suit{k_j/n-\overline{F}_j(\tau_j)} \\
    =& \int_{U_j(n/k_j)}^{\tau_j} \frac{\overline{F}_j(y)}{y}dy - \overline{F}_j(\tau_j) \suit{\gamma_j +3\gamma_j\log (k_j+p)} -\gamma_j \suit{k_j/n-\overline{F}_j(\tau_j)} \\
    =& \frac{k_j}{n}\int_{1}^{\frac{\tau_j}{U_j(n/k_j)}} \frac{\overline{F}_{j}(yU_j(n/k_j))}{y\overline{F}_j(U_j(n/k_j))}dy -\gamma_j\frac{k_j}{n} -3\gamma_j \log (k_j+p) \frac{k_j}{n}\frac{\overline{F}_j(\tau_j) }{\overline{F}_j(U_j(n/k_j))} \\
    =& \frac{k_j}{n} \int_{1}^{\frac{\tau_j}{U_j(n/k_j)}} y^{-1/\gamma_j-1}dy -\gamma_j \frac{k_j}{n} -3\gamma_j\suit{\frac{\tau_j}{U_j(n/k_j)}}^{-1/\gamma_j}\frac{k_j}{n}\log (k_j+p)+O(1)\abs{A_j(n/k_j)}\frac{k_j}{n} \\
    =& O(1)\frac{k_j}{n} \set{\abs{A_j(n/k_j)} + k_j^{-3}\log (k_j+p)},
\end{aligned}
$$
as $n\to\infty$, 
where the $O(1)$ is uniform for all $1\le j \le p$.   
By  Condition \ref{condition:k:choice:boot}, we have that, as $n\to\infty$, 
\begin{equation}\label{s:mean:tw}
	\bE  \suit{\sqrt{n} \tW_i^{(j)}} = O(1)\set{\sqrt{k_j}\abs{A_j(n/k_j)} + k_j^{-3/2}\log p } =  o(1),
\end{equation}
uniformly for all $1\le j \le p$.

Moreover, by  Condition \ref{condition:k:choice:boot},  we have that, as $n\to\infty$,
\begin{equation}\label{s:cal:var:bound}
	\begin{aligned}
     &\bE \suit{ \log X_{i}^{(j)} - \log  U_j(n/k_j)-\gamma_j}^2\mI\suit{ U_j(n/k_j)<X_i^{(j)}\le \tau_j } \\
     \le & \bE \suit{ \log X_{i}^{(j)} - \log  U_j(n/k_j)-\gamma_j}^2\mI\suit{ U_j(n/k_j)<X_i^{(j)}  } \\
     =& \int_{U_j(n/k_j)}^\infty \log^2 ydF_j(y) - 2\set{\log  U_j(n/k_j)+\gamma_j } \int_{U_j(n/k_j)}^\infty \log ydF_j(y) +\set{\log  U_j(n/k_j)+\gamma_j }^2 \frac{k_j}{n} \\
     =& 2\int_{U_j(n/k_j)}^\infty \frac{\overline{F}_j(y)\log y  }{y}dy + \suit{\log U_j(n/k_j) }^2\frac{k_j}{n}   -2\set{\log  U_j(n/k_j)+\gamma_j } \int_{U_j(n/k_j)}^\infty \frac{\overline{F}_j(y)  }{y}dy \\
     &\quad - 2\frac{k_j}{n}\set{\log  U_j(n/k_j)+\gamma_j } \log U_j(n/k_j) +\set{\log  U_j(n/k_j)+\gamma_j }^2 \frac{k_j}{n} \\
     =& 2\frac{k_j}{n} \int_{1}^\infty \frac{\overline{F}_j(yU_j(n/k_j)) }{\overline{F}_j(U_j(n/k_j)) }\frac{\log y + \log U_j(n/k_j)}{y}dy  \\
     &-2\set{\log  U_j(n/k_j)+\gamma_j }\frac{k_j}{n} \int_{1}^\infty  \frac{\overline{F}_j(yU_j(n/k_j)) }{\overline{F}_j(U_j(n/k_j)) }\frac{1}{y} dy +\gamma_j^2  \frac{k_j}{n} \\
     =& 2\frac{k_j}{n}\int_{1}^\infty \frac{\overline{F}_j(y U_j(n/k_j))}{\overline{F}_j( U_j(n/k_j))y} \log ydy-2\gamma_j\frac{k_j}{n}\int_{1}^\infty \frac{\overline{F}_j(y U_j(n/k_j))}{\overline{F}_j( U_j(n/k_j))y}dy+\gamma_j^2 \frac{k_j}{n} \\
     =& O(1)\frac{k_j}{n},
\end{aligned}
\end{equation}
and hence 
\begin{equation}\label{s:eq:var:tw}
    \begin{aligned}
    \text{Var}(\sqrt{n}\tW_i^{(j)}) =O(1)n,
\end{aligned}
\end{equation}
uniformly for all $1\le j \le p$. 

Also, note that,  for sufficiently large $n$, 
\begin{equation}\label{s:eq:W:upper}
    \abs{\tW_i^{(j)} }\le \sqrt{\frac{n}{k_j}}\max\suit{ \gamma_j,  \log \frac{\tau_j^{(j)}}{U_j(n/k_j) e^{\gamma_j}} } \le 3\gamma_j \sqrt{\frac{n}{k_j}} \log (k_j+p).
\end{equation}
Then, by  Lemma \ref{Bernstein}, \eqref{s:mean:tw}   and  \eqref{s:eq:W:upper},   we have that, for any constant $C>0$, 
$$
\begin{aligned}
    &\Pr\suit{ \max_{1\le j\le p}\abs{\frac{1}{n}\sum_{i=1}^n \sqrt{n}\tW_i^{(j)}} > C\sqrt{\log p} } \\
    \le  &   p\max_{1\le j\le p} \Pr\suit{ \abs{\frac{1}{n}\sum_{i=1}^n \sqrt{n}\tW_i^{(j)}} > C\sqrt{\log p} }\\
    \le &2p\max_{1\le j\le p} \exp\suit{ - \frac{nC^2\log p}{\text{Var}(\sqrt{n}\tW_i^{(j)}) +2\gamma_j\sqrt{\frac{n}{k_j}}\log (k_j+p)C \sqrt{\log p} } }.
\end{aligned}
$$
By Condition \ref{condition:k:choice:boot} and  \eqref{s:eq:var:tw},   we have that,     for sufficiently large $n$,
$$
 \frac{1}{\log p}\frac{nC^2\log p}{\text{Var}(\sqrt{n}\tW_i^{(j)}) +2\gamma_j\sqrt{\frac{n}{k_j}}\log (k_j+p)C \sqrt{\log p} } \ge \eta >1, 
$$
for some constant $\eta>1$, 
and hence
$$
\Pr\suit{ \max_{1\le j\le p}\abs{\frac{1}{\sqrt{n}}\sum_{i=1}^n \tW_i^{(j)}} > C\sqrt{\log p} } \to 0.
$$
Thus, \eqref{s:eq:bound:tw} holds and the proof is complete.
\end{proof}

\begin{lemma}\label{lemma:to:iidsum}
    Assume that Conditions \ref{condition:SOC} and \ref{condition:k:choice:boot}  hold. Then, as $n\to\infty$, 
    $$
        \max_{1\le j\le p}\abs{\frac{1}{\sqrt{n}}\sum_{i=1}^n V_i^{(j)} -  \frac{1}{\sqrt{n}}\sum_{i=1}^n W_i^{(j)} }= o_P(1/\sqrt{\log p}).
    $$
\end{lemma}
\begin{proof}[Proof of Lemma \ref{lemma:to:iidsum}]
Write 
$$
\begin{aligned}
    \frac{1}{\sqrt{n}}\sum_{i=1}^n V_i^{(j)}  =&  \sqrt{k_j}\set{\frac{1}{k_j}\sum_{i=1}^n \log \frac{X_{i}^{(j)}}{X_{n-k_j,n}^{(j)}} \mI\suit{X_i^{(j)}>X_{n-k_j,n}^{(j)} }-\gamma_j} \\
    =& \sqrt{k_j}\suit{\widehat{\gamma}_j(k_j)-\gamma_j},
\end{aligned}
$$
and 
$$
\begin{aligned}
    \frac{1}{\sqrt{n}}\sum_{i=1}^n W_i^{(j)} =& \sqrt{k_j} \frac{1}{k_j}\set{ \sum_{i=1}^n \log \frac{X_{i}^{(j)}}{ U_j(n/k_j)} \mI\suit{X_i^{(j)}>U_j(n/k_j) } -\gamma_j \sum_{i=1}^n \mI\suit{X_i^{(j)}>U_j(n/k_j) }  } \\
    =& \sqrt{k_j} \frac{\sum_{i=1}^n \mI\suit{X_i^{(j)}>U_j(n/k_j) }}{k_j}\set{\frac{\sum_{i=1}^n \log \frac{X_{i}^{(j)}}{U_j(n/k_j)} \mI\suit{X_i^{(j)}>U_j(n/k_j) }}{\sum_{i=1}^n \mI\suit{X_i^{(j)}>U_j(n/k_j) }}-\gamma_j  } \\
    =& \frac{\widetilde{k}_j }{k_j}\sqrt{k_j}\suit{\widetilde{\gamma}_j(k_j)-\gamma_j},
\end{aligned}
$$
where 
$$
\begin{aligned}
    \widetilde{\gamma}_j = &\frac{\sum_{i=1}^n \log \frac{X_{i}^{(j)}}{U_j(n/k_j)} \mI\suit{X_i^{(j)}>U_j(n/k_j) }}{\sum_{i=1}^n \mI\suit{X_i^{(j)}>U_j(n/k_j) }}, \\
    \widetilde{k}_j = & \sum_{i=1}^n \mI\suit{X_i^{(j)}>U_j(n/k_j)}.
\end{aligned}
$$
It follows that, 
$$
\begin{aligned}
    \frac{1}{\sqrt{n}}\sum_{i=1}^n V_i^{(j)} -\frac{1}{\sqrt{n}}\sum_{i=1}^n W_i^{(j)} =& \sqrt{k_j}\suit{\widehat{\gamma}_j(k_j)-\widetilde{\gamma}_j(k_j)} +\sqrt{k_j}\suit{\widetilde{\gamma}_j(k_j)-\gamma_j}\suit{ \frac{ \widetilde{k}_j }{k_j}-1}.
\end{aligned}
$$
Note that Lemmas \ref{lemma:gamma:diff} and \ref{lemma:bound:k} still hold when Condition \ref{condition:k:choice} is replaced by Condition \ref{condition:k:choice:boot}.  Thus, 
we have that, as $n\to\infty$,
\begin{align}
    \sqrt{k_j}\suit{\widehat{\gamma}_j(k_j)-\widetilde{\gamma}_j(k_j)} =& o_P(1/\sqrt{\log p}), \label{s:eq:tilde:gamma}\\
    \sqrt{k_j}\suit{ \frac{ \widetilde{k}_j }{k_j}-1} =& O_P(\sqrt{\log p}) \label{s:eq:tilde:k},
\end{align}
uniformly for all $1\le j\le p$.
Combining Lemma \ref{lemma:bound:W},  \eqref{s:eq:tilde:k} and Condition \ref{condition:k:choice:boot}, we have that, as $n\to\infty$, 
	\begin{align}
    \widetilde{\gamma}_j(k_j)-\gamma_j =& \frac{\sum_{i=1}^n \log \frac{X_{i}^{(j)}}{U_j(n/k_j)} \mI\suit{X_i^{(j)}>U_j(n/k_j) }}{\sum_{i=1}^n \mI\suit{X_i^{(j)}>U_j(n/k_j) }} \notag\\
    =& \frac{1}{\widetilde{k}_j}\sum_{i=1}^n \suit{\log  X_{i}^{(j)} - \log U_j(n/k_j) -\gamma_j} \mI\suit{X_i^{(j)}>U_j(n/k_j) } \notag \\
    =& \frac{k_j}{\widetilde{k}_j}\frac{1}{\sqrt{k_j}} \frac{1}{\sqrt{n}}\sum_{i=1}^n W_i^{(j)} \notag\\
    =& \suit{1+o_P(1) } \frac{1}{\sqrt{k_j}} O_P(1)\sqrt{\log p} \notag \\
    =& o_P(1)\frac{1}{\log p},  \label{s:upper:bound:tilde:gamma}
\end{align}
uniformly for all $1\le j \le p$.
Combining  \eqref{s:upper:bound:tilde:gamma} with \eqref{s:eq:tilde:gamma} and \eqref{s:eq:tilde:k}, we conclude that, as $n\to\infty$, 
$$
\begin{aligned}
    \abs{\frac{1}{\sqrt{n}}\sum_{i=1}^n V_i^{(j)} -\frac{1}{\sqrt{n}}\sum_{i=1}^n W_i^{(j)}} = o_p(1/\sqrt{\log p}),
\end{aligned}
$$
uniformly for all $1\le j \le p$. The proof is then complete.
\end{proof}

\begin{lemma}\label{lemma:to:covariance}
    Assume that Conditions \ref{condition:SOC}  and \ref{condition:k:choice:boot} and.  Then, as $n\to\infty$, 
    $$
    \max_{1\le j,\ell\le p}\abs{\frac{1}{n}\sum_{i=1}^n V_i^{(j)}V_i^{(\ell)} -  \bE \suit{W_i^{(j)}W_i^{(\ell)}}} =o_P(1/\log^2p).
    $$
\end{lemma}
\begin{proof}[Proof of Lemma \ref{lemma:to:covariance}]
    We prove the lemma by showing that, as $n\to\infty$,
\begin{align}
\max_{1\le j,\ell \le p}\abs{\frac{1}{n}\sum_{i=1}^n V_i^{(j)} V_i^{(\ell)}- \frac{1}{n}\sum_{i=1}^n  \tV_i^{(j)} \tV_i^{(\ell)}} =&o_P(1/\log^2 p) \label{s:var:1},\\
 \max_{1\le j,\ell \le p}\abs{\frac{1}{n}\sum_{i=1}^n  \tV_i^{(j)} \tV_i^{(\ell)} -\frac{1}{n}\sum_{i=1}^n \tW_i^{(j)} \tW_i^{(\ell)}} =&  o_P(1/\log^2 p),  \label{s:var:2}\\
 \max_{1\le j,\ell\le p}\abs{\frac{1}{n}\sum_{i=1}^n \tW_i^{(j)} \tW_i^{(\ell)} - \bE \tW_i^{(j)} \tW_i^{(\ell)}}=&  o_P(1/\log^2 p),  \label{s:var:3} \\
 \max_{1\le j,\ell\le p}\abs{\bE \tW_i^{(j)} \tW_i^{(\ell)} - \bE W_i^{(j)} W_i^{(\ell)}}=&  o(1/\log^2 p) \label{s:var:4},
\end{align}
where 
$$
\tV_i^{(j)} = \sqrt{\frac{n}{k_j}} \suit{ \log X_{i}^{(j)} - \log X_{n-k_j,n}^{(j)}-\gamma_j}\mI\suit{ X_{n-k_j,n}^{(j)}< X_i^{(j)}\le \tau_j }.
$$

First, we handle  \eqref{s:var:1}. By \eqref{eq:set}, we have that, for any $\varepsilon>0$, as $n\to\infty$, 
$$
\begin{aligned}
	&\Pr\suit{\max_{1\le j,\ell \le p}\abs{\frac{1}{n}\sum_{i=1}^n V_i^{(j)} V_i^{(\ell)}- \frac{1}{n}\sum_{i=1}^n  \tV_i^{(j)} \tV_i^{(\ell)}} > \varepsilon/\log^2 p} \\
	\le &  \Pr\suit{ \max_{1\le i\le n}  X_i^{(j)}>\tau_j, \quad \text{for some} \  j\in  \set{1,\dots,p} }  \\
	\to & 0.
\end{aligned}
$$
Thus,  \eqref{s:var:1} holds. 

Then, we prove \eqref{s:var:4}. Note that, 
$$
  \abs{\bE \tW_i^{(j)} \tW_i^{(\ell)} - \bE W_i^{(j)} W_i^{(\ell)}}  \le \abs{\bE W_i^{(j)} W_i^{(\ell)} \mI(X_i^{(j)}>\tau_j ) } +\abs{\bE W_i^{(j)} W_i^{(\ell)}\mI(X_i^{(\ell)}>\tau_{\ell} ) }.
$$
By the generalized H\"older inequality, and some straightforward calculation, we have that, (S22) holds. 

  Next, we prove \eqref{s:var:3}. 
    By   Lemma \ref{Bernstein}, we have that,  for any constant $\varepsilon>0$, 
$$
\begin{aligned}
  &  \Pr\suit{ \max_{1\le j\le p}\abs{\frac{1}{n}\sum_{i=1}^n \suit{\tW_i^{(j)} \tW_i^{(\ell)} -\bE \tW_i^{(j)} \tW_i^{(\ell)}}}> \varepsilon/\log^2 p } \\
  \le &p \max_{1\le j\le p} \Pr\suit{ \abs{\frac{1}{n}\sum_{i=1}^n \suit{\tW_i^{(j)} \tW_i^{(\ell)} -\bE \tW_i^{(j)} \tW_i^{(\ell)}}}> \varepsilon/\log^2 p } \\
\le & 2p \max_{1\le j\le p}\exp\suit{ -\frac{n\varepsilon^2/\log^4p}{2\text{Var}(\tW_i^{(j)} \tW_i^{(\ell)})+\frac{2}{3}M_n \varepsilon/\log ^2p  } },
\end{aligned}
$$
where $M_n$ is an upper bound of $\tW_i^{(j)} \tW_i^{(\ell)}$, defined as follows.
By \eqref{s:eq:W:upper}, we have that, for some $C>0$, 
$$
\begin{aligned}
	\tW_i^{(j)} \tW_i^{(\ell)} \le & 4\gamma_j\gamma_{\ell} \frac{n}{\sqrt{k_jk_{\ell}}}\log (k_j+p) \log (k_{\ell}+p) \\
	\le & C \frac{n}{k_{\min}} \log^2 (k_{\max}+p)  \\
	=& :M_n.
\end{aligned}
$$
By \eqref{s:eq:var:tw},  we have that, as $n\to\infty$,
$$
\begin{aligned}
   \text{Var}\suit{\tW_i^{(j)}\tW_i^{(\ell)}}\le  \bE \suit{\tW_i^{(j)}\tW_i^{(\ell)}}^2 \le  \sqrt{\bE \suit{\tW_i^{(j)}}^4 \bE \suit{\tW_i^{(\ell)}}^4}. 
   \end{aligned}
$$
By  a  similar calculation as in  \eqref{s:cal:var:bound}, we obtain that, as $n\to\infty$, 
$$
\bE \suit{\tW_i^{(j)}}^4 =  O(1) \frac{n}{k_j} = O(1) \frac{n}{k_{\min}}.
$$
and
$$
\text{Var}\suit{\tW_i^{(j)}\tW_i^{(\ell)}}  = O(1) \frac{n}{k_{\min}}.
$$
Thus, by  Condition \ref{condition:k:choice:boot}, we have that, as $n\to\infty$,  
$$
\frac{n\varepsilon^2/\log^4p}{2\text{Var}(\tW_i^{(j)} \tW_i^{(\ell)})+\frac{2}{3}M_n \varepsilon/\log ^2p  }  \frac{1}{\log p}\to\infty,
$$
and hence \eqref{s:var:3} holds. 

Finally, we prove \eqref{s:var:2}.
    Note that,  
$$
\begin{aligned}
    \abs{\frac{1}{n}\sum_{i=1}^n  \tV_i^{(j)} \tV_i^{(\ell)} -\frac{1}{n}\sum_{i=1}^n \tW_i^{(j)} \tW_i^{(\ell)} }
\le &  \abs{ \frac{1}{n}\sum_{i=1}^n \suit{\tV_i^{(j)} - \tW_i^{(j)} }\tW_i^{(\ell)}  } +\abs{ \frac{1}{n}\sum_{i=1}^n \suit{\tV_i^{(\ell)} - \tW_i^{(\ell)} }\tW_i^{(j)}  } \\
&+ \abs{ \frac{1}{n}\sum_{i=1}^n \suit{\tV_i^{(j)} - \tW_i^{(j)}} \suit{\tV_i^{(\ell)} - \tW_i^{(\ell)}}}.
\end{aligned}
$$
We prove \eqref{s:var:2} by  showing that, as $n\to\infty$, 
\begin{align}
    \max_{1\le j, \ell \le p}\abs{ \frac{1}{n}\sum_{i=1}^n \suit{\tV_i^{(j)} - \tW_i^{(j)} }\tW_i^{(\ell)}  }  = & o_p(1/\log^2 p), \label{s:eq:jl1} \\
    \max_{1\le j, \ell \le p}\abs{ \frac{1}{n}\sum_{i=1}^n \suit{\tV_i^{(\ell)} - \tW_i^{(\ell)} }\tW_i^{(j)} } = & o_p(1/\log^2 p), \label{s:eq:jl2} \\
    \max_{1\le j, \ell \le p}  \abs{ \frac{1}{n}\sum_{i=1}^n \suit{\tV_i^{(j)} - \tW_i^{(j)}} \suit{\tV_i^{(\ell)} - \tW_i^{(\ell)}}} = & o_p(1/\log^2 p). \label{s:eq:jl3}
\end{align}
We start with \eqref{s:eq:jl1}.  Write
$$
\begin{aligned}
   &\tV_i^{(j)} - \tW_i^{(j)} \\
    =  &     \sqrt{\frac{n}{k_j}} \suit{ \log X_{i}^{(j)} - \log  X_{n-k_j,n}^{(j)}-\gamma_j}\mI\suit{X_{n-k_j,n}^{(j)}<X_i^{(j)}\le \tau_j} \\
   &- \sqrt{\frac{n}{k_j}} \suit{ \log X_{i}^{(j)} - \log  U_j(n/k_j)-\gamma_j}\mI\suit{X_{n-k_j,n}^{(j)}<X_i^{(j)}\le \tau_j}  \\
   &+ \sqrt{\frac{n}{k_j}} \suit{ \log X_{i}^{(j)} - \log  U_j(n/k_j)-\gamma_j}\mI\suit{X_{n-k_j,n}^{(j)}<X_i^{(j)}\le \tau_j}\\
   &-\sqrt{\frac{n}{k_j}} \suit{ \log X_{i}^{(j)} - \log  U_j(n/k_j)-\gamma_j}\mI\suit{U_j(n/k_j)< X_i^{(j)}\le \tau_j  } \\
   =& - \sqrt{\frac{n}{k_j}} \log \frac{X_{n-k_j,n}^{(j)}}{U_j(n/k_j)} \mI\suit{X_{n-k_j,n}^{(j)}<X_i^{(j)}\le \tau_j}  \\
   &+ \sqrt{\frac{n}{k_j}} \suit{ \log X_{i}^{(j)} - \log  U_j(n/k_j)} \mI\suit{X_i^{(j)}\le \tau_j}\set{\mI\suit{X_{n-k_j,n}^{(j)}< X_i^{(j)}} -\mI\suit{U_j(n/k_j)< X_i^{(j)}}} \\
   &-\sqrt{\frac{n}{k_j}} \gamma_j \mI\suit{X_i^{(j)}\le \tau_j}\set{\mI\suit{X_{n-k_j,n}^{(j)}< X_i^{(j)}} -\mI\suit{U_j(n/k_j)< X_i^{(j)}}}.
\end{aligned}
$$

It follows that, 
\begin{equation}\label{s:bound:v_w}
    \begin{aligned}
   \abs{\tV_i^{(j)} - \tW_i^{(j)}} 
\le & \sqrt{\frac{n}{k_j}}  \abs{\log \frac{X_{n-k_j,n}^{(j)}}{U_j(n/k_j)} } \mI\suit{X_{n-k_j,n}^{(j)}<X_i^{(j)} }   \\
&+ \sqrt{\frac{n}{k_j}}   \abs{\log \frac{X_{n-k_j,n}^{(j)}}{U_j(n/k_j)} }   \mI\suit{X_i^{(j)} \in (U_j(n/k_j),X_{n-k_j,n}^{(j)}) }  \\
&+\sqrt{\frac{n}{k_j}} \gamma_j   \mI\suit{X_i^{(j)} \in (U_j(n/k_j),X_{n-k_j,n}^{(j)}) } \\
\le & \sqrt{\frac{n}{k_j}}  \abs{\log \frac{X_{n-k_j,n}^{(j)}}{U_j(n/k_j)} } \set{ 2\mI\suit{X_i^{(j)}>U_j(n/k_j)} + 2\mI\suit{X_i^{(j)}>X_{n-k_j,n}^{(j)}} }   \\
&+\sqrt{\frac{n}{k_j}} \gamma_j  \mI\suit{X_i^{(j)} \in (U_j(n/k_j),X_{n-k_j,n}^{(j)}) }.
\end{aligned}
\end{equation}

Here,  the notation $X_i^{(j)} \in (U_j(n/k_j),X_{n-k_j,n}^{(j)})$ refers to  
$$
 \min\suit{U_j(n/k_j),X_{n-k_j,n}^{(j)}} \le X_i^{(j)} \le \max\suit{U_j(n/k_j),X_{n-k_j,n}^{(j)}}.
$$
Combining with the upper bound of $\tW_i^{(\ell)}$ (see \eqref{s:eq:W:upper}), we obtain that,
$$
\begin{aligned}
   &\abs{\frac{1}{n}\sum_{i=1}^n \suit{\tV_i^{(j)} - \tW_i^{(j)}} \tW_i^{(\ell)}}\\
\le &  2\log(p+k_{\ell})\sqrt{\frac{n}{k_{\ell}}} \sqrt{\frac{n}{k_j}}\abs{\log \frac{X_{n-k_j,n}^{(j)}}{U_j(n/k_j)} }\set{ \frac{1}{n}\sum_{i=1}^n \mI\suit{X_i^{(j)}> U_j(n/k_j)} + \frac{1}{n}\sum_{i=1}^n \mI\suit{X_i^{(j)}>X_{n-k_j,n}^{(j)} }} \\
&+\gamma_j \log(p+k_{\ell})\sqrt{\frac{n}{k_{\ell}}} \sqrt{\frac{n}{k_j}}  \frac{1}{n}\sum_{i=1}^n  \mI\suit{X_i^{(j)} \in (U_j(n/k_j),X_{n-k_j,n}^{(j)}) } \\
=& 2\log(p+k_{\ell})\sqrt{\frac{n}{k_{\ell}}} \sqrt{\frac{n}{k_j}}\abs{\log \frac{X_{n-k_j,n}^{(j)}}{U_j(n/k_j)} }\set{\frac{k_j}{n}   + \frac{1}{n}\sum_{i=1}^n \mI\suit{X_i^{(j)}> U_j(n/k_j)} } \\
&+\gamma_j \log(p+k_{\ell})\sqrt{\frac{n}{k_{\ell}}} \sqrt{\frac{n}{k_j}}  \frac{1}{n}\sum_{i=1}^n  \mI\suit{X_i^{(j)} \in (U_j(n/k_j),X_{n-k_j,n}^{(j)}) } \\ 
=:& I_{1,j,l}+ I_{2,j,l}.
\end{aligned}
$$

By \eqref{s:eq:tilde:gamma} and \eqref{s:eq:tilde:k}, we have that, as $n\to\infty$, 
    \begin{align*}
    I_{1,j,l} =&  O_P(1)\log(p+k_{\ell})\sqrt{\frac{n}{k_{\ell}}}\sqrt{\frac{n}{k_j}} \frac{\sqrt{\log p}}{\sqrt{k_j}} \frac{k_j}{n}  \\
    =&O_P(1)\log(p+k_{\ell}) \frac{\sqrt{\log p}}{\sqrt{k_{\ell}}} \\
    =& o_P(1)\frac{1}{\log^2p},
\end{align*}
uniformly for all $1\le j,\ell \le p$. Here, the last step follows from   
 Condition \ref{condition:k:choice:boot}.

Next, we handle $I_{2,j,\ell}$. Note that, if $X_{n-k_j,n}^{(j)}> U_j(n/k_j)$, then
$$
\begin{aligned}
     \frac{1}{n}\sum_{i=1}^n  \mI\suit{X_i^{(j)} \in (U_j(n/k_j),X_{n-k_j,n}^{(j)}) }  =& \frac{1}{n}\sum_{i=1}^n \mI\suit{X_i^{(j)}>U_j(n/k_j)} - \frac{1}{n}\sum_{i=1}^n \mI\suit{X_i^{(j)}>X_{n-k_j,n}^{(j)}} \\
     =&  \frac{1}{n}\sum_{i=1}^n \mI\suit{X_i^{(j)}>U_j(n/k_j)} -\frac{k_j}{n}.
\end{aligned}
$$
If $X_{n-k_j,n}^{(j)}\le  U_j(n/k_j)$, then 
$$
 \frac{1}{n}\sum_{i=1}^n  \mI\suit{X_i^{(j)} \in (U_j(n/k_j),X_{n-k_j,n}^{(j)}) } = \frac{k_j}{n}-\frac{1}{n}\sum_{i=1}^n \mI\suit{X_i^{(j)}>U_j(n/k_j)}.
$$
Thus, we conclude that, 
$$
 \frac{1}{n}\sum_{i=1}^n  \mI\suit{X_i^{(j)} \in (U_j(n/k_j),X_{n-k_j,n}^{(j)}) } = \abs{\frac{1}{n}\sum_{i=1}^n \mI\suit{X_i^{(j)}>U_j(n/k_j)} -\frac{k_j}{n}} =\frac{k_j}{n}\abs{\frac{\widetilde{k_j}}{k_j}-1} .
$$
By \eqref{s:eq:tilde:k} and Condition \ref{condition:k:choice:boot}, we have that, as $n\to\infty$, 
$$
I_{2,j,\ell}= O_P(1) \log(p+k_{\ell})\sqrt{\frac{n}{k_{\ell}}} \sqrt{\frac{n}{k_j}}  \frac{k_j}{n} \frac{\sqrt{\log p}}{\sqrt{k_j}} = O_P(1) \frac{\sqrt{\log p} \log(p+k_{\ell})}{\sqrt{k_{\ell}}}=o_P(1)\frac{1}{\log^2p},
$$
uniformly for all $1\le j,\ell \le p$.

Combining the results of $I_{1,j,\ell}$ and $I_{2,j,\ell}$, 
we have then proved \eqref{s:eq:jl1}. 
The proof of \eqref{s:eq:jl2} is similar and thus omitted.

Finally, we handle \eqref{s:eq:jl3}.  By  \eqref{s:bound:v_w},  we have that, 
$$
\begin{aligned}
    &\frac{1}{n}\sum_{i=1}^n \abs{\tV_i^{(j)} - \tW_i^{(j)}} \abs{\tV_i^{(\ell)} - \tW_i^{(\ell)}} \\
\le & 4\sqrt{\frac{n}{k_j}}  \sqrt{\frac{n}{k_{\ell}}}  \abs{\log \frac{X_{n-k_j,n}^{(j)}}{U_j(n/k_j)}} \abs{\log \frac{X_{n-k_{\ell},n}^{(\ell)}}{U_{\ell}(n/k_{\ell})}}  \times  \\
&\quad \frac{1}{n}\sum_{i=1}^n \set{ 2\mI\suit{X_i^{(j)}>U_j(n/k_j)} + 2\mI\suit{X_i^{(j)}>X_{n-k_j,n}^{(j)}} }\\
&+ 4\gamma_{\ell}\sqrt{\frac{n}{k_j}}  \sqrt{\frac{n}{k_{\ell}}}  \abs{\log \frac{X_{n-k_j,n}^{(j)}}{U_j(n/k_j)}} \frac{1}{n}\sum_{i=1}^n \mI\suit{X_i^{(\ell)} \in (U_{\ell}(n/k_{\ell}),X_{n-k_{\ell},n}^{(\ell)}) } \\
&+  4\gamma_j\sqrt{\frac{n}{k_j}}  \sqrt{\frac{n}{k_{\ell}}}  \abs{\log \frac{X_{n-k_{\ell},n}^{(\ell)}}{U_{\ell}(n/k_{\ell})}} \frac{1}{n}\sum_{i=1}^n\mI\suit{X_i^{(j)} \in (U_{j}(n/k_j),X_{n-k_j,n}^{(j)}) }\\
&+ \gamma_j \gamma_{\ell} \sqrt{\frac{n}{k_j}}  \sqrt{\frac{n}{k_{\ell}}}\frac{1}{n}\sum_{i=1}^n\mI\suit{X_i^{(j)} \in (U_{j}(n/k_j),X_{n-k_j,n}^{(j)}) } \\
=&:J_{1,j,\ell}+ J_{2,j,\ell}+J_{3,j,\ell}+J_{4,j,\ell}.
\end{aligned}
$$

By \eqref{s:eq:tilde:gamma}, \eqref{s:eq:tilde:k} and Condition \ref{condition:k:choice:boot}, we have that, as $n\to\infty$, 
$$
\begin{aligned}
  J_{1,j,\ell}= &  O_P(1) \sqrt{\frac{n}{k_j}}  \sqrt{\frac{n}{k_{\ell}}} \frac{\sqrt{\log p}}{\sqrt{k_j}} \frac{\sqrt{\log p}}{\sqrt{k_{\ell}}} \frac{k_j}{n} = \frac{\log p}{k_{\ell}} =o_P(1/\log^2 p),\\
  J_{2,j,\ell} =& O_P(1) \sqrt{\frac{n}{k_j}}  \sqrt{\frac{n}{k_{\ell}}} \frac{\sqrt{\log p}}{\sqrt{k_j}} \frac{k_{\ell}}{n} \frac{\sqrt{\log p}}{\sqrt{k_{\ell}}}= \frac{\log p}{k_{j}} =o_P(1/\log^2 p), \\
  J_{3,j,\ell} =& O_P(1) \sqrt{\frac{n}{k_j}}  \sqrt{\frac{n}{k_{\ell}}} \frac{\sqrt{\log p}}{\sqrt{k_{\ell}}} \frac{k_j}{n} \frac{\sqrt{\log p}}{\sqrt{k_j}}= \frac{\log p}{k_{\ell}} =o_P(1/\log^2 p),
\end{aligned}
$$
uniformly for all $1\le j,\ell \le p$. 
For $J_{4,j,\ell}$, by \eqref{s:eq:tilde:k} and Condition \ref{condition:k:choice:boot},  we have that, as $n\to\infty$,
$$
\begin{aligned}
    J_{4,j,\ell} = & \gamma_j \gamma_{\ell} \sqrt{\frac{n}{k_j}}  \sqrt{\frac{n}{k_{\ell}}} \frac{k_j}{n}\abs{\frac{\widetilde{k_j}}{k_j}-1}  \\
    =& O_P(1) \sqrt{\frac{n}{k_j}}  \sqrt{\frac{n}{k_{\ell}}}  \frac{k_j}{n} \frac{\sqrt{\log p}}{\sqrt{k_j}} = \frac{\sqrt{\log p}}{\sqrt{k_{\ell}}} = o_P(1/\log^2 p),
\end{aligned}
$$
uniformly for all $1\le j,\ell \le p$. 
Combining the  four terms, we have proved \eqref{s:eq:jl3}. The proof is then complete.
\end{proof}

\begin{proof}[Proof of Theorem \ref{Theorem:size:boot}]

  We intend to apply Lemma 1 of \cite{chernozhukov2023high} to approximate the distribution of $\frac{1}{\sqrt{n}}\sum_{i=1}^n W_i^{(j)}$. To this end, we verify the conditions in  Lemma 1 of \cite{chernozhukov2023high}.
By  Conditions \ref{condition:SOC} and \ref{condition:k:choice:boot}, we have that, as $n\to\infty$, uniformly for all $1\le j\le p$, 
$$
\begin{aligned}
    \bE\suit{ \frac{1}{\sqrt{n}}\sum_{i=1}^{n}W_i^{(j)}}=&  \frac{n}{\sqrt{k_j}}\int_{U_j(n/k_j)}^{\infty} \log \frac{y}{U_j(n/k_j)} d F_j(y)-\gamma_j\sqrt{k_j}\\
    =& \frac{n}{\sqrt{k_j}} \int_{U_j(n/k_j)}^{\infty} \frac{\overline{F}_j(y)}{y}dy -\gamma_j\sqrt{k_j} \\
    =& \sqrt{k_j}\int_{1}^{\infty} \frac{\overline{F}_{j}(yU_j(n/k_j))}{\overline{F}_j(U_j(n/k_j))}dy -\gamma_j\sqrt{k_j} \\
    =& O(1)\sqrt{k_j}A_j(n/k_j) \\
    =&o(1/\sqrt{\log p}),
\end{aligned}
$$
and 
$$
\begin{aligned}
    \bE \suit{W^{(j)}}^2 \to & \gamma_j^2,\\
 \bE \suit{W^{(j)}}^4 = & O(1)\frac{n}{k_j},
\end{aligned}
$$

Take 
$
B_n =c^{-1} \sqrt{n/k_{min}},
$
where $c$ is a small constant satisfying $0<c<\log (1.5)/\max_{1\le j\le p}\gamma_j$. Then, we have that,  
$
\bE \set{\suit{X_{+}^{(j)}}^c}<\infty,
$
where $X_{+}^{(j)} = \max(X^{(j)},0)$. 
It follows that, for sufficiently large $n$, 
$$
\begin{aligned}
    \bE \msuit{ \exp\suit{\abs{W_i^{(j)}}/B_n}}\le & \bE\msuit{ \exp\set{ \abs{\log \suit{\frac{X^{(j)}}{U_j(n/k_j)e^{\gamma_j}}}^c }\mI\suit{X^{(j)}>U_j(n/k_j) } }  }  \\
    \le & \bE\msuit{\suit{\frac{X^{(j)}}{U_j(n/k_j)e^{\gamma_j}}}^c\mI\suit{X^{(j)}>e^{\gamma_j}U_j(n/k_j) }   } \\
    &+ \bE\msuit{\suit{\frac{U_j(n/k_j)e^{\gamma_j}}{X^{(j)} }}^c\mI\suit{U_j(n/k_j)<X^{(j)}\le e^{\gamma_j}U_j(n/k_j) }   } \\
    \le & \bE\msuit{\suit{\frac{X_{+}^{(j)}}{U_j(n/k_j)e^{\gamma_j}}}^c}+ e^{c\gamma_j}\\
    \le & \bE\msuit{\suit{\frac{X_{+}^{(j)}}{U_j(n/k_j)e^{\gamma_j}}}^c}+ 1.5
    \le  2,
\end{aligned}
$$
uniformly for all $1\le j\le p$. 
Let $\boldsymbol{\Sigma}$ be a $p\times p$ matrix with elements 
$$
\boldsymbol{\Sigma}_{j\ell} = \bE\suit{ W_i^{(j)}W_i^{(\ell)}}, \quad 1\le j,\ell \le p.
$$
By Theorem 1 of \cite{chernozhukov2023high} and  Condition \ref{condition:k:choice:boot}, we have that,  as $n\to\infty$, 
$$
    \begin{aligned}
            &\Pr\set{\suit{ \frac{1}{\sqrt{n}}\sum_{i=1}^n W_i^{(j)},\dots, \frac{1}{\sqrt{n}}\sum_{i=1}^n W_i^{(j)}}^\top \in A}- \Pr\suit{N(0,\boldsymbol{\Sigma}) \in A} \\
    =& O(1) \suit{\frac{\log^5(pn)}{k_{\min}}}^{1/4} = o(1),
    \end{aligned}
$$
uniformly for all $A\in \mathcal{A}$, where $\mathcal{A}$ is the collection of closed rectangles in $\mathbb{R}^p$, 
$$
\mathcal{A} = \set{\prod_{j=1}^{p} [a_j,b_j]: -\infty\le a_j\le b_j\le \infty, \quad j=1,\dots,p}.
$$
By Lemma \ref{lemma:to:iidsum}, we have that,  as $n\to\infty$,
  $$
 \frac{1}{\sqrt{n}}\sum_{i=1}^n V_i^{(j)} =  \frac{1}{\sqrt{n}}\sum_{i=1}^n W_i^{(j)} +o_P(1/\sqrt{\log p}),
  $$
uniformly for all $1\le j\le p$. Thus, 
By Lemma 1 of \cite{chernozhukov2023high} , we have that,  as $n\to\infty$, 
\begin{equation}\label{s:eq:clt1}
    \begin{aligned}
            \Pr\set{\suit{ \frac{1}{\sqrt{n}}\sum_{i=1}^n V_i^{(j)},\dots, \frac{1}{\sqrt{n}}\sum_{i=1}^n V_i^{(j)}}^\top \in A}- \Pr\suit{N(0,\boldsymbol{\Sigma}) \in A} = o(1).
    \end{aligned}
\end{equation}

Note that, given the data $\mbX$, 
$$
\suit{  \frac{1}{\sqrt{n}}\sum_{i=1}^n \xi_iV_i^{(1)},\dots, \frac{1}{\sqrt{n}} \sum_{i=1}^n \xi_i V_i^{(p)}}^\top \sim N\suit{0,\widehat{\boldsymbol{\Sigma}}},
$$
where $\widehat{\boldsymbol{\Sigma}}$ is a $p\times p$ matrix with elements 
$$
\widehat{\boldsymbol{\Sigma}}_{j\ell} =\frac{1}{n} \sum_{i=1}^n V_i^{(j)}V_i^{(\ell)}, \quad 1\le j, \ell \le p.
$$
Combining Proposition 2.1 of \cite{chernozhuokov2022improved} with Lemma \ref{lemma:to:covariance}, we have that, as $n\to\infty$, 
$$
\Pr\suit{N(0,\widehat{\boldsymbol{\Sigma}}) \in A} - \Pr\suit{N(0,\boldsymbol{\Sigma}) \in A} =o(1).
$$
Combining with \eqref{s:eq:clt1}, we have that, as $n\to\infty$,
$$
\Pr\set{\suit{ \frac{1}{\sqrt{n}}\sum_{i=1}^n V_i^{(1)},\dots, \frac{1}{\sqrt{n}}\sum_{i=1}^n V_i^{(p)}}^\top \in A}- \Pr\suit{N(0,\widehat{\boldsymbol{\Sigma}}) \in A} =o(1).
$$
The proof is then complete.
\end{proof}

\section{Proofs of Theorem  \ref{theorem:power:i} and \ref{theorem:power} }

\begin{proof}[Proof of Theorem \ref{theorem:power:i}]
Write 
\begin{align*}
\sqrt{k_j}\suit{\frac{\hatgamma_j}{\gamma_j^0}-1} =& \sqrt{k_j}\suit{\frac{\hatgamma_j}{\gamma_j} \frac{\gamma_j}{\gamma_j^0}-1} \\
=& \sqrt{k_j}\suit{\frac{\hatgamma_j}{\gamma_j}(1+\delta_j)-1 }\\
=& \sqrt{k_j}\suit{\frac{\hatgamma_j}{\gamma_j}-1} +\sqrt{k_j}\delta_j \frac{\hatgamma_j}{\gamma_j}
\end{align*}
Denote 
$$
 \ell= \arg\max_{j} \abs{\sqrt{k_j}\delta_j}.
$$ 
 Without loss of generality, we assume that, $\delta_{\ell}>0$. 
It follows that, 
$$
\begin{aligned}
    &\Pr\suit{ \max_{1\le j \le p}\abs{\sqrt{k_j}\suit{\frac{\hatgamma_j}{\gamma_j^0}-1} }>c_{\alpha} } \\
\ge &  \Pr\suit{ \abs{ \sqrt{k_{\ell}}\suit{\frac{\hatgamma_{\ell}}{\gamma_{\ell}^0}-1} }>c_{\alpha} } \\
= & 1-\Pr \suit{\abs{ \sqrt{k_{\ell}}\suit{\frac{\hatgamma_{\ell}}{\gamma_{\ell}}-1} + \sqrt{k_{\ell}}\delta_{\ell} \frac{\hatgamma_{\ell}}{\gamma_{\ell}}    }\le c_{\alpha}  } \\
\ge & 1-\Pr \suit{\sqrt{k_{\ell}}\suit{\frac{\hatgamma_{\ell}}{\gamma_{\ell}}-1} \le c_{\alpha} -\sqrt{k_{\ell}}\delta_{\ell}\frac{\hatgamma_{\ell}}{\gamma_{\ell}}  }.
\end{aligned}
$$
Denote 
$$
\mathcal{F}_{\varepsilon} = \set{ \abs{\frac{\hatgamma_{\ell}}{\gamma_{\ell}}-1}\le \varepsilon}. 
$$
Note that, for any $\varepsilon>0$, $\Pr(\mathcal{F}_{\varepsilon}) \to 1$. On the set $\mathcal{F}_{\mathcal{\varepsilon}}$, we have that, as $n\to\infty$,
\begin{align*}
	c_{\alpha} -\sqrt{k_{\ell}}\delta_{\ell}\frac{\hatgamma_{\ell}}{\gamma_{\ell}} \le \sqrt{2\log p -\log \log p +q_{\alpha}} - \sqrt{\lambda \log p} (1-\varepsilon) \to -\infty. 
\end{align*} 

Moreover,  note that, as $n\to\infty$,
$$
\sqrt{k_{\ell}}\suit{\frac{\hatgamma_{\ell}}{\gamma_{\ell}}-1} \stackrel{d}{\to} N(0,1).
$$
 Thus, we have that, as $n\to\infty$,
$$
\Pr \suit{\sqrt{k_{\ell}}\suit{\frac{\hatgamma_{\ell}}{\gamma_{\ell}}-1} \le c_{\alpha} -\sqrt{k_{\ell}}\delta_{\ell}\frac{\hatgamma_{\ell}}{\gamma_{\ell}}  } \to 0,
$$
and hence 
$$
\Pr\suit{ \max_{1\le j \le p}\abs{\sqrt{k_j}\suit{\frac{\hatgamma_j}{\gamma_j^0}-1} }>c_{\alpha} } \to 1. 
$$
The proof is then complete.
\end{proof}

\begin{proof}[Proof of Theorem \ref{theorem:power}]
Define 
$$
V_i^{(j,*)} = \sqrt{\frac{n}{k_j}} \suit{ \log X_{i}^{(j)} - \log X_{n-k_j,n}^{(j)}-\gamma_j^0}\mI\suit{X_i^{(j)}>X_{n-k_j,n}^{(j)} }.
$$
Then, given the data $\mbX_1,\dots,\mbX_n$, 
$$
\suit{  \frac{1}{\sqrt{n}}\sum_{i=1}^n \xi_iV_i^{(1,*)},\dots, \frac{1}{\sqrt{n}} \sum_{i=1}^n \xi_n V_i^{(p,*)}}^\top \sim N\suit{0,\widehat{\boldsymbol{\Sigma}}^*},
$$
where $\widehat{\boldsymbol{\Sigma}}^*$ is a $p\times p$ matrix with elements 
$$
\widehat{\boldsymbol{\Sigma}}_{j\ell}^* =\frac{1}{n} \sum_{i=1}^n V_i^{(j,*)}V_i^{(\ell,*)}, \quad 1\le j, \ell \le p.
$$ 
By using the  standard result on Gaussian maximum (see e.g. (B.31) of \cite{chang2017simulation}), we have 
    $$
        c_{\alpha}^B \le \set{\msuit{1+\set{2\log p}^{-1}}\sqrt{2\log p} +\sqrt{2\log (1/\alpha)}}\suit{\max_{1\le j\le p} \frac{\widehat{\boldsymbol{\Sigma}}_{jj}^*}{\suit{\gamma_j^0}^2}}^{1/2}. 
    $$
Write 
$$
\begin{aligned}
    V_i^{(j,*)} = &\sqrt{\frac{n}{k_j}} \suit{ \log X_{i}^{(j)} - \log X_{n-k_j,n}^{(j)}-\gamma_j^0}\mI\suit{X_i^{(j)}>X_{n-k_j,n}^{(j)} } \\
    =& V_i^{(j)} + \sqrt{\frac{n}{k_j}}\suit{\gamma_j-\gamma_j^0}\mI\suit{X_i^{(j)}>X_{n-k_j,n}^{(j)} } 
\end{aligned}
$$
and hence 
$$
\begin{aligned}
    &\frac{1}{n} \sum_{i=1}^n V_i^{(j,*)}V_i^{(\ell,*)}  \\
     =& \frac{1}{n}\sum_{i=1}^n V_i^{(j)}V_i^{(j)} +2 \sqrt{\frac{n}{k_j}}\suit{\gamma_j-\gamma_j^0}\frac{1}{n}\sum_{i=1}^n V_i^{(j)} +\frac{n}{k_j} \suit{\gamma_j-\gamma_j^0}^2  \frac{1}{n}\sum_{i=1}^n \mI\suit{X_i^{(j)}>X_{n-k_j,n}^{(j)} }.
\end{aligned}
$$
By  Lemmas \ref{lemma:bound:W},    \ref{lemma:to:iidsum} and   \ref{lemma:to:covariance}, we have that, as $n\to\infty$, 
$$
\begin{aligned}
&\frac{1}{\suit{\gamma_j^0}^2}\frac{1}{n}\sum_{i=1}^n V_i^{(j)}V_i^{(j)}-\suit{ 1+\delta_j}^2 =o_P(1), \\
&    \frac{1}{\suit{\gamma_j^0}^2} \sqrt{\frac{n}{k_j}}\suit{\gamma_j-\gamma_j^0}\frac{1}{n }\sum_{i=1}^n V_i^{(j)} = O_P(1) \sqrt{\frac{n}{k_j}} \frac{\delta_j}{\gamma_j^0}\frac{\sqrt{\log p}}{\sqrt{n}} =o_P(1), \\
 &   \frac{n}{k_j} \suit{\gamma_j-\gamma_j^0}^2  \frac{1}{n\suit{\gamma_j^0}^2}\sum_{i=1}^n \mI\suit{X_i^{(j)}>X_{n-k_j,n}^{(j)} }= \delta_j^2, 
\end{aligned}
$$
uniformly for all $1\le j\le p$.  Then, we have that, for any $\varepsilon>0$ and 
 sufficiently large $n$, with probability tending to 1, 
\begin{equation}\label{s:bound:c:alpha:boot}
	c_{\alpha}^B \le (1+\varepsilon)\sqrt{2\log p}\sqrt{ (1+\delta)^2 +\delta^2  },
\end{equation}
where 
$\delta  = \max_{1\le j\le p} \abs{\delta_j}.$

In addition, note that, 
$$
\begin{aligned}
  \frac{1}{\gamma_j^0\sqrt{n}}\sum_{i=1}^n V_i^{(j,*)} =&   \frac{1}{\sqrt{n}\gamma_j^0}\sum_{i=1}^n V_i^{(j)} + \frac{1}{\sqrt{n}\gamma_j^0}\sum_{i=1}^n \sqrt{\frac{n}{k_j}}\mI\suit{X_i>X_{n-k_j,n}^{(j)}}\suit{\gamma_j-\gamma_j^0} \\
  =&   \frac{1}{\sqrt{n}\gamma_j^0}\sum_{i=1}^n V_i^{(j)} + \sqrt{k_j}\delta_j.
\end{aligned}
$$
Denote
$
\ell = \arg\max_{j} \abs{\delta_j}.
$
Without loss of generality, we assume that, $\delta_{\ell}>0$. 
It follows that, 
$$
\begin{aligned}
    &\Pr\suit{ \max_{1\le j \le p}\abs{\frac{1}{\sqrt{n}\gamma_j^0}\sum_{i=1}^n V_i^{(j,*)}}>c_{\alpha}^B } \\
\ge &  \Pr\suit{ \abs{\frac{1}{\sqrt{n}\gamma_{\ell}^0}\sum_{i=1}^n V_i^{(\ell,*)}}>c_{\alpha}^B } \\
= & 1-\Pr \suit{\abs{\frac{1}{\sqrt{n}\gamma_{\ell}^0}\sum_{i=1}^n V_i^{(\ell)} + \sqrt{k_{\ell}}\delta_{\ell}}\le c_{\alpha}^B  } \\
\ge & 1-\Pr \suit{\frac{1}{\sqrt{n}\gamma_{\ell}^0}\sum_{i=1}^n V_i^{(\ell)} \le c_{\alpha}^B -\sqrt{k_{\ell}}\delta_{\ell}}\\
=& 1-\Pr \suit{\frac{1}{\sqrt{n} \gamma_{\ell}}\sum_{i=1}^n V_i^{(\ell)} \le -\frac{\gamma_{\ell}^0}{\gamma_{\ell}}\suit{\sqrt{k_{\ell}}\delta_{\ell}-c_{\alpha}^B}  }\\
\end{aligned}
$$

By \eqref{s:bound:c:alpha:boot}, we have that, 
 with probability tending to $1$, 
$$
\begin{aligned}
  \sqrt{k_{\ell}}\delta_{\ell}  -c_{\alpha}^B \ge  &  \sqrt{k_{\ell}}\delta-(1+\varepsilon)\sqrt{2\log p}\sqrt{ (1+\delta)^2 +\delta^2}  \\
  =&  \sqrt{k_{\ell}}\delta \suit{ 1-(1+\varepsilon) \frac{\sqrt{2\log p}}{\sqrt{k_{\ell}} }\sqrt{2+\frac{2}{\delta}+\frac{1}{\delta^2}} } \\
  \ge &  \sqrt{k_{\ell}}\delta \set{ 1- (1+\varepsilon) \frac{\sqrt{2\log p}}{\sqrt{k_{\ell}}} \suit{2 +2\sqrt{\frac{k_{\min}}{\lambda \log p}}+\frac{k_{\min}}{\lambda\log p} }^{1/2} }.
\end{aligned}
$$
Note that, for sufficiently large $n$,  
$$
2 +2\sqrt{\frac{k_{\min}}{\lambda \log p}}+\frac{k_{\min}}{\lambda\log p} \le (1+\varepsilon)  \frac{k_{\min}}{\lambda\log p},
$$
and hence 
$$
  \sqrt{k_{\ell}}\delta_{\ell}  -c_{\alpha}^B \ge  \sqrt{k_{\ell}}\delta\set{1-(1+\varepsilon)^{3/2} \frac{\sqrt{2}}{\sqrt{\lambda}} }.
$$
By taking $\varepsilon$ sufficiently small, we have that, 
$$
1-(1+\varepsilon)^{3/2} \frac{\sqrt{2}}{\sqrt{\lambda}} >0
$$
and hence with probability tending to 1, 
$$
  \sqrt{k_{\ell}}\delta_{\ell}  -c_{\alpha}^B \to\infty.
$$

In addition, 
by Lemma \ref{lemma:to:iidsum}, we have that, as $n\to\infty$,
$$
\frac{1}{\sqrt{n} \gamma_{\ell}}\sum_{i=1}^n V_i^{(\ell)} \stackrel{d}{\to} N(0,1). 
$$
Thus, we have that, as $n\to\infty$,
$$
\Pr \suit{\frac{1}{\sqrt{n} \gamma_{\ell}}\sum_{i=1}^n V_i^{(\ell)} \le -\frac{\gamma_{\ell}^0}{\gamma_{\ell}}\suit{\sqrt{k_{\ell}}\delta_{\ell} - c_{\alpha}^B } }\to 0, 
$$
and hence
$$
\Pr\suit{ \max_{1\le j \le p}\abs{\frac{1}{\sqrt{n}\gamma_j^0}\sum_{i=1}^n V_i^{(j,*)}}>c_{\alpha}^B  } \to 1. 
$$
The proof is then finished.
\end{proof}

\section{Proofs for Section \ref{sec:equal:index}}
\begin{proof}[Proof of Proposition \ref{theorem:average}]
By Lemma \ref{lemma:gamma:diff}, we have that, as $n\to\infty$,
    $$
    \begin{aligned}
       \frac{\overline{\gamma}}{\gamma_0} -1
       =& \frac{1}{p}\sum_{j=1}^p \suit{\frac{\widehat{\gamma}_j(k_j)}{\gamma_0}-1}
        \\
        =& \frac{1}{p}\sum_{j=1}^p \suit{\frac{\widetilde{\gamma}_j(k_j)}{\gamma_0}-1} +\frac{1}{p} \sum_{j=1}^p   \frac{ \widehat{\gamma}_j(k_j) - \widetilde{\gamma}_j(k_j)}{\gamma_0} \\
        =&  \frac{1}{p}\sum_{j=1}^p \suit{\frac{\widetilde{\gamma}_j(k_j)}{\gamma_0}-1}  +o_P(1/\sqrt{k\log p}).
    \end{aligned} 
    $$

    Moreover, we have that,

    $$
    \begin{aligned}
        &\frac{1}{p}\sum_{j=1}^p \suit{\frac{\widetilde{\gamma}_j(k_j)}{\gamma_0}-1} \\
        =&   \frac{1}{p}\sum_{j=1}^p  \suit{\frac{\frac{1}{\gamma_j}\sum_{i=1}^n \log \set{X_i^{(j)}/U_j(n/k_j)} \mI\suit{X_i^{(j)}>U_j(n/k_j)} }{\sum_{i=1}^n \mI\suit{X_i^{(j)}>U_j(n/k_j)}} -1}\\
        =&  \frac{1}{p}\sum_{j=1}^p  \frac{\frac{1}{\gamma_j}\frac{1}{k_j}\sum_{i=1}^n \set{\log \suit{X_i^{(j)}/U_j(n/k_j)}-1} \mI\suit{X_i^{(j)}>U_j(n/k_j)}  }{\widetilde{k}_j/k_j}
    \end{aligned} 
    $$

    By Lemma \ref{lemma:bound:k}, we have that, as $n\to\infty$,
    $$
    \max_{1\le j \le p}\abs{\hatk_j/k_j- 1}=o_P(1).
    $$
Denote 
$$
\begin{aligned}
    G_n:= & \suit{ k  \log p}^{1/2}  \frac{1}{p}\sum_{j=1}^p  \frac{1}{k_j}\sum_{i=1}^n \set{\frac{1}{\gamma_j}\log \suit{X_i^{(j)}/U_j(n/k_j)}-1} \mI\suit{X_i^{(j)}>U_j(n/k_j)}.
\end{aligned}
$$
It suffices to show that, as $n\to\infty$,
    $
    G_n =o_P(1).
    $

    We are going to complete the proof by showing that $\bE G_n =o(1)$ and $\Var (G_n) =o(1)$, as $n\to\infty$.
Similar to the calculation in the proof of Lemma \ref{lemma:remark:covariance},   
 we have that, as $n\to\infty$, 
    $$
    \begin{aligned}
        \bE G_n =&\suit{ k \log p}^{1/2}  \frac{1}{p}\sum_{j=1}^p  \frac{n}{k_j} \int_{U_j(n/k_j)}^{\infty} \frac{1}{\gamma_j} \suit{\log \frac{x}{U_j(n/k_j)}-1} dF_j(x) \\
        =& O(1)  \suit{ k  \log p}^{1/2} \max_{1\le j \le p} \abs{A_j(n/k)},\\ 
        = &o(1),
    \end{aligned}
    $$
   by   Conditions \ref{condition:SOC} and \ref{condition:k:choice}. Moreover, by Conditions  \ref{Eigens} and \ref{eq:k:choice:other}, we have that, as $n\to\infty$,
    $$
    \begin{aligned}
        \Var (G_n) =&   k  \frac{\log p}{p^2n} \Var\suit{\sum_{j=1}^p \frac{Y_i^{(j)}}{\sqrt{k_j/n}}}  \\
        =&  k   \frac{\log p}{p^2n } \sum_{i=1}^p  \Cov\suit{\frac{Y^{(i)}}{\sqrt{k_i/n}}, \sum_{j=1}^p \frac{Y^{(j)}}{\sqrt{k_j/n}} }  \\
        \le & \frac{ k  }{ \min_{1\le j\le p}k_j }   \frac{\log p}{p^2} \sum_{i=1}^p  \Cov(Y^{(i)}, \sum_{j=1}^p Y^{(j)} )  \\
        \le & \frac{ k  }{ \min_{1\le j\le p}k_j }\frac{\log p}{p} \max_{1\le i\le p} \sum_{j=1}^p\sigma_{ij}  \\
        \le &  \frac{ k  }{ \min_{1\le j\le p}k_j } \frac{\log p}{p}\sqrt{pC}, \\
        \to & 0.
    \end{aligned}
    $$
The proof is then complete.
    
    \end{proof}

    \begin{proof}[Proof of Theorem \ref{theorem:identical:size}]
     By  Proposition \ref{theorem:average}, we have that, as $n\to\infty$,
    $$
    \begin{aligned}
        &\mbT^2_*(k_1,\dots,k_p)\\
        =& \max_{1\le j\le p} k_j \suit{\frac{\hatgamma_j(k_j)}{\overline{\gamma}}-1}^2 \\
        =& \frac{1}{\overline{\gamma}^2}\max_{1\le j\le p} k_j\suit{\hatgamma_j(k_j)-\gamma_0+\gamma_0-\overline{\gamma}}^2 \\
        =& \frac{\gamma_0^2}{\overline{\gamma}^2} 
         \max_{1\le j\le p} k_j \set{ \suit{\frac{\hatgamma_j(k_j)}{\gamma_0}-1}^2 -2\suit{ \frac{\hatgamma_j(k_j)}{\gamma_0}-1} \suit{\frac{\overline{\gamma}}{\gamma_0}-1}+\suit{ \frac{\overline{\gamma}}{\gamma_0}-1}^2 }\\ 
        =& \suit{1+o_P(1)\frac{1}{\log p} }
         \max_{1\le j\le p} k_j \set{ \suit{\frac{\hatgamma_j(k_j)}{\gamma_0}-1}^2 -2\suit{ \frac{\hatgamma_j(k_j)}{\gamma_0}-1} \suit{\frac{\overline{\gamma}}{\gamma_0}-1}+\suit{ \frac{\overline{\gamma}}{\gamma_0}-1}^2 }.\\
    \end{aligned}
    $$
    Denote
    $$
    \begin{aligned}
        \max_{1\le j\le p}  k_j \abs{ \suit{ \frac{\hatgamma_j(k_j)}{\gamma_0}-1} \suit{\frac{\overline{\gamma}}{\gamma_0}-1}} 
    \le  \msuit{ \max_{1\le j\le p}\sqrt{k_j } \abs{ \frac{\overline{\gamma}}{\gamma_0} -1}}\set{\max_{1\le j\le p} \sqrt{k_j}\abs{\frac{\hatgamma_j(k_j)}{\gamma_0}-1}}
    =&: I_1 I_2.
    \end{aligned}
    $$
    By  Proposition \ref{theorem:average} and Theorem \ref{theorem:max:i}, we have that, as $n\to\infty$,
    $$
    \begin{aligned}
        I_1 = o_P(1)\frac{1}{\sqrt{\log p}}, \quad 
      I_2 = O_P(1)\sqrt{\log p}, 
    \end{aligned}
    $$
    which leads to
    $$
    \max_{1\le j\le p}  k_j \abs{ \suit{ \frac{\hatgamma_j(k_j)}{\gamma_0}-1} \suit{\frac{\overline{\gamma}}{\gamma_0}-1}}=o_P(1).
    $$
    By Proposition \ref{theorem:average}, we have that,  as $n\to\infty$,
    $$
     \max_{1\le j\le p} k_j\suit{ \frac{\overline{\gamma}}{\gamma_0}-1}^2  =o_P(1).
    $$
  Combining with Theorem \ref{theorem:max:i}, Theorem \ref{theorem:identical:size} is proved.

    \end{proof}

\begin{proof}[Proof of Theorem \ref{theorem:identical:size:boot}]
Define 
$$
V_i^{(j,*)} = \sqrt{\frac{n}{k_j}} \suit{ \log X_{i}^{(j)} - \log X_{n-k_j,n}^{(j)}-\overline{\gamma}}\mI\suit{X_i^{(j)}>X_{n-k_j,n}^{(j)} }.
$$
Then, given the data $\mbX$, 
$$
\suit{  \frac{1}{\sqrt{n}}\sum_{i=1}^n \xi_iV_i^{(1,*)},\dots, \frac{1}{\sqrt{n}} \sum_{i=1}^n \xi_n V_i^{(p,*)}}^\top \sim N\suit{0,\widehat{\boldsymbol{\Sigma}}^*},
$$
where $\widehat{\boldsymbol{\Sigma}}^*$ is a $p\times p$ matrix with elements 
$$
\widehat{\boldsymbol{\Sigma}}_{j\ell}^* =\frac{1}{n} \sum_{i=1}^n V_i^{(j,*)}V_i^{(\ell,*)}, \quad 1\le j, \ell \le p.
$$ 
Write 
$$
\begin{aligned}
    V_i^{(j,*)} = &\sqrt{\frac{n}{k_j}} \suit{ \log X_{i}^{(j)} - \log X_{n-k_j,n}^{(j)}-\overline{\gamma}}\mI\suit{X_i^{(j)}>X_{n-k_j,n}^{(j)} } \\
    =& V_i^{(j)} + \sqrt{\frac{n}{k_j}}\suit{\gamma_j-\overline{\gamma}}\mI\suit{X_i^{(j)}>X_{n-k_j,n}^{(j)} } 
\end{aligned}
$$
and hence 
$$
\begin{aligned}
    &\frac{1}{n} \sum_{i=1}^n V_i^{(j,*)}V_i^{(\ell,*)}  \\
     =& \frac{1}{n}\sum_{i=1}^n V_i^{(j)}V_i^{(j)} +2 \sqrt{\frac{n}{k_j}}\suit{\gamma_j-\overline{\gamma}}\frac{1}{n}\sum_{i=1}^n V_i^{(j)} +\frac{n}{k_j} \suit{\gamma_j-\overline{\gamma}}^2  \frac{1}{n}\sum_{i=1}^n \mI\suit{X_i^{(j)}>X_{n-k_j,n}^{(j)} }.
\end{aligned}
$$

By Lemma \ref{lemma:gamma:diff} and \eqref{s:upper:bound:tilde:gamma}, we have that, 
$$
\max_{1\le j\le p}\sqrt{k_j}\abs{\widehat{\gamma}_j-\gamma_j} =O_P(\sqrt{\log p}),
$$
uniformly for all $1\le j\le p$. Under $H_0^*$, by Condition \ref{eq:k:choice:other}, we have that, 
\begin{align*}
\overline{\gamma} -\gamma_j = \frac{1}{p}\sum_{j=1}^p \suit{\widehat{\gamma}_j-\gamma_j}  = O_P(1) \frac{\sqrt{\log p}}{\sqrt{k}}.
\end{align*}
Then, by  Condition \ref{condition:k:choice:boot}, we have that,  as $n\to\infty$,
$$
\begin{aligned}
	&\sqrt{\frac{n}{k_j}}\suit{\gamma_j-\overline{\gamma}}\frac{1}{n}\sum_{i=1}^n V_i^{(j)}  = O_P(1)  \sqrt{\frac{n}{k_j}}\frac{\sqrt{\log p}}{\sqrt{k }}\frac{1}{\sqrt{n}}\sqrt{\log p} =o_P(1/\log^2 p), \\
&	\frac{n}{k_j} \suit{\gamma_j-\overline{\gamma}}^2  \frac{1}{n}\sum_{i=1}^n \mI\suit{X_i^{(j)}>X_{n-k_j,n}^{(j)} } = o_P(1/\log^2 p),
\end{aligned}
$$
uniformly for all $1\le j\le p$. 
Therefore, 
$$
\widehat{\boldsymbol{\Sigma}}_{j\ell}^* =\frac{1}{n} \sum_{i=1}^n V_i^{(j,*)}V_i^{(\ell,*)} = \frac{1}{n} \sum_{i=1}^n V_i^{(j)}V_i^{(\ell)}+ o_P(1/\log^2 p),
$$
uniformly for all $1\le j\le p$. 
Moreover, we have that, as $n\to\infty$, uniformly for all $1\le j\le p$,  
$$
\begin{aligned}
\frac{1}{\sqrt{n}}\sum_{i=1}^n V_i^{(j,*)} = \frac{1}{\sqrt{n} }\sum_{i=1}^n V_i^{(j)} + \sqrt{k_j}\suit{\overline{\gamma}-\gamma_j} =  \frac{1}{\sqrt{n} }\sum_{i=1}^n V_i^{(j)} +o_P(1/\sqrt{\log p}).
\end{aligned}
$$
The rest of the proof is similar to that of Theorem \ref{Theorem:size:boot}.

\end{proof}

\section{Sufficient conditions for Condition \ref{Eigens} }
In this Section, we provide sufficient conditions for \ref{Eigens} related to the pairwise tail dependence matrix of $\mbX$ \citep{cooley2019decompositions, kiriliouk2022estimating}.   
        
 Assume that, there exist positive    functions $\widetilde{A}_{ij}(t) \to 0$ as $t\to\infty$, and   non-negative functions $R_{ij}$, for $1\le i, j\le p$, 
         such that, for some $T_0>0, \nu>0$, as $t\to\infty$,
\begin{equation}\label{eq:tail:dependence}
          \sup_{x,y \in (0,T_0]^2} (xy)^{-\nu}\abs{R_{t,i,j}(x,y) - R_{ij}(x,y)} = O(1)\widetilde{A}_{ij}(t),
         \end{equation}
         where 
         $$
         R_{t,i,j}(x,y) =t\Pr\set{1-F_i\suit{X^{(i)}}\le  x/t,1-F_j\suit{X^{(j)}}\le  y/t}.
         $$
         and the $O(1)$ term is uniform for all $1\le i, j \le p$. The functions $ R_{ij}$ are called {\it tail dependence functions}.
     Note that for one given pair $(i,j)$, the assumption in \eqref{eq:tail:dependence} is a standard second order condition often assumed in bivariate extreme value statistics, see e.g. \cite{drees1998best} and \cite{beirlant2006statistics}. The quantity $R_{ij}(1,1)$ is called the tail dependence coefficient, which measures the tail dependence between dimensions $i$ and $j$.  Furthermore, we  impose the following conditions on  the pairwise tail dependence matrix $\suit{R_{ij}(1,1)}_{1\leq i, j\leq p}$ and $\widetilde{A}_{ij}(t)$.
        
         \begin{enumerate}[label=(B\arabic*)]
             \setcounter{enumi}{0}
             \item\label{eq:condition:tail:dependence} 
             For some constant $0<c_0<1, C_0>0$,
             $\max_{1\le i<j \le p} R_{i,j}(1,1)\le c_0$ and $\max_{1\le i \le p} \sum_{j=1}^p	 \suit{R_{ij}(1,1)}^2<C_0$, for sufficiently large $n$. 
             \item \label{eq:choice:k:alternative} Choose $k_j, j=1,\dots,p$, such that  $k_j = c_jk$, where $c_j>0$ are positive constants and $0<\min_{1\le j \le p} c_j\le  \max_{1\le j \le p} c_j<\infty$, and $k$ is an intermediate sequence such that as $n\to\infty$, $k \to\infty, k/n\to 0$.
             \item \label{eq:converge:alternative} As $n\to\infty$, $\max_{1\le i\le p}\sum_{j=1}^p \suit{A_i^2(n/k)+A_j^2(n/k)+\widetilde{A}_{ij}^2(n/k)}  = O(1)$.
         \end{enumerate}
        
 Condition \ref{eq:condition:tail:dependence} requires that the pairwise tail dependence matrix  $\suit{R_{ij}(1,1)}_{p\times p}$  is sparse. Condition \ref{eq:choice:k:alternative}  requires choosing $k_j$, $j=1,\dots, p$ at the same order. Condition \ref{eq:converge:alternative} imposes a technical constraint on the functions $A_i$ and $\widetilde{A}_{ij}$, $1\le i, j\le p$.  
The following lemma shows the sufficient conditions for the validity of Condition \ref{Eigens}.
         \begin{lemma}\label{lemma:remark:covariance}
             Condition \ref{Eigens} holds provided that Conditions \ref{condition:SOC}, \ref{eq:condition:tail:dependence}, \ref{eq:choice:k:alternative}, and \ref{eq:converge:alternative} hold.
         \end{lemma}

\begin{proof}[Proof of Lemma \ref{lemma:remark:covariance}]

    We start with calculating the expectation and covariance matrix of $\mbY$.
    The proof is similar to that of Theorem 4 in \cite{stupfler2019relationship}, but additional analysis is required to establish the rate of convergence.
    By Condition \ref{condition:SOC}, we have that,  as $n\to\infty$,
        $$
        \begin{aligned}
            \bE Y^{(j)} = & \sqrt{\frac{n}{k_j}}\bE \set{ \frac{1}{\gamma_j}\log \suit{\frac{X_i^{(j)}}{U_j(n/k_j)}} \mI\suit{X_i^{(j)}>U_j(n/k_j)}}  -\sqrt{\frac{k_j}{n}}  \\
            =& \sqrt{\frac{n}{k_j}} \int_{U_j(n/k_j)}^{\infty} \frac{1}{\gamma_j} \log \suit{\frac{x}{U_j(n/k_j)}} d \overline{F}_j(x) -\sqrt{\frac{k_j}{n}}  \\
            =& \sqrt{\frac{n}{k_j}} \frac{1}{\gamma_j} \int_{1}^{\infty}(1-F_j(xU_j(n/k_j))) \frac{dx}{x}-\sqrt{\frac{k_j}{n}}  \\
            =& \sqrt{\frac{k_j}{n}} \frac{1}{\gamma_j}\int_{1}^{\infty}\frac{1-F_j(xU_j(n/k_j))}{1-F_j(U_j(n/k_j))} \frac{dx}{x}-\sqrt{\frac{k_j}{n}}  \\
            =& \sqrt{\frac{k_j}{n}} \frac{1}{\gamma_j}\int_{1}^{\infty}x^{-1/\gamma_j} \frac{dx}{x} -\sqrt{\frac{k_j}{n}}  + O(1)\sqrt{\frac{k_j}{n}}  A_j(n/k_j) \\
            =& O(1)\sqrt{\frac{k}{n}}  A_j(n/k),
        \end{aligned}
        $$
    where the $O(1)$ term  is uniform for $1\le j \le p$.  
   Similarly, we have that, 
    $$
    \begin{aligned}
        &\bE Y^{(i)}Y^{(j)}  \\
         =&  \sqrt{\frac{n^2}{k_ik_j}}  \frac{1}{\gamma_i\gamma_j}\bE \set{\log \frac{X^{(i)}}{U_i(n/k_i)} \log \frac{X^{(j)}}{U_j(n/k_j)} \mI\suit{X^{(i)}>U_i(n/k_i)}  \mI\suit{X^{(j)}>U_j(n/k_j)}} \\
        &-\frac{1}{\gamma_j}  \sqrt{\frac{n^2}{k_ik_j}} \bE \set{\mI\suit{X^{(i)}>U_i(n/k_i)} \log(X^{(j)}/U_j(n/k_j)) \mI\suit{X^{(j)}>U_j(n/k_j)}} \\
        &-\frac{1}{\gamma_i}\sqrt{\frac{n^2}{k_ik_j}}\bE \set{\mI\suit{X^{(j)}>U_j(n/k_j)} \log(X^{(i)}/U_i(n/k_i)) \mI\suit{X^{(i)}>U_i(n/k_i)}}\\
        &+ \sqrt{\frac{n^2}{k_ik_j}} \Pr\suit{X^{(i)}>U_i(n/k_i),X^{(j)}>U_j(n/k_j)} \\
        =&: I_1-I_2-I_3+I_4.
    \end{aligned}
    $$

    We start with $I_1$.
  Write
    $$
    \begin{aligned}
        I_1 
        =& \frac{1}{\gamma_i\gamma_j} \sqrt{\frac{n^2}{k_ik_j}} \int_{U_j(n/k_j)}^{\infty} \int_{U_i(n/k_i)}^{\infty}\Pr \suit{X^{(i)}>x_i,X^{(j)}>x_j } \frac{1}{x_ix_j} dx_idx_j \\
        =&\frac{1}{\gamma_i\gamma_j}\sqrt{\frac{n^2}{k_ik_j}} \int_{1}^{\infty}\int_{1}^{\infty} \Pr \suit{X^{(i)}>x_iU_i(n/k_i),X^{(j)}>x_jU_j(n/k_j)} \frac{1}{x_ix_j} dx_idx_j \\
        =& \frac{1}{\gamma_i\gamma_j}\sqrt{\frac{n^2}{k_ik_j}} \int_{0}^1 \int_{0}^1 \Pr \suit{X^{(i)}>\frac{U_i(n/k_i)}{x_i},X^{(j)}>\frac{U_j(n/k_j)}{x_j}} \frac{1}{x_ix_j} dx_idx_j \\
        =& \frac{1}{\gamma_i\gamma_j}\sqrt{\frac{k^2}{k_ik_j}} \int_{0}^1 \int_{0}^1 \frac{n}{k}\Pr \set{X^{(i)}> U_i\suit{\frac{n}{ks_i(x_i)}},X^{(j)}>U_j\suit{\frac{n}{ks_j(x_j)}} }\frac{1}{x_ix_j} dx_idx_j \\
        =& \frac{1}{\sqrt{c_ic_j}} \frac{1}{\gamma_i\gamma_j} \int_{0}^1 \int_{0}^1 \frac{n}{k}\Pr \set{X^{(i)}> U_i\suit{\frac{n}{ks_i(x_i)}},X^{(j)}>U_j\suit{\frac{n}{ks_j(x_j)}} }\frac{1}{x_ix_j} dx_idx_j
    \end{aligned}
    $$
    where 
    $$
    \begin{aligned}
        s_i(x_i) = \frac{n}{k} \overline{F}_i\set{\frac{1}{x_i}U_i\suit{\frac{n}{k_i}} } = x_i^{1/\gamma_i}c_i \set{1+O(1)A_i(n/k)}, \\
        s_j(x_j) = \frac{n}{k} \overline{F}_j\set{\frac{1}{x_j}U_j\suit{\frac{n}{k_j}} } = x_j^{1/\gamma_j}c_j \set{1+O(1)A_j(n/k)}, \\
    \end{aligned}
    $$
as $n\to\infty$.

By Condition \eqref{eq:tail:dependence} and Lipschitz continuous of the $R_{ij}$ function, we have that,  
$$
\begin{aligned}
    &\int_{0}^1 \int_{0}^1 \frac{n}{k}\Pr \set{X^{(i)}> U_i\suit{\frac{n}{ks_i(x_i)}},X^{(j)}>U_j\suit{\frac{n}{ks_j(x_j)}} }\frac{1}{x_ix_j} dx_idx_j \\
=& \int_{0}^1 \int_{0}^1  R(s_i(x_i), s_j(x_j)) \frac{1}{x_ix_j} dx_idx_j \\
& +O(1) \sup_{x,y\in (0,T_0]^2} \frac{\abs{R_{n/k,i,j}(x,y) -R_{ij}(x,y)}}{(xy)^{\nu}} \int_{0}^1 \int_0^1  (s_i(x_i) s_j(x_j))^{\nu}\frac{1}{x_i}\frac{1}{x_j} dx_idx_j \\
=&  \int_{0}^1 \int_{0}^1  R(x_i^{1/\gamma_i}c_i, x_j^{1/\gamma_j}c_j) \frac{1}{x_ix_j} dx_idx_j +O(1)\suit{ \abs{\widetilde{A}_{ij}(n/k)}+\abs{A_i(n/k)} +\abs{A_j(n/k) }}.
\end{aligned}
$$
By the homogeneity of $R_{ij}$, we have that, 
$$
\begin{aligned}
    & \frac{1}{\gamma_i\gamma_j}\int_{0}^{1} \int_{0}^{1} R_{ij}(c_ix_i^{1/\gamma_i}, c_jx_j^{^{1/\gamma_j}})\frac{1}{x_ix_j} dx_idx_j \\
= & \int_{0}^1 \int_{0}^{1} \frac{R_{ij}(c_iu,c_jv)}{uv}dudv \\
=& \int_{0}^1 \int_{0}^v \frac{R_{ij}(c_iu,c_jv)}{uv}dudv  + \int_{0}^1 \int_{0}^u \frac{R_{ij}(c_iu,c_jv)}{uv}dvdu  \\
=& \int_{0}^1 \int_{0}^1 \frac{R_{ij}(c_iu,c_j)}{u}dudv + \int_{0}^1 \int_{0}^1 \frac{R_{ij}(c_i,c_jv)}{v}dvdu\\
=& \int_{0}^1 \frac{R_{ij}(c_iu,c_j)}{u}du +\int_{0}^1 \frac{R_{ij}(c_i,c_jv)}{v}dv.
\end{aligned}
$$
Thus, we conclude that, 
$$
I_1 = \frac{1}{\sqrt{c_ic_j}} \suit{\int_{0}^1 \frac{R_{ij}(c_iu,c_j)}{u}du +\int_{0}^1 \frac{R_{ij}(c_i,c_jv)}{v}dv} +O(1)\suit{ \abs{\widetilde{A}_{ij}(n/k)}+\abs{A_i(n/k)} +\abs{A_j(n/k) }}.
$$

    Similarly, we can show that, as $n\to\infty$,
    $$
    \begin{aligned}
        I_2 = &  \frac{1}{\sqrt{c_ic_j}}\int_{0}^1 \frac{R_{ij}(c_iu,c_j)}{u}du +O(1)\suit{ \abs{\widetilde{A}_{ij}(n/k)}+\abs{A_i(n/k)} +\abs{A_j(n/k) }},\\
        I_3 = &\frac{1}{\sqrt{c_ic_j}} \int_{0}^1 \frac{R_{ij}(c_i,c_jv)}{v}dv +O(1)\suit{ \abs{\widetilde{A}_{ij}(n/k)}+\abs{A_i(n/k)} +\abs{A_j(n/k) }},\\
        I_4 = & \frac{1}{\sqrt{c_ic_j}}R_{i,j}(c_i,c_j) + O(1)\suit{ \abs{\widetilde{A}_{ij}(n/k)}+\abs{A_i(n/k)} +\abs{A_j(n/k) }}.
    \end{aligned}
    $$

  Combining the results for $I_1, I_2, I_3$ and $I_4$,  we have that, as $n\to\infty$,
    $$
    \begin{aligned}
        \text{Cov}\suit{ Y^{(i)},Y^{(j)}}  =& \frac{1}{\sqrt{c_ic_j}}R_{i,j}(c_i,c_j)+ O(1)\suit{ \abs{\widetilde{A}_{ij}(n/k)}+\abs{A_i(n/k)} +\abs{A_j(n/k) }}. 
    \end{aligned}
    $$
    
The proof can be completed by noting that, 
$$
\begin{aligned}
    \frac{1}{\sqrt{c_ic_j}}R_{ij}(c_i,c_j) = & \frac{\max(c_i,c_j)}{\sqrt{c_ic_j}}  R_{ij}\suit{\frac{c_i}{\max(c_i,c_j)}, \frac{c_j}{\max(c_i,c_j)}} \\ 
    \le &\frac{\max(c_i,c_j)}{\sqrt{c_ic_j}} R_{i,j}(1,1) \\
    \le & \frac{\max_{1\le j \le p} c_j}{\min_{1\le j \le p} c_j}R_{i,j}(1,1).
\end{aligned}
$$

\end{proof}

\bibliographystyle{apalike}

\bibliography{mybib}

@article{sibuya1960bivariate,
  title={Bivariate extreme statistics},
  author={Sibuya, Masaaki},
  journal={Annals of the Institute of Statistical Mathematics},
  volume={11},
  number={2},
  pages={195--210},
  year={1960},
  publisher={Tokyo}
}

@article{tang2022conditional,
  title={Conditional marginal test for high dimensional quantile regression},
  author={Tang, Yanlin and Wang, Yinfeng and Wang, Huixia Judy and Pan, Qing},
  journal={Statistica Sinica},
  volume={32},
  number={2},
  pages={869--892},
  year={2022},
  publisher={JSTOR}
}

@article{feng2022high,
  title={High-dimensional test for alpha in linear factor pricing models with sparse alternatives},
  author={Feng, Long and Lan, Wei and Liu, Binghui and Ma, Yanyuan},
  journal={Journal of Econometrics},
  volume={229},
  number={1},
  pages={152--175},
  year={2022},
  publisher={Elsevier}
}

@article{ma2021global,
  title={Global and simultaneous hypothesis testing for high-dimensional logistic regression models},
  author={Ma, Rong and Tony Cai, T and Li, Hongzhe},
  journal={Journal of the American Statistical Association},
  volume={116},
  number={534},
  pages={984--998},
  year={2021},
  publisher={Taylor \& Francis}
}

@article{xue2020distribution,
  title={Distribution and correlation-free two-sample test of high-dimensional means},
  author={Xue, Kaijie and Yao, Fang},
  journal={Annals of Statistics},
  volume={48},
  number={3},
  pages={1304--1328},
  year={2020},
  publisher={JSTOR}
}

@article{chang2017comparing,
  title={Comparing large covariance matrices under weak conditions on the dependence structure and its application to gene clustering},
  author={Chang, Jinyuan and Zhou, Wen and Zhou, Wen-Xin and Wang, Lan},
  journal={Biometrics},
  volume={73},
  number={1},
  pages={31--41},
  year={2017},
  publisher={Wiley Online Library}
}

@article{tang2024high,
  title={High-Dimensional Extreme Quantile Regression},
  author={Tang, Yiwei and Wang, Judy Huixia and Li, Deyuan},
  journal={arXiv preprint arXiv:2411.13822},
  year={2024}
}

@article{chang2023testing,
  title={Testing the martingale difference hypothesis in high dimension},
  author={Chang, Jinyuan and Jiang, Qing and Shao, Xiaofeng},
  journal={Journal of Econometrics},
  volume={235},
  number={2},
  pages={972--1000},
  year={2023},
  publisher={Elsevier}
}

@article{sasaki2024high,
  title={High-Dimensional Tail Index Regression: with An Application to Text Analyses of Viral Posts in Social Media},
  author={Sasaki, Yuya and Tao, Jing and Wang, Yulong},
  journal={arXiv preprint arXiv:2403.01318},
  year={2024}
}

@article{chernozhukov2023high,
  title={High-dimensional data bootstrap},
  author={Chernozhukov, Victor and Chetverikov, Denis and Kato, Kengo and Koike, Yuta},
  journal={Annual Review of Statistics and Its Application},
  volume={10},
  number={1},
  pages={427--449},
  year={2023},
  publisher={Annual Reviews}
}

@article{chang2017simulation,
  title={Simulation-based hypothesis testing of high dimensional means under covariance heterogeneity},
  author={Chang, Jinyuan and Zheng, Chao and Zhou, Wen-Xin and Zhou, Wen},
  journal={Biometrics},
  volume={73},
  number={4},
  pages={1300--1310},
  year={2017},
  publisher={Wiley Online Library}
}

@article{engelke2021learning,
  title={Learning extremal graphical structures in high dimensions},
  author={Engelke, Sebastian and Lalancette, Micha{\"e}l and Volgushev, Stanislav},
  journal={arXiv preprint arXiv:2111.00840},
  year={2021}
}

@article{lederer2023extremes,
  title={Extremes in high dimensions: Methods and scalable algorithms},
  author={Lederer, Johannes and Oesting, Marco},
  journal={arXiv preprint arXiv:2303.04258},
  year={2023}
}

@article{engelke2025extremal,
  title={Extremal graphical modeling with latent variables via convex optimization},
  author={Engelke, Sebastian and Taeb, Armeen},
  journal={Journal of Machine Learning Research},
  volume={26},
  number={42},
  pages={1--68},
  year={2025}
}

@article{boulin2025high,
  title={High-dimensional variable clustering based on maxima of a weakly dependent random process},
  author={Boulin, Alexis and Di Bernardino, Elena and Lalo{\"e}, Thomas and Toulemonde, Gwladys},
  journal={Journal of the American Statistical Association},
  pages={1--21},
  year={2025},
  note ={https://doi.org/10.1080/01621459.2025.2459443}

}

@article{butsch2025estimation,
  title={Estimation of the number of principal components in high-dimensional multivariate extremes},
  author={Butsch, Lucas and Fasen-Hartmann, Vicky},
  journal={arXiv preprint arXiv:2505.22437},
  year={2025}
}

@article{giessing2023bootstrap,
  title={A bootstrap hypothesis test for high-dimensional mean vectors},
  author={Giessing, Alexander and Fan, Jianqing},
  journal={arXiv preprint arXiv:2309.01254},
  year={2023}
}

@article{chernozhuokov2022improved,
  title={Improved central limit theorem and bootstrap approximations in high dimensions},
  author={Chernozhuokov, Victor and Chetverikov, Denis and Kato, Kengo and Koike, Yuta},
  journal={Annals of Statistics},
  volume={50},
  number={5},
  pages={2562--2586},
  year={2022},
  publisher={Institute of Mathematical Statistics}
}

@article{cai2013two,
  title={Two-sample covariance matrix testing and support recovery in high-dimensional and sparse settings},
  author={Cai, Tony and Liu, Weidong and Xia, Yin},
  journal={Journal of the American Statistical Association},
  volume={108},
  number={501},
  pages={265--277},
  year={2013},
  publisher={Taylor \& Francis}
}

@article{tony2014two,
  title={Two-sample test of high dimensional means under dependence},
  author={Cai, T. Tony and Liu, Weidong and Xia, Yin},
  journal={Journal of the Royal Statistical Society Series B: Statistical Methodology},
  volume={76},
  number={2},
  pages={349--372},
  year={2014},
  publisher={Oxford University Press}
}

@article{cai2011estimation,
  title={Estimation of extreme risk regions under multivariate regular variation},
  author={Cai, J and Einmahl, JHJ and de Haan, LFM},
  journal={Annals of Statistics},
  volume={39},
  number={3},
  pages={1803--1826},
  year={2011}
}

@article{zaitsev1987gaussian,
  title={{On the Gaussian approximation of convolutions under multidimensional analogues of SN Bernstein's inequality conditions}},
  author={Zaitsev, A Yu},
  journal={Probability Theory and Related Fields},
  volume={74},
  number={4},
  pages={535--566},
  year={1987},
  publisher={Springer-Verlag Berlin/Heidelberg}
}

@article{hill1975simple,
  title={A simple general approach to inference about the tail of a distribution},
  author={Hill, Bruce M},
  journal={Annals of Statistics},
    volume={3},
  number={5},
  pages={1163--1174},
  year={1975}
}

@book{haan2006extreme,
  title={Extreme Value Theory: an Introduction},
  author={de Haan, Laurens and Ferreira, Ana},
  year={2006},
  publisher={Springer}
}

@article{stupfler2019relationship,
  title={On a relationship between randomly and non-randomly thresholded empirical average excesses for heavy tails},
  author={Stupfler, Gilles},
  journal={Extremes},
  volume={22},
  number={4},
  pages={749--769},
  year={2019},
  publisher={Springer}
}

@book{shorack1986empirical,
  title={Empirical Processes with Applications to Statistics},
  author={Shorack, Galen R and Wellner, Jon A},
  year={1986},
  publisher={Wiley}
}

@article{potter1942mean,
  title={{The mean values of certain Dirichlet series, II}},
  author={Potter, HSA},
  journal={Proceedings of the London Mathematical Society},
  volume={2},
  number={1},
  pages={1--19},
  year={1942},
  publisher={Wiley Online Library}
}

@article{daouia2023optimal,
  title={Optimal weighted pooling for inference about the tail index and extreme quantiles},
  author={Daouia, Abdelaati and Padoan, Simone A and Stupfler, Gilles and others},
  journal={Bernoulli},
  volume         = {20},
  number         = {2},
   pages={1287--1312},
  year={2024}
}

@book{beirlant2006statistics,
  title={Statistics of extremes: theory and applications},
  author={Beirlant, Jan and Goegebeur, Yuri and Segers, Johan and Teugels, Jozef L},
  year={2006},
  publisher={John Wiley \& Sons}
}

@article{feng2022asymptotic,
  title={Asymptotic independence of the sum and maximum of dependent random variables with applications to high-dimensional tests},
  author={Feng, Long and Jiang, Tiefeng and Li, Xiaoyun and Liu, Binghui},
  journal={arXiv preprint arXiv:2205.01638},
  year={2022}
}

@article{engelke2022structure,
  title={Structure learning for extremal tree models},
  author={Engelke, Sebastian and Volgushev, Stanislav},
  journal={Journal of the Royal Statistical Society Series B: Statistical Methodology},
  volume={84},
  number={5},
  pages={2055--2087},
  year={2022},
  publisher={Oxford University Press}
}

@article{wan2023graphical,
  title={Graphical lasso for extremes},
  author={Wan, Phyllis and Zhou, Chen},
  journal={arXiv preprint arXiv:2307.15004},
  year={2023}
}

@inproceedings{fisher1928limiting,
  title={Limiting forms of the frequency distribution of the largest or smallest member of a sample},
   author={Fisher, RA and Tippett, LHC},
  booktitle={Mathematical Proceedings of the Cambridge Philosophical Society},
  pages={180--190},
  year={1928},
  organization={Cambridge University Press}
}

@article{gnedenko1943distribution,
  title={Sur la distribution limite du terme maximum d'une serie aleatoire},
  author={Gnedenko, B.V.},
  journal={Annals of Mathematics},
  volume          = {44},
  number          = {3},
  pages={423--453},
  year={1943},
  publisher={JSTOR}
}

@article{cooley2019decompositions,
  title={Decompositions of dependence for high-dimensional extremes},
  author={Cooley, Daniel and Thibaud, Emeric},
  journal={Biometrika},
  volume={106},
  number={3},
  pages={587--604},
  year={2019},
  publisher={Oxford University Press}
}

@article{kiriliouk2022estimating,
  title={Estimating probabilities of multivariate failure sets based on pairwise tail dependence coefficients},
  author={Kiriliouk, Anna and Zhou, Chen},
  journal={arXiv preprint arXiv:2210.12618},
  year={2022}
}

@article{kinsvater2016regional,
  title={Regional extreme value index estimation and a test of tail homogeneity},
  author={Kinsvater, Paul and Fried, Roland and Lilienthal, Jona},
  journal={Environmetrics},
  volume={27},
  number={2},
  pages={103--115},
  year={2016},
  publisher={Wiley Online Library}
}

@article{mainik2015portfolio,
  title={Portfolio optimization for heavy-tailed assets: Extreme Risk Index vs. Markowitz},
  author={Mainik, Georg and Mitov, Georgi and R{\"u}schendorf, Ludger},
  journal={Journal of Empirical Finance},
  volume={32},
  pages={115--134},
  year={2015},
  publisher={Elsevier}
}

@book{resnick2008extreme,
  title={Extreme Values, Regular Variation, and Point Processes},
  author={Resnick, Sidney I},
  year={2008},
  publisher={Springer Science \& Business Media}
}

@article{buishand2008spatial,
  title={On Spatial Extremes: with Application to a Rainfall Problem},
  author={Buishand, TA and de Haan, L and Zhou, C},
  journal={Annals of Applied Statistics},
  volume          = {2},
  number          = {2},
  pages={624--642},
  year={2008},
  publisher={JSTOR}
}

@article{fuentes2013nonparametric,
  title={Nonparametric spatial models for extremes: Application to extreme temperature data},
  author={Fuentes, Montserrat and Henry, John and Reich, Brian},
  journal={Extremes},
  volume={16},
  pages={75--101},
  year={2013},
  publisher={Springer}
}

@article{drees1998best,
  title={Best attainable rates of convergence for estimators of the stable tail dependence function},
  author={Drees, Holger and Huang, Xin},
  journal={Journal of Multivariate Analysis},
  volume={64},
  number={1},
  pages={25--46},
  year={1998},
  publisher={Elsevier}
}

@article{kiefer1967bahadur,
  title={{On Bahadur's representation of sample quantiles}},
  author={Kiefer, Jack},
  journal={Annals of Mathematical Statistics},
  volume={38},
  number={5},
  pages={1323--1342},
  year={1967},
  publisher={JSTOR}
}

@article{Hector2023dis,
author = {Emily C. Hector and Brian J. Reich},
title = {Distributed Inference for Spatial Extremes Modeling in High Dimensions},
journal = {Journal of the American Statistical Association},
pages = {1297--1308},
year = {2024},
volume = {119},
number = {546}
}

@article{Huang2022review,
title = {An overview of tests on high-dimensional means},
journal = {Journal of Multivariate Analysis},
volume = {188},
pages = {104813},
year = {2022},
author = {Yuan Huang and Changcheng Li and Runze Li and Songshan Yang},
}

@article{skorski2023bernstein,
  title={Bernstein-type bounds for beta distribution},
  author={Skorski, Maciej},
  journal={Modern Stochastics: Theory and Applications},
  volume={10},
  number={2},
  pages={211--228},
  year={2023},
  publisher={VTeX: Solutions for Science Publishing}
}

@article{fan2008sure,
  title={Sure independence screening for ultrahigh dimensional feature space},
  author={Fan, Jianqing and Lv, Jinchi},
  journal={Journal of the Royal Statistical Society Series B: Statistical Methodology},
  volume={70},
  number={5},
  pages={849--911},
  year={2008},
  publisher={Oxford University Press}
}

@article{wang2012quantile,
  title={Quantile Regression for Analyzing Heterogeneity in Ultra-high Dimension.},
  author={Wang, L and Wu, Y and Li, R},
  journal={Journal of the American Statistical Association},
  volume={107},
  number={497},
  pages={214--222},
  year={2012}
}

@article{gribkova2012bahadur,
  title={{On a Bahadur--Kiefer representation of von Mises statistic type for intermediate sample quantiles}},
  author={Gribkova, NV and Helmers, Roelof},
  journal={Probability and Mathematical Statistics},
  volume={32},
  number={2},
  pages={255--279},
  year={2012}
}

@article{hoeffding1963probability,
  title={Probability Inequalities for Sums of Bounded Random Variables},
  author={Hoeffding, Wassily},
  journal={Journal of the American Statistical Association},
  volume={58},
  number={301},
  pages={13--30},
  year={1963},
  publisher={Taylor \& Francis}
}

\end{document}